\documentclass[twocolumn]{aastex62}%twocolumn
\usepackage{CJK}
\usepackage{natbib}
\usepackage{graphicx}
\usepackage {amsmath,amssymb,eqnarray}
\usepackage{enumerate}
\usepackage{enumitem}
\usepackage{mathtools}
\usepackage{lineno}

\newcommand{\hi}{{\rm H}{\textsc i}}

\newcommand{\degree}{\ensuremath{^\circ}}

\begin{document}
\begin{CJK*}{UTF8}{gbsn}

%\linenumbers
\title{ WALLABY Pilot Survey: the diversity of ram pressure stripping of the galactic HI gas in the Hydra Cluster  }

\author[0000-0002-6593-8820]{Jing Wang (王菁)}
\affiliation{ Kavli Institute for Astronomy and Astrophysics, Peking University, Beijing 100871, China}

\author{Lister Staveley-Smith}
\affiliation{International Centre for Radio Astronomy Research, University of Western Australia, 35 Stirling Highway, Crawley, WA 6009, Australia}
\affiliation{ARC Centre of Excellence for All-Sky Astrophysics in 3 Dimensions (ASTRO 3D), Australia}

\author{Tobias Westmeier}
\affiliation{International Centre for Radio Astronomy Research, University of Western Australia, 35 Stirling Highway, Crawley, WA 6009, Australia}
\affiliation{ARC Centre of Excellence for All-Sky Astrophysics in 3 Dimensions (ASTRO 3D), Australia}

\author{Barbara Catinella}
\affiliation{International Centre for Radio Astronomy Research, University of Western Australia, 35 Stirling Highway, Crawley, WA 6009, Australia}
\affiliation{ARC Centre of Excellence for All-Sky Astrophysics in 3 Dimensions (ASTRO 3D), Australia}

\author[0000-0003-2015-777X]{Li Shao (邵立)}
\affiliation{National Astronomical Observatories, Chinese Academy of Sciences, 20A Datun Road, Chaoyang District, Beijing, China}

\author{T.N. Reynolds}
\affiliation{International Centre for Radio Astronomy Research, University of Western Australia, 35 Stirling Highway, Crawley, WA 6009, Australia}
\affiliation{ARC Centre of Excellence for All-Sky Astrophysics in 3 Dimensions (ASTRO 3D), Australia}

\author{Bi-Qing For}
\affiliation{International Centre for Radio Astronomy Research, University of Western Australia, 35 Stirling Highway, Crawley, WA 6009, Australia}3
\affiliation{ARC Centre of Excellence for All-Sky Astrophysics in 3 Dimensions (ASTRO 3D), Australia}

\author{Bumhyun Lee}
\affiliation{ Kavli Institute for Astronomy and Astrophysics, Peking University, Beijing 100871, China}

\author{Ze-zhong Liang (梁泽众)}
\affiliation{Department of Astronomy, School of Physics, Peking University, Beijing 100871, People's Republic of China}

\author{Shun Wang (王舜)}
\affiliation{ Kavli Institute for Astronomy and Astrophysics, Peking University, Beijing 100871, China}
\affiliation{Department of Astronomy, School of Physics, Peking University, Beijing 100871, People's Republic of China}

%%%%

%%%%###
\author{A. Elagali}
\affiliation{Telethon Kids Institute, Perth Children’s Hospital, Perth, Australia}

\author{H. D{\'e}nes}
\affiliation{ASTRON - The Netherlands Institute for Radio Astronomy, 7991 PD Dwingeloo, The Netherlands}

\author{D. Kleiner}
\affiliation{INAF - Osservatorio Astronomico di Cagliari, Via della Scienza 5, I-09047 Selargius (CA), Italy}

\author{B\"arbel S. Koribalski}
\affiliation{Australia Telescope National Facility, CSIRO Astronomy and Space Science, P.O. Box 76, NSW 1710, Epping, Australia}
\affiliation{Western Sydney University, Locked Bag 1797, Penrith, NSW 2751, Australia}

\author{K. Lee-Waddell}
\affiliation{International Centre for Radio Astronomy Research, University of Western Australia, 35 Stirling Highway, Crawley, WA 6009, Australia}
\affiliation{CSIRO Astronomy and Space Science (CASS), PO Box 1130, Bentley, WA 6102, Australia}

\author{S-H. Oh}
\affiliation{Department of Physics and Astronomy, Sejong University, 209 Neungdong-ro, Gwangjin-gu, Seoul, Republic of Korea}

\author{J. Rhee}
\affiliation{International Centre for Radio Astronomy Research, University of Western Australia, 35 Stirling Highway, Crawley, WA 6009, Australia}
\affiliation{ARC Centre of Excellence for All-Sky Astrophysics in 3 Dimensions (ASTRO 3D), Australia}

\author{P. Serra}
\affiliation{INAF - Osservatorio Astronomico di Cagliari, Via della Scienza 5, I-09047 Selargius (CA), Italy}

\author{K. Spekkens}
\affiliation{Royal Military College of Canada, PO Box 17000, Station Forces, Kingston, Ontario, Canada K7K7B4}

\author{O. I. Wong}
\affiliation{CSIRO Astronomy and Space Science (CASS), PO Box 1130, Bentley, WA 6102, Australia}
\affiliation{International Centre for Radio Astronomy Research, University of Western Australia, 35 Stirling Highway, Crawley, WA 6009, Australia}
\affiliation{ARC Centre of Excellence for All-Sky Astrophysics in 3 Dimensions (ASTRO 3D), Australia}

%%%%###
\author{K. Bekki}
\affiliation{International Centre for Radio Astronomy Research, University of Western Australia, 35 Stirling Highway, Crawley, WA 6009, Australia}
\affiliation{ARC Centre of Excellence for All-Sky Astrophysics in 3 Dimensions (ASTRO 3D), Australia}

\author{F. Bigiel}
\affiliation{Argelander-Institut f\"ur Astronomie, Universit\"at Bonn, Auf dem H\"ugel 71, 53121 Bonn, Germany}

\author{ H.M. Courtois}
\affiliation{ Univ Lyon, Univ Claude Bernard Lyon 1, IUF, IP2I Lyon, F-69622, Villeurbanne, France}

\author{Kelley M. Hess}
\affiliation{ASTRON, the Netherlands Institute for Radio Astronomy, Postbus 2, 7990 AA, Dwingeloo, The Netherlands}
\affiliation{Kapteyn Astronomical Institute, University of Groningen, P.O. Box 800, 9700 AV Groningen, The Netherlands} 

\author{B.W. Holwerda}
\affiliation{University of Louisville, Department of Physics and Astronomy, 102 Natural Science Building, 40292 KY Louisville, USA}

\author{Kristen B.W. McQuinn}
\affiliation{Rutgers University, Department of Physics and Astronomy, 136 Frelinghuysen Road, Piscataway, NJ 08854, USA}

\author{ M. Pandey-Pommier}
\affiliation{University Claude Bernard Lyon 1, 43 Boulevard du 11 Novembre 1918, 69100 Villeurbanne, France}

\author{J.M. van der Hulst}
\affiliation{ Kapteyn Astronomical Institute, University of Groningen}

\author{L. Verdes-Montenegro}
\affiliation{ Instituto de Astrofísica de Andalucía (CSIC) }

\begin{abstract}
This study uses $\hi$ image data from the WALLABY pilot survey with the ASKAP telescope, covering the Hydra cluster out to 2.5$r_{200}$.
We present the projected phase-space distribution of $\hi$-detected galaxies in Hydra, and identify that nearly two thirds of the galaxies within $1.25r_{200}$ may be in the early stages of ram pressure stripping.
 More than half of these may be only weakly stripped, with the ratio of strippable $\hi$ (i.e., where the galactic restoring force is lower than the ram pressure in the disk) mass fraction (over total $\hi$ mass) distributed uniformly below 90\%. Consequently, the $\hi$ mass is expected to decrease by only a few 0.1 dex after the currently strippable portion of $\hi$ in these systems has been stripped. A more detailed look at the subset of galaxies that are spatially resolved by WALLABY observations shows that, while it typically takes less than 200 Myr for ram pressure stripping to remove the currently strippable portion of $\hi$, it may take more than 600 Myr to significantly change the total $\hi$ mass. Our results provide new clues to understanding the different rates of $\hi$ depletion and star formation quenching in cluster galaxies. 

\end{abstract}


\section{Introduction} %600
\label{sec:introduction}
Galaxies evolve in their morphologies, kinematics and stellar population, a process which is accelerated when they are in clusters \citep{Boselli06}. Neutral atomic hydrogen ($\hi$) is a major part of the interstellar medium (ISM) in a galaxy \citep[e.g.,][]{Catinella18, Wang20a}, and is a crucial component in the kinematic and thermal cooling of baryons \citep{Putman12}. It is also an important step in the baryonic mass flow, and therefore an important key for understanding galaxy evolution. Clusters provide an environment where gravitational and hydrodynamic effects efficiently remove the $\hi$ in their constituent galaxies \citep[e.g.,][]{Boselli06, Stevens19b}, with ram pressure stripping identified as one of the most important mechanisms at low redshift. 

Based on the analytical model of \citet{GunnGott72}, the strength of ram pressure stripping can be quantified by comparing the ram pressure from the intra-cluster medium (ICM) against the localized gravitational anchoring force of the galactic disk. We can thus expect that there is a diversity of the way that galaxies experience ram pressure stripping, as infalling galaxies have different distributions of mass and gas, and travel along different orbits and with different disk inclinations against the ICM wind. Hydrodynamical simulations of cluster systems confirm these complexities \citep{Tonnesen19, Lotz19, Bekki14, Jachym09, Roediger07, Roediger06, Vollmer01, Abadi99} and further elaborate on the influence of other factors such as the multi-phase nature of gas \citep{Stevens20, Lee20, Tonnesen10, Tonnesen09}, magnetic fields \citep{RamosMartnez18,Tonnesen14}, and sub-structures in clusters \citep{Ruggiero19, Tonnesen08}. Probing the observational dependence of ram pressure stripping on galaxy properties requires statistically mapping the $\hi$ in cluster galaxies with high resolution. Ram pressure stripping is also studied with other tracers like the ionized gas \citep{Jaffe18} and the radio continuum \citep{Chen20}, but this paper focuses on the ram pressure stripping of $\hi$ which is the reservoir for star formation. 

Galaxies displaying $\hi$ tails that resemble expected ram pressure stripping morphologies have been identified in nearby clusters \citep[e.g.][]{Kenney04, Chung07, Chung09, Koribalski20b}. The shape, length and column density of those $\hi$ tails provide information regarding the progress of gas removal, orbits of infall, time since infall, and local ICM structures, particularly when they are combined with multi-wavelength information like the star formation rate \citep{Jaffe16,Vollmer12}, gas in other phases \citep{Lee17, Abramson11, Moretti20}, dust \citep{Crowl05}, radio polarization \citep{Vollmer13}, and numerical simulations \citep{Vollmer01, Tonnesen10}. 
Statistically, ram pressure stripping provides a good explanation for the observed distributions of star formation rates (SFR) and $\hi$-richness in clusters, including trends as a function of the cluster-centric projected distance \citep{Gavazzi06, Woo13, Hess15}, cluster-centric radial velocity offset \citep{Jaffe15, Mahajan11}, and galactic stellar mass \citep{Zhang13}. However, detailed studies of resolved systems have been mostly limited to a few systems, whilst statistical studies have mostly been based on spatially unresolved $\hi$ data.

Contiguously mapping the $\hi$ in clusters is essential to bridge the gap between signatures of ram pressure stripping in individual galaxies and the role that ram pressure stripping plays in cosmological galaxy evolution. The need for a cosmological context is because the majority of the galaxies infalling for the first time into massive clusters may have been pre-processed as satellites \citep{Fujita04, Cybulski14, Bahe19}, or undergone ``mass quenching'' as centrals \citep{Kauffmann03} in less massive groups. It has been found that the timescale for quenching the SFR anti-correlates with the stellar mass of satellite galaxies, not because the more massive galaxies are more vulnerable to the current environment, but because they were fully or partly pre-processed or mass quenched for a longer time in their previous environments \citep{DeLucia12, Wetzel13, Oman16, Rhee20}. It has also been found that 20-50\% of low-mass satellite galaxies may have already or partly been quenched in star formation, with gas depleted to some extent, in another dark matter halo before being accreted into the current cluster \citep{Wetzel13, Hess13, Haines15, Jung18}. 
 The galactic properties shaped by the past evolution strongly affect the strength and importance of environmental processing in the current cluster \citep{Jung18}. A complete census of cluster galaxies is needed in order to properly account for the diversity of initial conditions at infall. 

However, it has been hard to achieve both high completeness and high resolution in $\hi$ observations for clusters, mostly because covering a large area with interferometric 21 cm observations has been unfeasible. Extensive statistical studies based on more complete samples of galaxies detected in the nearby clusters Virgo, Coma, Abell 1367 and others have been conducted using $\hi$ data from blind single dish $\hi$ surveys, particularly ALFALFA \citep{Haynes18} and HIPASS \citep{Meyer04}. They provide the bench mark for integral properties of $\hi$ in massive clusters, including the major scaling relations \citep{Cortese08, Cortese11, Denes14, Odekon16} and mass functions \citep{Gavazzi06, Jones16}. Interferometric $\hi$ images of selected galaxies from these $\hi$ samples were also obtained to gain details about ongoing physical processes \citep{Warmels88, Gavazzi89, Cayatte90, Scott10, Scott18}. Among them, the VIVA targeted survey of 50 late-type galaxies in the Virgo cluster \citep{Chung09} has been one of the largest resolved datasets. 
Further away, BUDHIES mapped two clusters \citep{Jaffe15} at relatively high redshifts ($z\sim0.2$) out to $3r_{200}$ (where $r_{200}$ is the radius within which the averaged density is 200 times the critical density of the universe) with interferometry, with poorer resolution and mass sensitivity. \citet{Wang20b} thus used the predicted $\hi$ radial distributions, in order to better exploit the low-resolution, wide-field $\hi$ data. Based on a relatively complete overlap between the ROSAT X-ray survey and the ALFALFA $\hi$ survey, for 26 massive clusters and around 200 galaxies, they showed that ram pressure stripping of the $\hi$ outer disks is prevalent out to 1.5 $r_{200}$ in Coma-like clusters, pointing out the potentially important role of weak ram pressure stripping on galaxy evolution in clusters. 

Taking advantage of the high survey efficiency of ASKAP, the pilot survey of Widefield ASKAP L-band Legacy All-sky Blind surveY (WALLABY \footnote{https://wallaby-survey.org/}, \citealt{Koribalski20a}) has been targeting nearby clusters and groups. In this study, we use its second internal data release of the Hydra cluster \footnote{https://research.csiro.au/casda/}. 

As we will show in this paper, WALLABY is biased against galaxies with $\hi$ masses less than a few times $10^8~M_{\odot}$ at the distance of Hydra. Galaxies with stellar masses below around $10^9~M_{\odot}$ may therefore not appear in the WALLABY catalogue if they are highly $\hi$ deficient (Sec.3.2), which are the most commonly used samples for ram pressure stripping studies (e.g., \citealt{Boselli14}). But the data fully cover the Hydra cluster out to 2 $r_{200}$ (Sec.3.1), and provide resolved $\hi$ distributions in a few galaxies, which we use as the test sample for predicting the $\hi$ distribution in the unresolved galaxies (Sec.2.3.2). Based on this, we attempt to quantify the instantaneous speed of $\hi$ depletion due to ram pressure stripping in the detected galaxies by quantifying the level of ram pressure, anchoring force, and mass of strippable $\hi$. The analysis thus provides statistics about the acceleration and speed of cluster processing through ram pressure stripping at a relatively early stage of $\hi$ depletion (i.e. before the $\hi$ deficiency level becomes high, as the sample is strongly biased toward $\hi$ rich galaxies). Although we do not know the initial conditions of galactic $\hi$ masses or distributions upon infall (passing 2$r_{200}$), the WALLABY coverage provides us with a snapshot of cluster processing, giving insight into the role of ram pressure stripping in removing $\hi$ from galaxies infalling into massive clusters.  
Observationally this is the first major study of a cluster beyond the local Virgo, Coma and A1676 clusters. The most extensively studied of these, Virgo, is not representative of clusters with similar masses as it is dynamically unrelaxed \citep{Boselli14}. Furthermore,  Virgo has an ICM which is more centrally concentrated compared to a dynamically mature cluster \citep{Roediger07}.

 Hydrodynamic \citep{Oman21, Ayromlou20, Stevens20, Lotz19, Bahe19, Jung18, Bahe15, Bahe13} and semi-analytical \citep{Xie20, DeLucia19, Stevens19b, Stevens17, Luo16} models have made progress in recent years narrowing the differences between predicted and observed distribution of $\hi$ density within galaxies and scaling relations of $\hi$ mass in galaxies. However, because galaxies are complex systems, many different mechanisms produce a similar trend \citep{Stevens17}. So far as we know, after the early works of \citet{Vollmer01, Boselli14} and others for the Virgo cluster galaxies, there has been very few observational studies that directly provide the full distribution of the ram pressure$/$restoring forces and the mass of strippable $\hi$ for all the detected galaxies in the cluster. This is needed to separate the amount of gas loss due to ram pressure stripping from that due to feedback or star formation. Furthermore, resolved $\hi$ images are preferable to simple catalogues of HI detections. As we show in the paper (Sec.2.3.2), the radial distribution of $\hi$ in cluster galaxies differs from that of galaxies in the field, though they do obey the same size-mass relation \citep{Wang16, Stevens19a}. As a result we apply a correction factor to the fraction of instantaneous strippable $\hi$ mass predicted with the typical radial profile of $\hi$ of field galaxies. Obviously 
 one individual cluster is not enough to provide strong constraints to cosmological simulations, but the analysis presented in this paper can be extended with future WALLABY observations covering multiple clusters.

The outline of this paper is as follows. We present the data in Sec.2, discuss the phase-space distribution of $\hi$-detected galaxies and ram pressure stripping affected galaxies in Sec.3, the distribution of $\hi$ richness and ram pressure stripping strength in Sec.4, the timescales for ram pressure stripping to remove the currently affected $\hi$ and the existing $\hi$ reservoir in Sec.5. We assume a $\Lambda$CDM cosmology, with $\Omega_{m}=0.3$, $\Omega_{\lambda}=0.7$ and $h=0.7$. We assume the \citet{Chabrier03} initial mass function when estimating the stellar mass. 

\section{Data and methodology} %800
\subsection{Data}
The Hydra cluster has a distance of 47.5$\pm3$ Mpc,  a center at $\alpha=159.0865\degree$ and $\delta=-27.5629\degree$, and a heliocentric velocity of 3686 km~s$^{-1}$ \citep{Kourkchi17}.  
Analysis based on X-ray data suggested a characteristic radius $r_{200}\sim$1.35 Mpc, and the mass within this radius, $M_{200}\sim3.02\times10^{14}~M_{\odot}$ \citep{Reiprich02}. Its velocity dispersion $\sigma_C=620~km~s^{-1}$ is derived from $M_{200}$ using the equation of \citet{Evrard08}.
The Hydra cluster is a dynamically highly mature system, with a smooth X-ray halo, and little substructure \citep{Fitchett88,LimaDias21}. There is a hint of two or three substructures in velocity near the cluster center (within $0.4r_{200}$, \citealt{Fitchett88}), but on the whole Hydra is dynamically much more settled than the Virgo cluster.
The Hydra cluster is connected to the Antlia cluster via a filament, and the latter was mapped in $\hi$ by the Karoo Array Telescope (KAT-7) with a mosaic of 4.4 deg$^2$ \citep{Hess15}. 

\begin{figure*} 
\centering
\includegraphics[width=18cm]{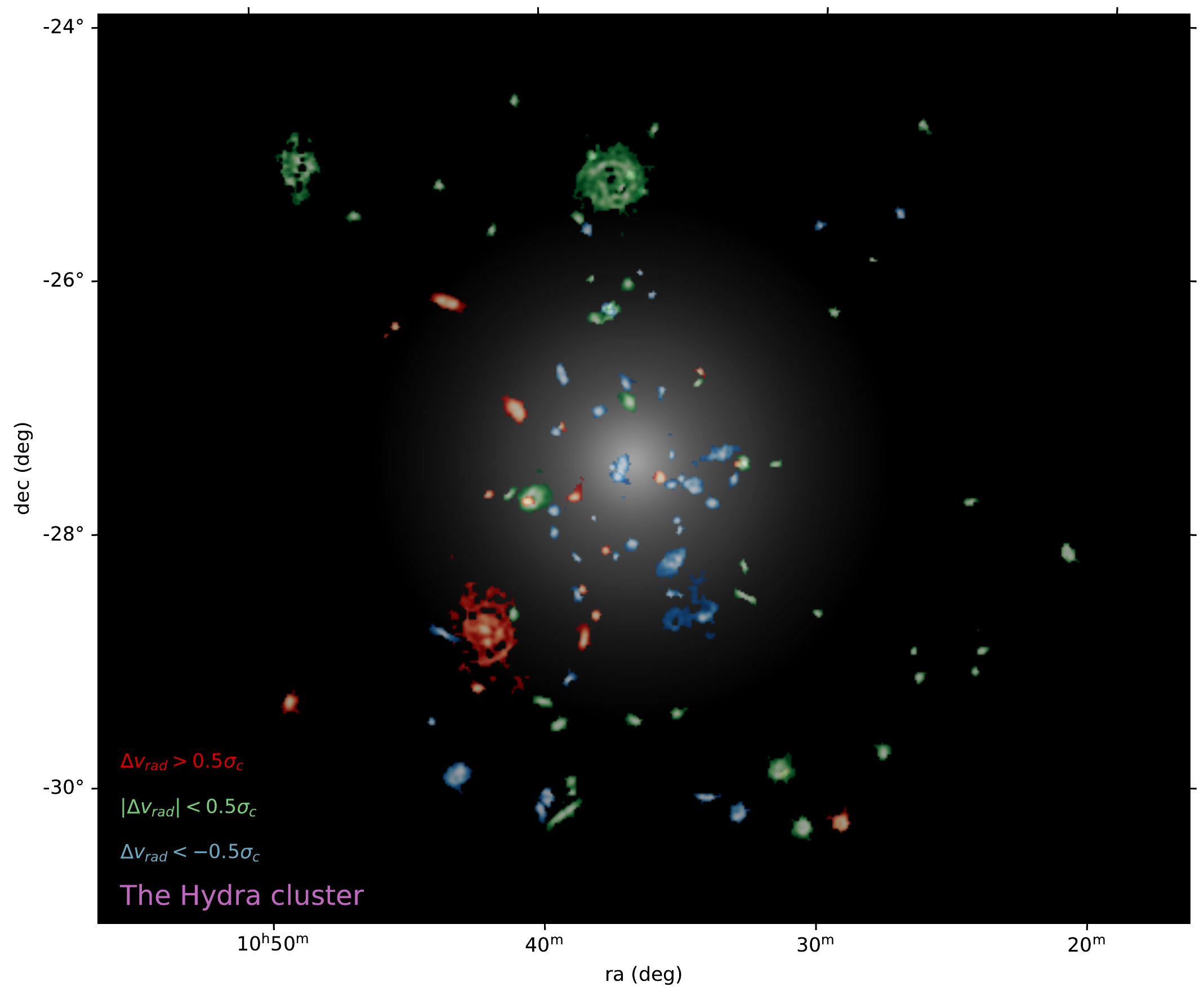}
\caption{ WALLABY detected galaxies in the Hydra cluster. The frame is $5.5r_{200}$ in width and $4.5r_{200}$ in height, where $r_{200}$ is the cluster centric radius where the enclosed averaged density is 200 times the cosmological critical density. The $\hi$ column density images are shown in different slices of velocity offset from the cluster center ($\Delta v_{rad}$) around the X-ray halo of Hydra (grey scale, based on the ICM model described in Sec.2.2). The column density images are enlarged by a factor of 5 for clarity. A similar image has been shown for the Virgo Cluster by \citep{Chung09} using targeted HI observations.  }
\label{fig:Hydra}
\end{figure*}

WALLABY mapped a 60 deg$^2$ area around its center, reaching out to 4$r_{200}$ and fully covering the region within 2$r_{200}$. The WALLABY observation reached a targeted sensitivity of $\sigma_{cube}=2.0\pm0.5$ mJy/beam, with a circular beam full-width half-maximum of $b_{maj}\sim$30 arcsec ($\sim$7 kpc at the distance of Hydra) and a velocity spectral resolution of 4 km~s$^{-1}$. The raw data was reduced with ASKAPsoft \citep{Whiting20}, and $\hi$ sources were extracted using SoFiA 2 \footnote{https://github.com/SoFiA-Admin/SoFiA-2} with a multi-kernel smooth$+$clip algorithm (\citealt{Serra15}, Westmeier et al. submitted) at a significance level of 3.5-$\sigma$. Details about the observation, data reduction and source finding are discussed in earlier WALLABY publications \citep{Koribalski20a, For19, Kleiner19, LeeWaddell19, Elagali19}. More details of the data used in this paper will be further elaborated in Westmeier et al. (in prep).
 In this analysis we focus on the detections (N$=$105) with projected distances within 2.5 $r_{200}$ and radial velocity differences within the upper limit of the escape velocity from the cluster center. We highlight an $\hi$ overview of this phase-space region around the Hydra center in Fig.~\ref{fig:Hydra}. 

We use the optical $g$ and $r$-band images from the second data release of PanSTARRS \citep{Waters20} to derive optical properties. We use the photometric pipeline described in \citet{Wang17, Wang18}, which includes standard procedures of background removal, masking, segmentation, and flux measurements. 
We use Kron magnitudes, and the radial profiles of the surface brightness. We correct for the Galactic extinction using the IRSA Dust Extinction Service. 
We use the formula of \citet{Zibetti09} to calculate the $r$-band stellar mass to light ratio based on the $g-r$ color (assuming the \citet{Chabrier03} initial mass function), and estimate the stellar mass $M_*$ based on the $r$-band luminosity. 

We exclude 9 systems which are likely to be strongly tidally interacting, have highly irregular optical and $\hi$ morphologies, or have $\hi$ bridges connecting two or multiply major galactic systems. We also exclude 7 galaxies which do not have good optical images from PanSTARRS, leaving 89 galaxies in our main sample. 

Although we attempted to exclude major merger candidates, we cannot exclude the influence from gravitational effects, including the harassment from surrounding galaxies \citep{Moore96} and the tidal force from the main cluster \citep{Byrd90}. Simulations suggests that, in addition to directly stripping the gas \citep{Merritt83, Lokas20} and inducing gas inflows \citep{Byrd90,  Blumenthal18, Moreno19, Patton20}, gravitational effects may assist ram pressure stripping by enhancing local ICM density \citep{McPartland16, Roediger11, Bekki10, Markevitch07} and relative velocities \citep{Ruggiero19}, and moving $\hi$ to regions of lower anchoring force \citep{Kapferer08}. There can be other hydrodynamic effects from the environment like viscous stripping \citep{Nulsen82}, but in this study we focus on the effect of ram pressuring stripping. 

\subsection{The ram pressure level}
We largely follow the procedure of W20 to estimate the ram pressure and the anchoring force. 
We use the equation from \citet{Ettori15} to convert the cluster dynamic mass to ICM mass, and then convert the double-beta model of the X-ray surface brightness radial profile (from \citealt{Eckert11}) to the ICM number density ($n_{{\rm ICM}}$) profile as a function of the cluster-centric radius $r$:
\begin{equation} \label{eq0}
\begin{split}
n_{{\rm ICM}}(r) / cm^{-3}= & 0.003 (1+(r/0.101~Mpc)^2  )^{-1}+  \\
                                 &  0.004  (1+ (r/0.027~Mpc )^2 )^{-1}  .
\end{split}
\end{equation}

 The ram pressure is estimated as $P_{ram}= \rho(d_{proj})~\Delta v_{rad}^2$ \citep{GunnGott72}, where $\rho(d_{proj})=1.4m_p n_{{\rm ICM}}(d_{proj})$, $m_p$ is the mass of proton and 1.4 accounts for mass contribution from the helium,  $d_{proj}$ is the projected distance from the cluster center, and $\Delta v_{rad}$ is the velocity difference from the cluster center. Because the observed $\Delta v_{rad}$ and $n_{{\rm ICM}}(d_{proj})$ are lower and upper limits of the real relative velocity and the ICM density, their under- and over-estimating effects to some extent cancel out in the estimate of $P_{ram}$, mitigating the uncertainties from projection effects. 
 
There may be additional uncertainties related to the assumptions of the isothermal and smooth nature of the ICM distribution, and the extrapolation of the ICM profile. These effects should be of lower importance than projection effects, as Hydra is a dynamically relaxed cluster. 
 
 \subsection{ Quantifying the strippable $\hi$}
The anchoring force (or restoring force) is estimated with a modified equation of the \citet{GunnGott72} model, as 
\begin{equation}
F_{anchor}=2 \pi G (\Sigma_*+\Sigma_{{\rm HI}} )\Sigma_{{\rm HI}}  , 
\label{eq:anchor_force1}
\end{equation}
where $G$ is the gravitational constant, and $\Sigma_*$ and $\Sigma_{{\rm HI}}$ are localized stellar and $\hi$ surface densities (projection corrected) within the galaxy. Below we describe two parallel methods of calculating $F_{anchor}$ and thereafter quantifying the ram pressure stripping strength by comparing $F_{anchor}$ with $P_{ram}$. The first method makes use of $\hi$ images and is most accurate for the resolved galaxies. The second method makes use of predicted $\hi$ radial distributions and works for the whole main sample. We will use results (e.g. identification of ram pressure stripping candidates, estimate of the strippable $\hi$ mass) from the first method to check the reliability of results from the second method.

 \subsubsection{Quantifying the strippable $\hi$ with HI moment-0 images}
We only consider the pixels of  $\hi$ intensity images with a column density $>10^{20}~cm^{-2}$, corresponding to a threshold of 2-$\sigma_{cube}$ with a velocity width of 20 $km~s^{-1}$. The $\hi$ disks with an area larger than 6.2 times the beam area are considered resolved, and others the unresolved disks. The threshold of 6.2 beam areas is equivalent to requiring a circle with the same area to have a radius $>1.5b_{maj}$, as the beam area is estimated as 1.134 $b_{maj}^2$. We use each of the pixels to estimate $\Sigma_{{\rm HI}}$ in a resolved $\hi$ disk, and only use the pixel with the peak value in an unresolved $\hi$ disk in order to have a conservative, high-limit estimate of $F_{anchor}$. We extrapolate the stellar mass density exponential profiles to match the radial extension of the $\hi$ disks. We correct both $\hi$ and stellar surface densities for projection effects by multiplying them by the optical axis ratio, assuming infinitely thin disks. We caution that, the inner $\Sigma_{\rm HI}$ can be both over-estimated and under-estimated when the galaxies are marginally resolved. A general approach needs to be devised in the future to correct for these beam smearing effects. By using the stellar mass density profiles instead of two dimensional distributions, we may average out outlying structures like faint spiral arms and cause a local under or over-estimate of the anchoring forces, but it is the most reliable way to derive $\Sigma_*$ in the low signal-to-noise ratio outer disks. 

We compare the ram pressure with the anchoring force to estimate the fraction of $\hi$ flux under stripping within a disk. For a resolved $\hi$ disk, we firstly produce an initial binary map for ram pressure stripped pixels with lower anchoring force than the ram pressure. We then undertake an ``erode'' and ``dilate'' process to remove unreliable pixels at the edge of disks and to ensure that the ram pressure stripped region covers at least one independent beam, which is 5 pixels across. This is done with two iterations of each operation using the python package scipy.ndimage. Such a treatment results in an identification bias against ram pressure stripped pixels along the minor axis of disks, but ensures a relatively conservative identification. We use this binary map to compute the fraction of $\hi$ flux being stripped, $f_{RPS}$.
 We consider those galaxies with $f_{RPS}>0.1$ as undergoing ram pressure stripping (referred to as RPS candidates). We refer to the resolved galaxies which have $f_{RPS}<0.1$ as the non-RPS candidates. By doing so, we attempt to catch the galaxies at an early stage of continuous stripping, and do not include the galaxies which were already severely stripped in the past, left with a heavily truncated disk of $\hi$ incapable of being stripped any more, and are extreme $\hi$ deficient at the present. We show later in Sec.2, that the sample is strongly biased toward $\hi$ rich galaxies, and more suitable to study the former type of galaxies than the latter. 
 
Because we only consider the peak pixel, $f_{RPS}$ is either 1 or 0 for an unresolved $\hi$ disk. We note that, $f_{RPS}$ for these unresolved RPS candidates are only used for marking them as RPS candidates, but not further analyzed. 

The fraction of strippable $\hi$ ($f_{RPS}$) is of course an underestimate as some $\hi$ will have already left the galaxy and no longer visible. Indeed, with $f_{RPS}$ we do not attempt to quantify the exact amount of stripped $\hi$ under the current or past ram pressure, but instead use it as an indicator for the instantaneous $\hi$ loss rate due to ram pressure stripping (or an indicator for the stripping strength). Two assumptions have been made when $f_{RPS}$ is interpreted in this way: the length of time for ram pressure to accelerate and remove a strippable $\hi$ cloud cannot be ignored, and the length of time for ram pressure to deplete the whole $\hi$ disk is relatively long compared to the removal of a strippable cloud. We show in Sec.5 that both assumptions are reasonable for at least the resolved galaxies. 

In total, there are 27 resolved $\hi$ disks, among which 10 are identified to be potentially under ram pressure stripping. As can be seen in Fig.~\ref{fig:HIprop}-c, the majority of the RPS candidates have $f_{RPS}>0.25$, so the sample and related results do not change much if we raise the criterion for identifying RPS candidates to $f_{RPS}>0.2$.
 From the $\hi$ and ram pressure stripping atlas of these 10 resolved RPS candidates (Fig.~\ref{fig:atlas} in the appendix), many of them display lopsidedness with respect to the optical disks, indicative of ram pressure stripped tails. We do not clearly observe tails in most of the galaxies, possibly because of the limited resolution ($\sim$7 kpc) and depth ($10^{20}~cm^{-2}$) of the data. On the other hand, the difficulty of identifying morphological features in weak RPS galaxies has been previously pointed out in hydrodynamic simulations \citep{Jung18}. We find that those tentative tails (or lopsidedness) do not always point away from the center of Hydra, which may result from the combined projection effect of the viewing angle and the disk orientation with respect to the orbits. The randomness in orientation of ram pressure stripped tails was also noticed in the literature, both in observations (e.g. \citealt{Kenney14}), and in hydro-dynamic simulations \citep{Yun19}.  
 Additionally there are 13 unresolved RPS candidates.

There are also several sources of uncertainties in the estimate of $F_{anchor}$, including the neglect of stellar bulges, dark matter halos and circumgalactic medium, and most importantly the assumption of face-on infall orientations. Uncertainties due to the neglect of the different galactic components are mitigated by the fact that the $\hi$-rich galaxies tend to have small bulges, and the high mass of Hydra tend to 
truncate dark matter halos and remove the circumgalactic medium from galaxies \citep{Bahe13, Bahe15}. Uncertainties due to the assumption of face-on infall are mitigated by previous hydrodynamic simulation results that orientations do not significantly affect $\hi$ mass loss due to ram pressure stripping unless the galaxies infall nearly edge-on (inclination$>60\degree$, \citealt{Roediger06}).
Nevertheless, given all these potential uncertainties, we caution that all results from the ram pressure stripping analysis in this paper should be understood in a statistical sense. 
 
\subsubsection{Quantifying the strippable $\hi$ with predicted HI radial distributions}
\begin{figure*} 
\centering
\includegraphics[width=5.5cm]{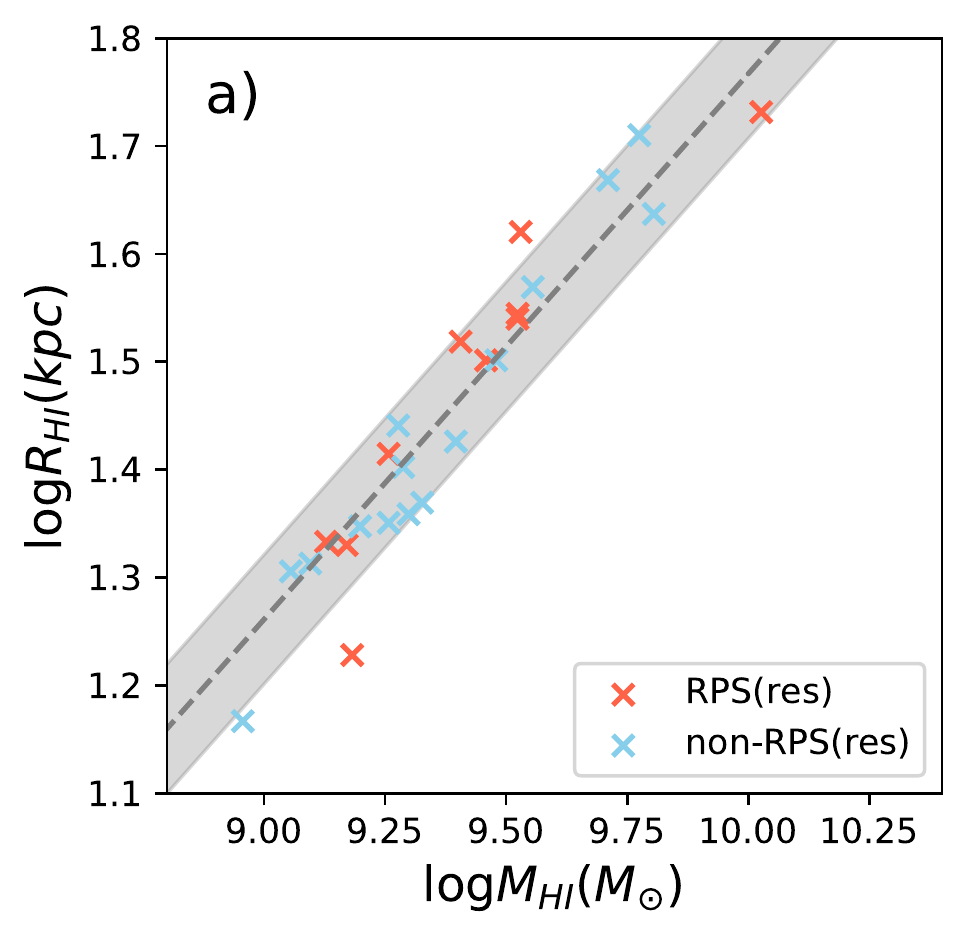}
\includegraphics[width=5.5cm]{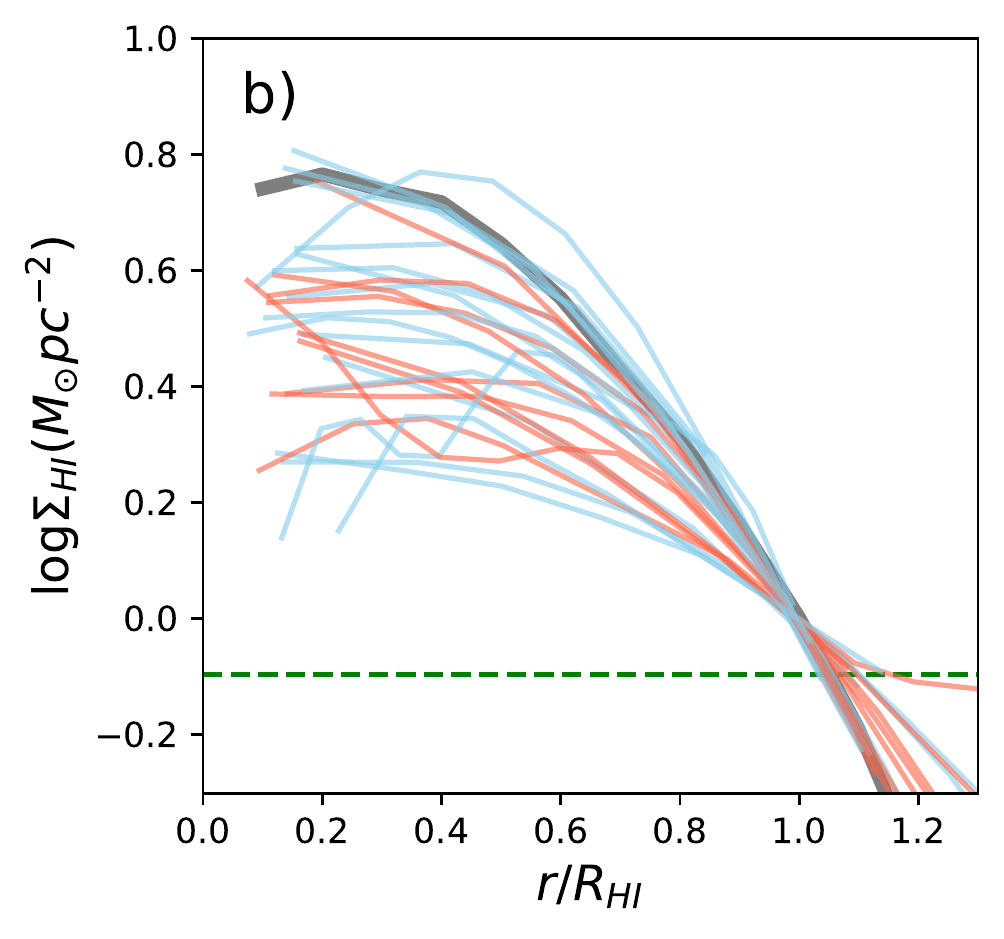}
\includegraphics[width=5.5cm]{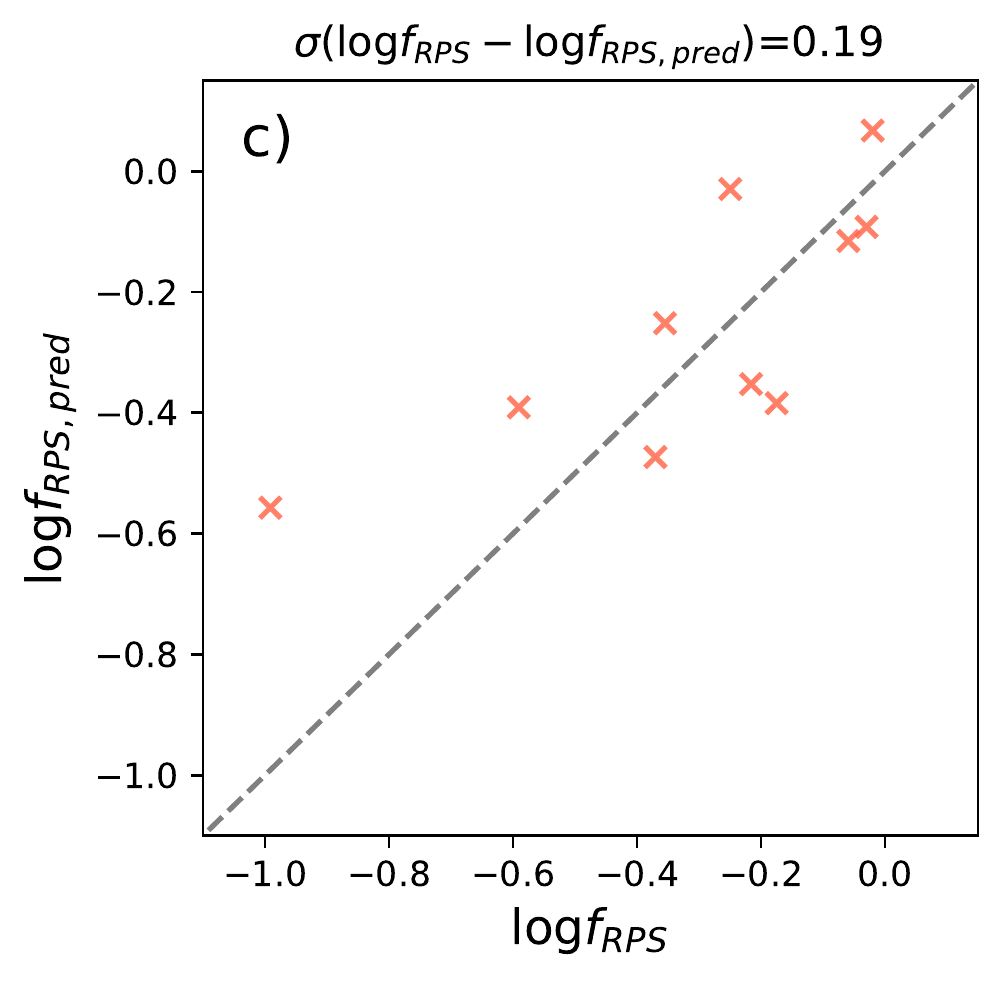}
\caption{Radial measurements of $\hi$ for the 27 well-resolved galaxies.
{\bf Panel a:} the size-mass relation of $\hi$. The sizes $R_{\rm HI}$ have been corrected for the effect of beam smearing as $R_{\rm HI}=\sqrt{R_{\rm HI,obs}^2-b_{maj}^2}$, where $R_{\rm HI,obs}$ is directly derived from the $\hi$ images as the semi-major axis of the ellipse with a de-projected surface density of 1$~M_{\odot}~pc^{-2}$. The mean relation and scatter from \citet{Wang16} is plotted as the grey line and shaded region. 
{\bf Panel b:} the radial profiles of $\hi$ surface density $\Sigma_{\rm HI}$. The projection correction has been applied assuming an infinitely thin disk and an axis ratio determined from the optical light. The radii are normalized by $R_{\rm HI}$. The median profile of 168 late-type galaxies \citep{Wang16,Wang20a} is plotted as a grey, thick curve. The green dashed line marks  $\Sigma_{\rm HI}=0.8~M_{\odot}$ corresponding to the detection limit of the WALLABY data.
{\bf Panel c:} predicted versus observed fraction of strippable $\hi$ mass over the total $\hi$ mass ($f_{RPS}$) for the resolved RPS candidates. The predicted fraction $f_{RPS,pred}$ has been corrected by multiplying by a factor of 1.4 (see text), in order to match the observed $f_{RPS}$ on the one-to-one line. The 1-$\sigma$ rms of the difference between the two quantities are 0.19 dex, which drops to 0.14 dex if we exclude the one outlier which has $f_{RPS}$ close to the threshold of 0.1 for identifying RPS candidates.  In all panels, RPS (non-RPS) candidates are in red (blue). }
\label{fig:HIprop}
\end{figure*}

\begin{figure*} 
\centering
\includegraphics[width=14cm]{ 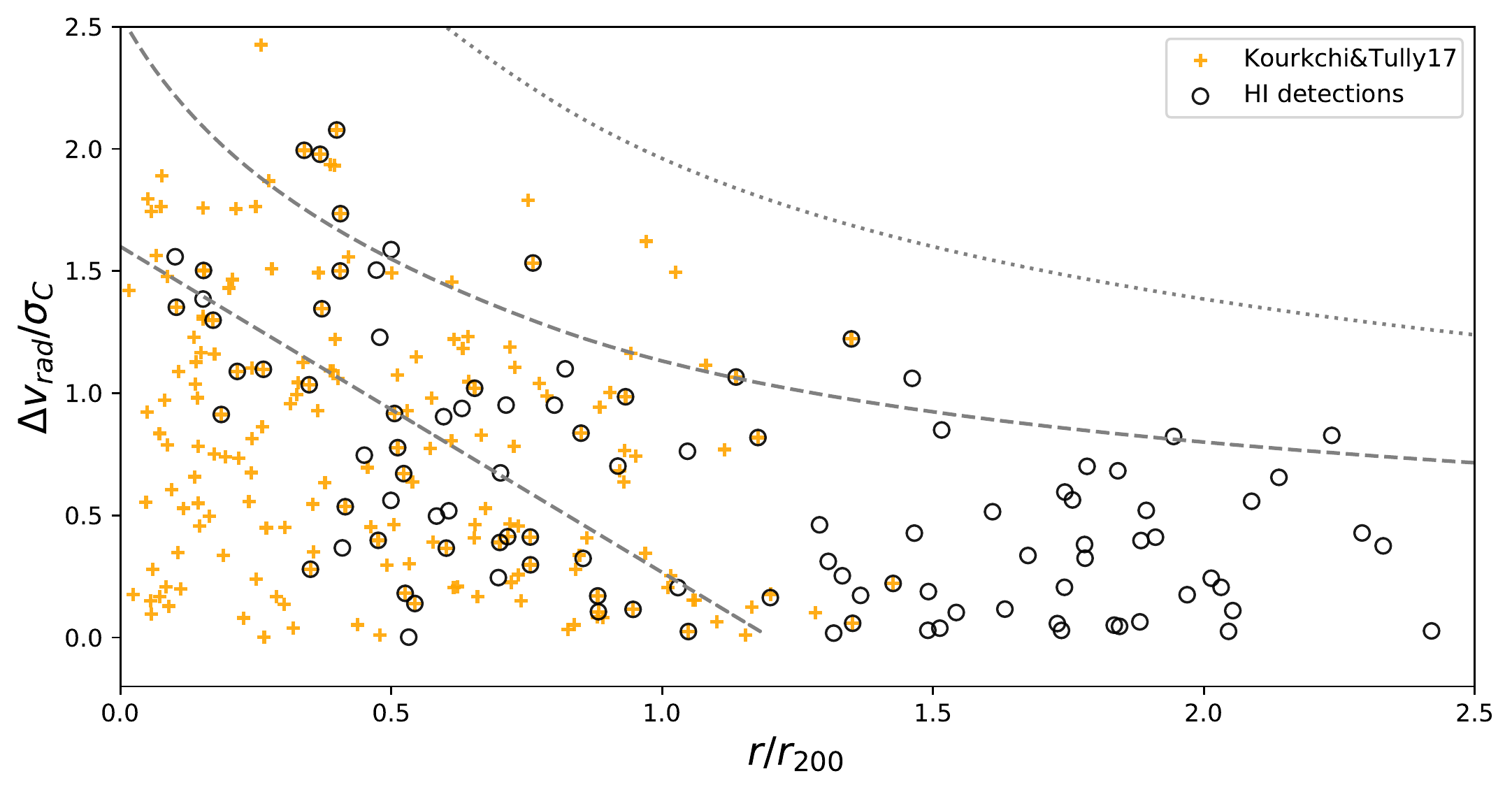}
\includegraphics[width=14cm]{ 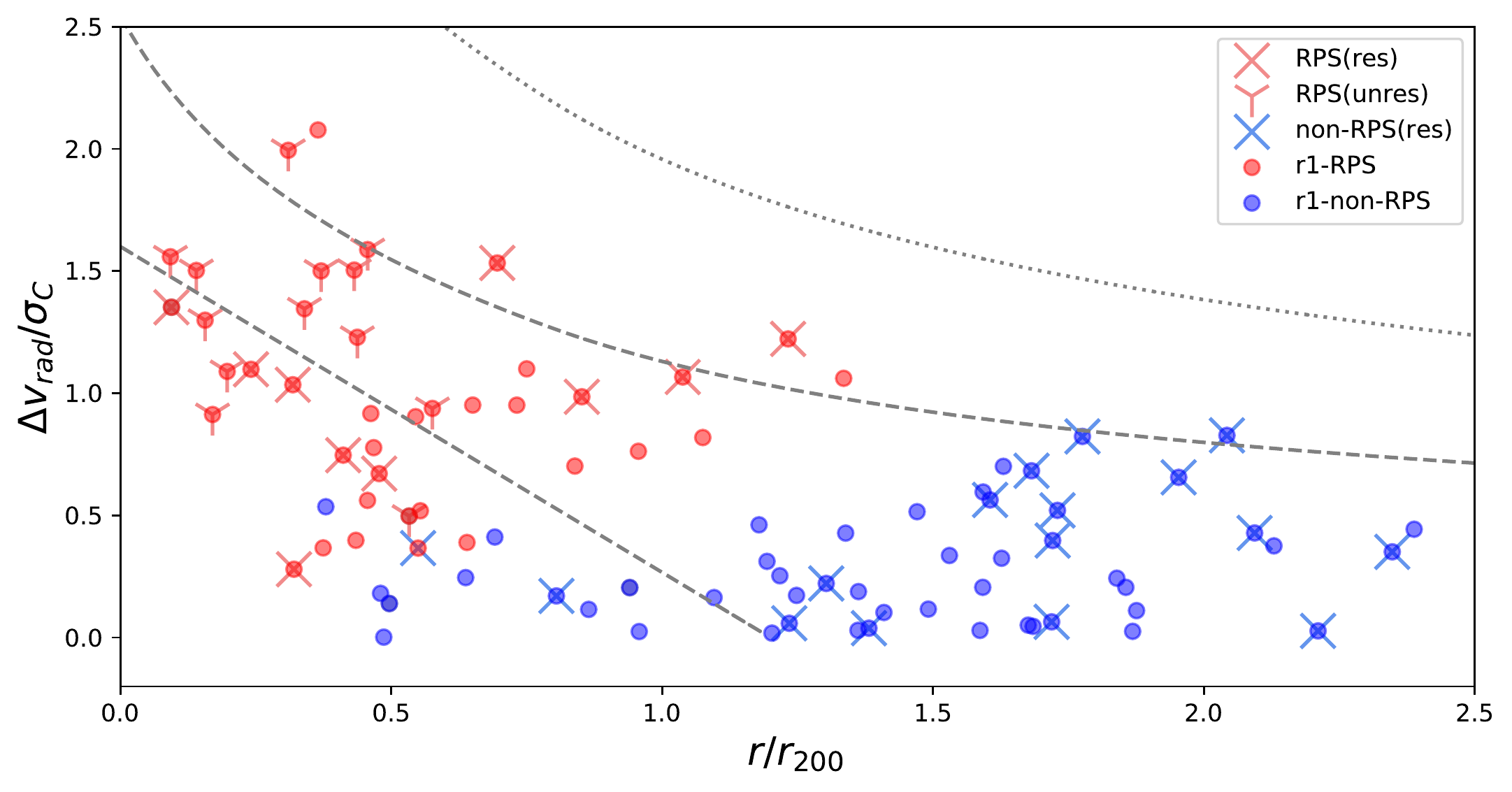}
\caption{ The projected phase-space diagram of the Hydra cluster. The dashed straight line marks the border of the virialized region, and the dashed (dotted) curve shows the averaged (upper limits of) projected escape velocities as a function of the projected cluster centric distance $d_{proj}$, which are expected for an NFW dark matter profile with a concentration of 4 \citep{Navarro97}. {\bf Top:} optical members of the Hydra cluster from \citet{Kourkchi17} in orange pluses, and all the $\hi$ detections of WALLABY in black circles. {\bf Bottom:} Only galaxies from the main sample are displayed. The resolved RPS (non-RPS) candidates are marked by red (blue) crosses respectively, and the unresolved RPS candidates in red ``tri-down'' symbols. The filled circles are colored red (blue) if they are in the r1-RPS (r1-non-RPS) type. }
\label{fig:phase-space diagram}
\end{figure*}

Since, as mentioned in the last section, when the HI disks are spatially unresolved we can only identify
the strongly stripped galaxies, we also use the method of \citet{Wang20b} based on the $\hi$ size-mass relation to identify RPS candidates in the main sample. We use the $\hi$ size-mass relation to estimate the radius $R_{{\rm HI}}$ where the $\hi$ surface density $\Sigma_{{\rm HI}}\sim 1~M_{\odot}~pc^{-2}$ \citep{Wang16}. 
We confirm in Fig.~\ref{fig:HIprop}-a that at least for the resolved galaxies, the size-mass relation still holds in the Hydra cluster, consistent with the theoretical predictions of \citet{Stevens19a}. 
We then compare the ram pressure with the anchoring force at $R_{{\rm HI}}$ for each galaxy, and identify the ram pressure stripping candidates (the r1-RPS candidates hereafter). We identify 40 r1-RPS candidates from the whole main sample. We check the reliability of r1-RPS identification in the bottom panel of Fig.~\ref{fig:phase-space diagram}, by over plotting the r1-(non-)RPS and (non-)RPS candidates. The r1-RPS identification is highly consistent with that based on $\hi$ images when the disks are resolved. Particularly, all the RPS (resolved and unresolved) candidates are successfully identified as r1-RPS candidates, and only one non-RPS candidates is mistakenly identified as a r1-RPS candidate. This galaxy (NGC 3336, WALLABY J104016-274630) has a strong bar and spiral arms, which might have concentrated the $\hi$ into a bright ring in the inner disk, and caused the deviation of the real $R_{\rm HI}$ from the predicted one. 

\begin{figure*} 
\centering
\includegraphics[width=14cm]{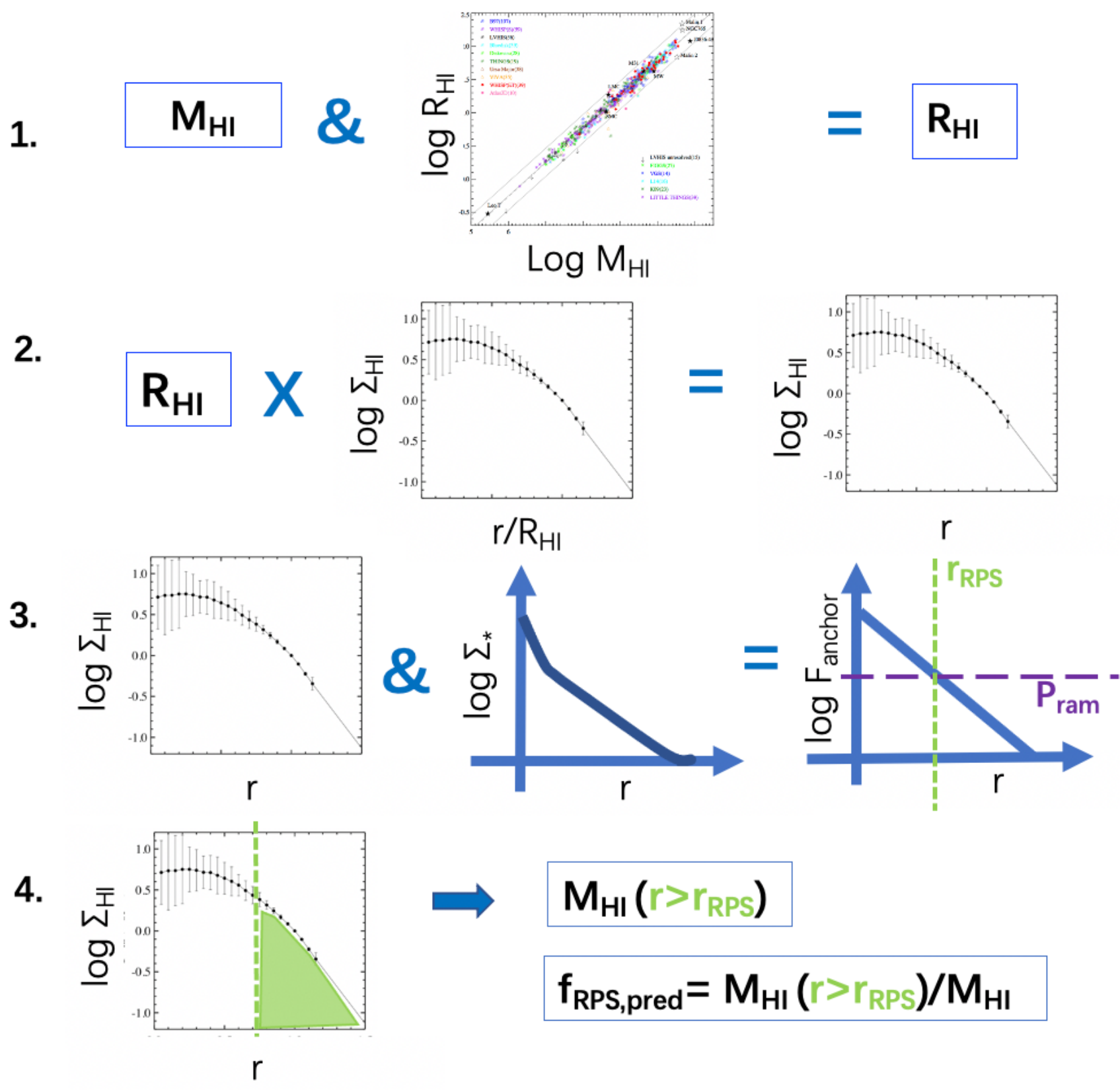}
\caption{ Procedure to estimate the fraction of strippable $\hi$ mass over the total $\hi$ mass ($f_{RPS,pred}$). We use the $\hi$ size-mass relation $\hi$ ($R_{\rm HI}$ versus $M_{\rm HI}$), and the median profile of $\hi$ surface density ($\Sigma_{\rm HI}$) as a function of $r/R_{\rm HI}$ from \citet{Wang16}. See text in Sec.2.3.2 for details. }
\label{fig:method}
\end{figure*}

Because different galaxies have similar profiles of $\Sigma_{{\rm HI}}$ as a function of $r/R_{{\rm HI}}$ in the outer region \citep{Wang16}, we use the median profile of 168 late-type galaxies from \citet{Wang16} to estimate the $\hi$ stripping fraction $f_{RPS,pred}$ for the r1-RPS candidates. For each galaxy, we have used the size-mass relation to estimate $R_{\rm HI}$ from $M_{\rm HI}$. We combine $R_{\rm HI}$ with the median profile of $\Sigma_{\rm HI}$ as a function of $r/R_{\rm HI}$ to predict the profile of $\Sigma_{\rm HI}$ as a function of radius. We estimate the radial profile of $F_{anchor}$ with the radial profiles of $\Sigma_{\rm HI}$ and $\Sigma_*$ based on Equ.2. We compare the $F_{anchor}$ profile with $P_{ram}$ to determine the radial range where $F_{anchor}<P_{ram}$. We then accumulate the $\Sigma_{\rm HI}$ profile within that radial range to estimate the mass of strippable $\hi$, and calculate $f_{RPS,pred}$. These steps are also summarized in Fig.~\ref{fig:method}.
 A similar technique was used in \citet{Wang20a} to successfully predict the $\hi$ mass within and beyond the optical radius of individual galaxies. 
Comparing the directly calculated $f_{RPS,pred}$ with $f_{RPS}$ for the resolved RPS candidates suggests a correction factor of 1.4 be multiplied to $f_{RPS,pred}$ in order to match $f_{RPS}$. We show in Fig.~\ref{fig:HIprop}-c that $f_{RPS,pred}$ is a good predictor of $f_{RPS}$ for the resolved-RPS candidates, after such a correction. Thus $f_{RPS,pred}$ hereafter has all been corrected in this way. A maximum value of unity is set for $f_{RPS,pred}$ after the correction. 

The correcting factor is likely because the median profile misses the $\hi$ tails (lopsidedness) in the outer disks caused by ram pressure stripping, and also misses the suppressed $\hi$ inner disks possibly as a result pre-processing \citep{Hess13, Bahe15}. Hints for these two features can be seen by comparing the $\hi$ radial profiles of the resolved galaxies to the median profile of late-type galaxies \citep{Wang16,Wang20a} (Fig.~\ref{fig:HIprop}-b). The deviation of shape in $\hi$ radial profiles cannot be fully explained as a result of beam smearing, as we find that the under-estimation of $f_{RPS}$ with $f_{RPS,pred}$ does not increase with decreasing $\hi$ disk sizes. We caution that applying this correcting factor manually sets a minimum value of 0.15 dex for $f_{RPS,pred}$, although it does not significantly affect our major conclusion, that for many of the r1-RPS candidates, the stripping efficiency is low. Note that such a treatment does not affect our identification of r1-RPS candidates, which were performed beforehand based on the size-mass relation of $\hi$.

In the rest of the paper, we analyze the RPS and r1-RPS candidates separately. We use the r1-RPS sample for statistical trends, counts and distributions in Sec. 3 and 4, though we also display corresponding results of the RPS sample for a reference. 
We further use the resolved sample for a detailed analysis of the ram pressure stripping history in Sec.5, through comparing the galactic radial profiles of anchoring forces with the ram pressure levels at different $d_{proj}$. 

Tab.~\ref{tab:cluster} lists the ram pressure stripping properties of galaxies derived above.

\section{Distribution of galaxies in the projected phase space diagram}
The main goal of this section is to obtain an overview of the distribution pattern of the $\hi$-detected galaxies and r1-RPS candidates in the Hydra cluster, and quantify the frequency of r1-RPS candidates among $\hi$-detected galaxies. 
The projected phase-space diagram is a good tool for statistically separating virialised cluster members, infalling galaxies and background/foreground galaxies \citep{Oman16}. Previous studies found good correspondence between the projected phase-space diagram positions and $\hi$-richness of galaxies \citep{Yoon17, Jaffe15, Solanes01}. An investigation of the projected phase space diagram of optically detected galaxies in Hydra was presented by \citet{LimaDias21}.

The top panel of Fig.~\ref{fig:phase-space diagram} displays the distribution of WALLABY-detected galaxies in the projected phase-space diagram, in comparison to that of optically identified Hydra members galaxies from \citet{Kourkchi17}.
It shows that the $\hi$-detected galaxies tend to avoid the inner part ($d_{proj}<0.5r_{200}$) of the virialized region in the projected phase-space diagram, in contrast to the optical members of the cluster. It suggests that the recent infallers are more likely $\hi$-rich, and is consistent with previous findings \citep{Jaffe15}.

In the bottom panel of Fig.~\ref{fig:phase-space diagram}, we compare the projected phase-space distributions of r1-RPS and r1-non-RPS candidates. The r1-RPS candidates tend to be within 1.25$r_{200}$ and have higher $\Delta v_{rad}$ than the other galaxies in the cluster, which is expected from the way that ram pressure strength is estimated \citep{GunnGott72}, but the amount of $\hi$ and the anchoring force also play a role in the exact distribution of the galaxies in the projected phase-space diagram. Out to a $d_{proj}\sim1.25r_{200}$, 70\% (44\%) of the $\hi$-detected galaxies in Hydra are r1-RPS (RPS) candidates.

\section{ HI richness and the influence from ram pressure stripping}
In this section, we present an overview of the $\hi$ richness of the detected galaxies, and assess the extent of changes the current level of ram pressure stripping will make to those $\hi$ masses ($M_{{\rm HI}}$).

\begin{figure*} 
\centering
\includegraphics[width=12cm]{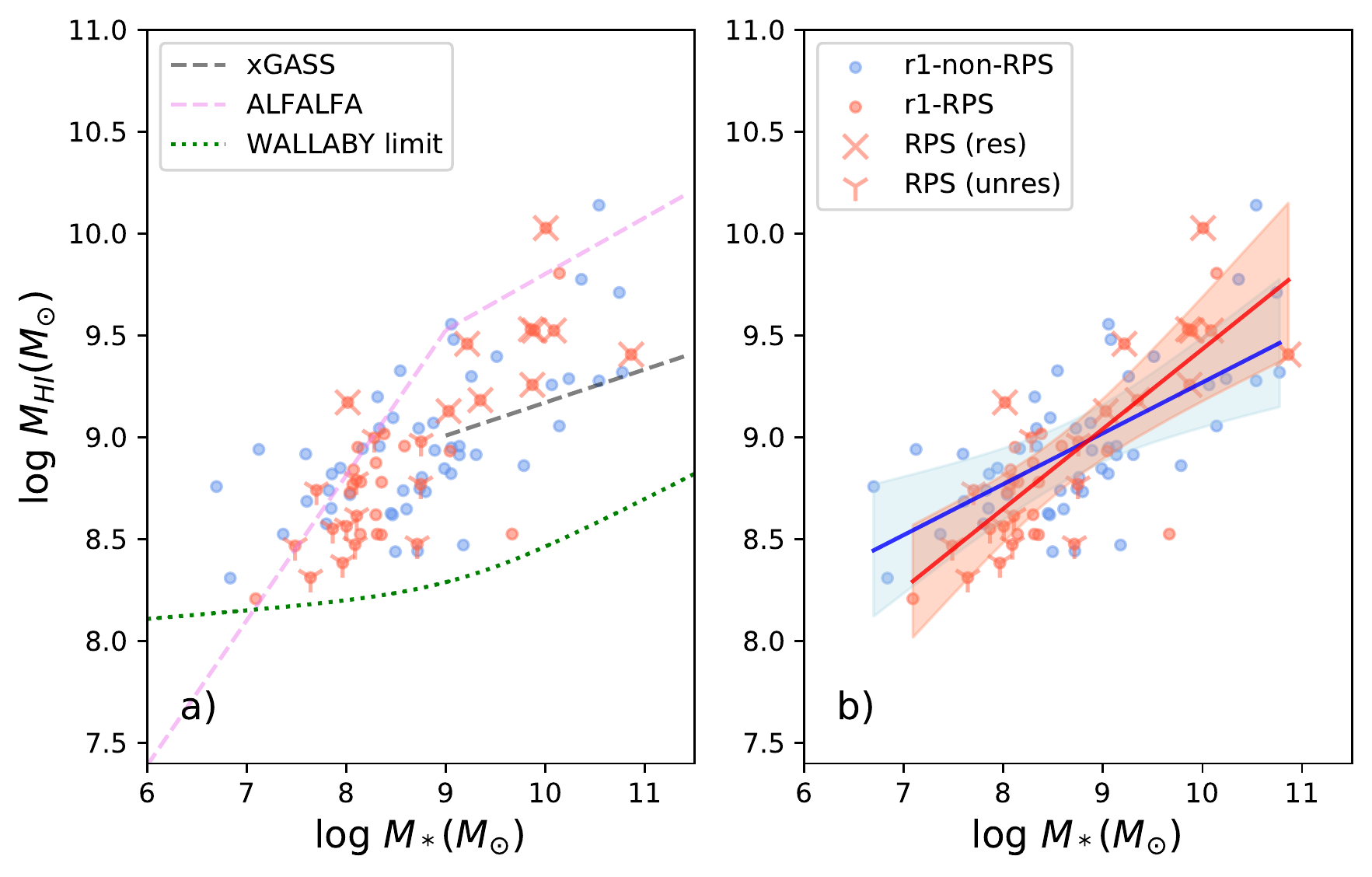}

\includegraphics[width=6cm]{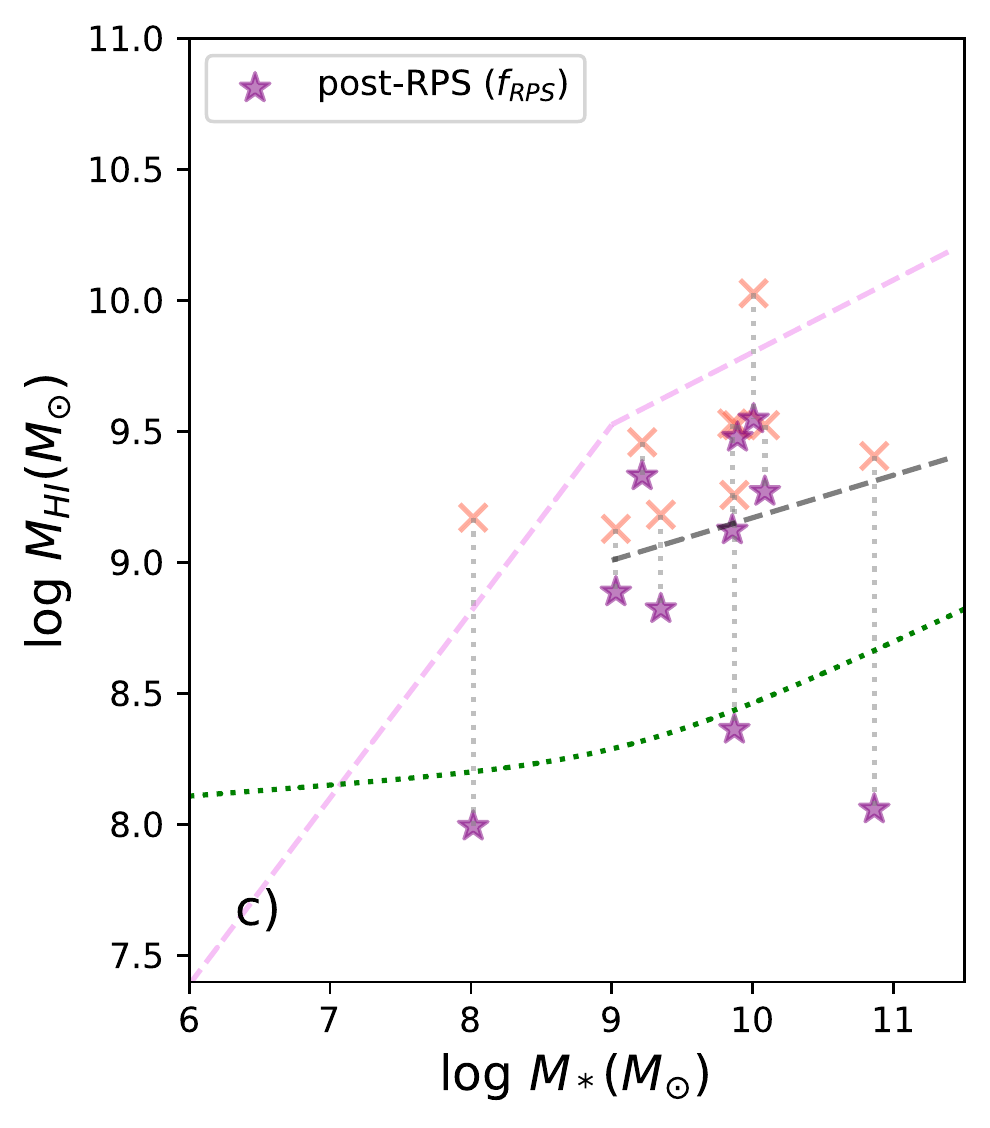}
\includegraphics[width=6cm]{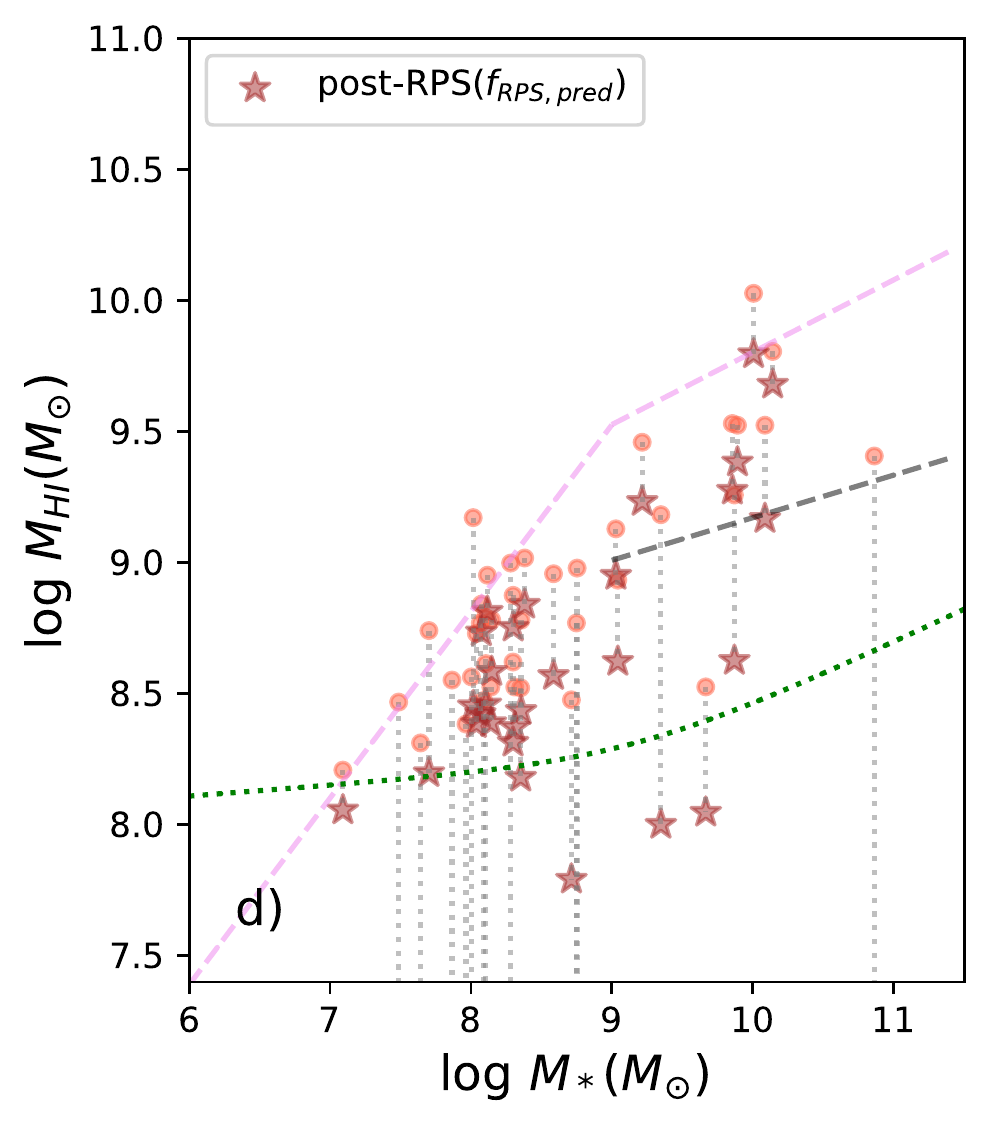}
\caption{ The relation between the $\hi$ mass ($M_{\rm HI}$) and the stellar mass ($M_*$) of the main sample. 
{\bf Panel a:} different types of galaxies are plotted in the same symbols as in the bottom panel of Fig.~\ref{fig:phase-space diagram}. We mark the mean relations from xGASS (grey dashed line) and ALFALFA (pink dashed line), and the detection limit of WALLABY (green dotted line, see text). 
{\bf Panel b:} the best-fit linear relations for the r1-RPS (red) and r1-non-PRS (blue) populations. The shaded regions mark the 1-$\sigma$ uncertainty of fitting.
{\bf Panel c:} brown stars plot the predicted position (post-RPS status) of galaxies after subtracting the strippable $\hi$, i.e. $M_{\rm HI}(1-f_{RPS})$ versus $M_*$. Dotted lines link the observed and post-RPS positions of each galaxy.
{\bf Panel d:} similar to panel c, but with the post-RPS $\hi$ mass estimated with $f_{RPS,pred}$ instead of $f_{RPS}$. As indicated by the dotted lines, many post-RPS positions are at $M_{\rm HI}$ lower than the displayed range.
 }
\label{fig:HIMS}
\end{figure*}

Fig.~\ref{fig:HIMS}-a presents the relation between $M_{\rm HI}$ and $M_*$ for different sub-samples.
We can see that the $M_{\rm HI}$ at a given $M_*$ is systematically lower than the mean relation of the galaxy population detected in the blind $\hi$ survey ALFALFA \citep{Huang12}, but systematically higher than the median relation of the targeted survey of $M_*$-selected galaxies in xGASS \citep{Catinella18}. 
The discrepancy between ALFALFA and WALLABY galaxies, which is more conspicuous at the high stellar mass end, is not majorly caused by the environment, but due to a volume effect. Although both surveys are relatively shallow, approximately flux-limited and have similar sensitivity, the WALLABY observations used here are limited to within $\sim$50 Mpc, whereas ALFALFA detections are within $\sim$200 Mpc \citep{Haynes11}. At larger distances (where more massive systems are more likely found), the bias toward the most gas-rich systems becomes more important, hence affects ALFALFA more strongly than this WALLABY volume. 
As the ALFALFA relation fully covers the $M_*$ range of our sample, and roughly traces the upper envelope of $M_{\rm HI}$ distributions at a given $M_*$ (also see \citealt{Maddox15}), we will use it as a reference line to calculate the relative $M_{\rm HI}$ of galaxies at the present and after stripping the currently strippable $\hi$ (the post-RPS status). The discrepancy of $M_{\rm HI}$ from the xGASS relation indicates that the sample misses the gas-poor, strongly depleted galaxies, due to the insufficient observational depth. Similarly, the $M_{\rm HI}$ distribution at a given $M_*$ does not differ significantly from that of relatively isolated galaxies targeted by the AMIGA project \citep{VerdesMontenegro05} either (not shown in Fig.~\ref{fig:HIMS}). So this study focuses on the onset and early stage of gas depletion in $\hi$-rich galaxies. 

The r1-RPS candidates have similar distributions of $M_{\rm HI}$ at a given $M_*$ compared to the r1-non-RPS candidates, partly due to the relatively narrow range of $M_{\rm HI}$ detectable in WALLABY. There is a hint that the r1-RPS candidates lie on a slightly steeper $M_{{\rm HI}}$-$M_*$ relation than the r1-non-RPS candidates, as indicated by the best-fit linear relations in Fig.~\ref{fig:HIMS}-b. The different slopes are mainly driven by the different $M_{{\rm HI}}$ distributions at low $M_*$. This supports the relatively greater effect of ram pressure stripping on low-mass galaxies.

\begin{figure*} 
\centering
\includegraphics[width=14cm]{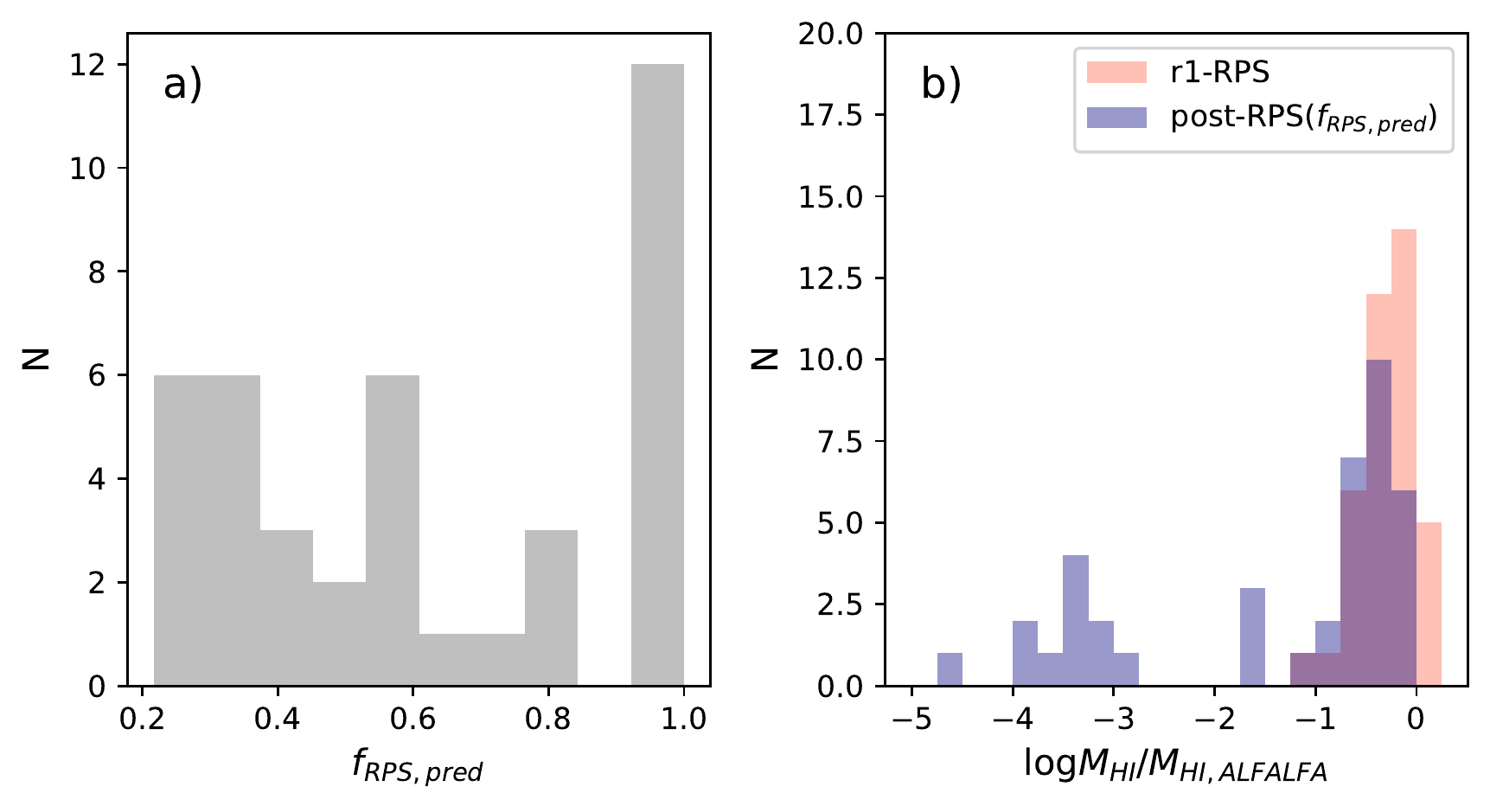}
\caption{ Distributions of the ram pressure stripping strengths for the r1-RPS candidates. Panel a plots the
predicted mass fraction of strippable $\hi$ ($f_{RPS,pred}$). Panel b plots the difference of $M_{\rm HI}$ at a given $M_*$ with respect to the ALFALFA mean relation of $M_{\rm HI}$ versus $M_*$ ($\log M_{\rm HI}/M_{\rm HI,ALFALFA}$). The distributions of $\log M_{\rm HI}/M_{\rm HI,ALFALFA}$ at the present and in the post-RPS status after removing the strippable $\hi$ (i.e. $M_{\rm HI}(1-f_{RPS,pred})/M_{\rm HI,ALFALFA}$) are plotted in red and blue respectively.  }
\label{fig:fRPS_distr}
\end{figure*}

To understand the influence of the observed ram pressure stripping on $M_{{\rm HI}}$, we investigate the distribution of $f_{RPS,pred}$ in Fig.~\ref{fig:fRPS_distr}-a. 
For 70\% of the r1-RPS candidates, $f_{RPS,pred}$ broadly falls below 0.9.  
We note that the 1-$\sigma$ scatter of the median relation of $M_{\rm HI}$ versus $M_*$ from xGASS is $\sim$0.38 dex \citep{Catinella18}, so the threshold $f_{RPS,pred}=$0.9 corresponds to a change in $M_{\rm HI}$ (i.e., $1-f_{RPS,pred}$) by around 2.5-$\sigma$ with respect to the median relation of $M_{\rm HI}$ versus $M_*$, after stripping the currently strippable $\hi$. If we decrease the $f_{RPS,pred}$ threshold to 0.8 (0.6), corresponding to a change in $M_{\rm HI}$ by 2 (1-)$\sigma$ with respect to the median relation of $M_{\rm HI}$ versus $M_*$, then 68\% (58\%) of the r1-RPS candidates still have $f_{RPS,pred}$ below the that.
The large fraction of galaxies with low values of $f_{RPS,pred}$ implies that the present ram pressure stripping may not strongly deplete $\hi$ in many of the affected galaxies (partly a result of sample bias toward the $\hi$-rich galaxies). 

The parameter $f_{RPS,pred}$ (and $f_{RPS}$) relates to the amount of strippable $\hi$, consisting of a spectrum of unresolved clouds with kinematics between those which are right to be accelerated by ram pressure and those which are right to disappear into the ICM. We investigate the post-RPS status quantified as $M_{{\rm HI}}(1-f_{RPS})$ and $M_{{\rm HI}}(1-f_{RPS, pred})$ for the RPS and r1-RPS candidates. 
For display purposes, we set the minimum post-RPS $M_{{\rm HI}}$ to be $10^{5.5}~M_{\odot}$. 

We compare the post-RPS $M_{\rm HI}$ at a given $M_*$ with the detection limit of WALLABY in Fig.~\ref{fig:HIMS}-c and d.
We estimate the WALLABY detection limit for $\hi$ flux,
\begin{equation}
f_{lim}/(Jy~km~s^{-1})= F_{thresh} F_{smooth} \sigma_{cube} 2v_{rot}
\end{equation}
where the line width $v_{rot}$ ($km~s^{-1}$) is the rotational velocity predicted from the baryonic Tully-Fisher relation \citep{McGaugh00} assuming an edge-on view. The baryonic mass is the sum of $M_*$ and 1.4 (to account for helium) times $M_{{\rm HI}}$ predicted from the mean relation of  $M_{{\rm HI}}$ versus $M_*$ taken from xGASS and extrapolated to the full $M_*$ range. The factor $F_{thresh}=3.5$ is the threshold for flux detection in units of the cube rms $\sigma_{cube}=2 mJy~beam^{-1}$. The factor $F_{smooth}$ is set to $1/7.75$, accounting for the maximum extent of smoothing in the channel maps (2 FWHM of Gaussian beams across), and in the velocity direction (15 channels) during the source finding. 
Our estimate for $f_{lim}$ and $M_{HI,lim}$ is only a rough approximation, for multiple smoothing kernels were used in the source finding by SoFiA, and SoFiA further exploits a reliability parameter to exclude unreliable threshold-based detections by comparing pixel distribution properties between the detected sources and pure noise \citep{Serra15, Serra12}. These steps are missing in our rough derivation of the detection limit, and may be partly responsible for missing detections at high-$M_*$ ($>10^9~M_{\odot}$) between the derived detection limit and the xGASS relation in Fig.~\ref{fig:HIMS}. 
Despite these complexities, the derived detection limit is close to the observed lower limit of detected fluxes at least for the low-$M_*$ galaxies (Fig.~\ref{fig:HIMS}-a), and is enough for the following analysis. 

In Fig.~\ref{fig:HIMS}-c and d, we examine whether the post-RPS status of the r1-RPS (RPS) candidates will be observable by WALLABY in order to check the efficiency of ram pressure stripping. It also helps assess whether it is feasible to study the early (relatively weak) stage of ram pressure stripping based on the relatively shallow data of WALLABY. In other words, if all the post-RPS $\hi$ masses were below the WALLABY detection threshold, then ram pressure stripping would be highly efficient in the Hydra cluster in depleting the total $\hi$, and WALLABY would not be very useful to study even the early stage of ram pressure stripping.
Based on the analysis of $f_{RPS}$ (Fig.~\ref{fig:HIMS}-c), 3 out of the 10 resolved RPS candidates will be undetected (below the WALLABY detection threshold) after the present stripping. Based on analysis of $f_{RPS,pred}$ (Fig.~\ref{fig:HIMS}-d), 39\% (23\%) of the r1-RPS candidates with $M_*$ below (above) $10^9~M_{\odot}$ will be undetected after the present stripping, but more than half of the r1-RPS candidates will only drop slightly in $M_{{\rm HI}}$ after the present removal, indicating the instantaneous ram pressure stripping to be weak in these galaxies. 

In Fig.~\ref{fig:HIMS}-d, we also compare these post-RPS $M_{{\rm HI}}$ at a given $M_*$ with the ALFALFA mean relation. From Fig.~\ref{fig:HIMS}-d, we can see that $M_{\rm HI}$ changes significantly (by $>1$dex) for a few of the r1-RPS candidates, but by a relatively small extent (a few 0.1 dex) for the remaining many r1-RPS candidates.
We calculate $M_{{\rm HI}}/M_{\rm HI,ALFALFA}$, the difference of $M_{\rm HI}$ from that indicated by the ALFALFA mean relation at the given $M_*$. Because many post-RPS $M_{\rm HI}$ drop below the displayed range in Fig.~\ref{fig:HIMS}-d, we show the distributions of $M_{{\rm HI}}/M_{\rm HI,ALFALFA}$ for the r1-RPS galaxies at the present and in the post-RPS status in Fig.~\ref{fig:fRPS_distr}-b. 
By doing so, the fraction of r1-RPS galaxies undergoing big or small changes in $M_{\rm HI}$ after the present stripping can be better seen. We can see that, after stripping the currently strippable $\hi$, most ($\sim$2/3) of the r1-RPS candidates will remain relatively gas rich with $\log M_{{\rm HI}}/M_{\rm HI,ALFALFA}>-1$, and only $\sim$1/3 of them will be severely depleted in $\hi$. 
We investigated the positions of the $1/3$ most stripped galaxies in the projected phase-space diagram (not displayed in this paper). Not surprisingly, they tend to locate in the region with $\Delta v_{rad}/\sigma_C>1$ and $d_{proj}/r_{200}<0.5$, i.e. the region with the highest level of $P_{ram}$. 

The distribution of the post-RPS $\hi$ richness (with respect to the WALLABY detection limit and with respect to the ALFALFA relation) helps explain the similarity of $M_{\rm HI}$ at a given $M_*$ between the r1-RPS and r1-non-RPS candidates. After a characteristic timescale for removing the currently strippable $\hi$ (the fraction indicated by $f_{RPS,pred}$), the strongly stripped galaxies will drop below the detection limit of WALLABY, the weakly stripped galaxies will only have a slightly decreased $M_{\rm HI}$, and some currently r1-non-RPS candidates will become r1-RPS candidates and recharge the population of $\hi$-rich r1-RPS population. In other words, the relatively narrow dynamical range of $M_{\rm HI}$ detectable by WALLABY, the cosmological context of galaxy infall, and the weak nature of a large fraction of the ram pressure stripping events may have worked together to produce a $M_{\rm HI}$-$M_*$ relation for the r1-RPS population that is very close to that of the r1-non-RPS population. 

To summarize, comparing the post-RPS $M_{\rm HI}$ to the WALLABY detection limit and to the ALFALFA $M_{\rm HI}$-$M_*$ relation both suggest that, the current strength of ram pressure stripping can significantly deplete some galaxies (strong ram pressure stripping, $f_{RPS}>0.9$), but only weakly reduce $M_{\rm HI}$ (weak ram pressure stripping, $f_{RPS}<0.9$) for a large fraction of the affected galaxies. 

At a first glance, a single temporal snapshot may not look particularly meaningful as the ram pressure will change (usually becoming strong) as galaxies move through the cluster; also, it may look obvious that $f_{RPS}$ should be low in many of these galaxies as they are not highly $\hi$ deficient. However, these results provide a first, comprehensive view of the weak ram-pressure regime in a nearby massive cluster, at a uniform detection limit. Although the instantaneous $f_{RPS,pred}$ is low in many cases, the compound effect is significant as galaxies travel through a large distance (and over a long time, $\sim$2 Gyr if traveling radially with a velocity of $\sigma_C$) from where ram pressure starts to strip $\hi$ ($\sim 1.25~r_{200}$), before reaching the core region ($<0.25 r_{200}$, see Sec.5.2, Fig.~\ref{fig:RPS_anchorforce}) where $\hi$ is stripped throughout the disks. Such a cumulative effect has been hinted at in past hydrodynamic simulations, as the gas tails appear early (e.g. \citealt{Tonnesen19}), and are as prevalent in cluster galaxies as for the Hydra cluster (e.g. \citealt{Yun19}) near the virial radius. \citet{Steinhauser16} showed in their hydrodynamic simulations that at least for the low-mass galaxy model ($M_*=8.25\times10^9~M_{\odot}$), the curve of growth of the stripped gas mass as a function of time is quite shallow (in contrast to an abrupt increase) during the ram pressure stripping history. Although the stripping of the outlying $\hi$ may not immediately affect the star-forming gas concentrated in the inner disk, it represents a shrinking of the gas reservoir for future star formation, and preludes the final gas depletion and star formation quenching. In order to break the degeneracy of HI depleting effects from ram pressure stripping, other environmental effects, and galactic internal mechanisms, it is necessary to consider the cumulative ram pressure stripping effect. In the future, many similar cluster observations will be provided by WALLABY, and combining them with cosmological simulations in a way like that of \citet{Rhee20, Oman16, Oman21} will eventually quantify the cumulative effects of weak ram pressure stripping.

\section{Ram pressure stripping of the resolved galaxies }
The $\hi$ images of resolved galaxies provide us with the opportunity to investigate how ram pressure stripping will deplete the $\hi$ in these galaxies. We note that these resolved galaxies typically (in 90\%) have $M_{\rm HI}>10^{9.1}~M_{\odot}$ and $M_*>10^{8.4}~M_{\odot}$, so they represent a relatively restricted sample. 

We investigate the capability of galaxies to restore their existing $\hi$ disks against ram pressure during the infall. 
We do a simplified experiment of putting each $\hi$ disk in our resolved sample at different projected distances $d$, and estimate the expected $\hi$ stripping fractions ($f_{RPS}(d)$). In the remaining part of this section, all estimates as a function of $d$ is denoted with ``(d)'', and the real galactic properties of the present are denoted with the subscript ``now'' to avoid confusion.
For each galaxy, we estimate the expected radial velocity offset at $d$ assuming the same infall acceleration profile as indicated by the projection averaged curve of $v_{esc}$ as a function of $d_{proj}$ (Fig.~\ref{fig:phase-space diagram}), 
\begin{equation}
\Delta v_{rad}(d)=(\Delta v_{rad,now}^2+(v_{esc}^2(d)-v_{esc,now}^2))^{0.5}.
\end{equation}
We then estimate the ram pressure $P_{ram}(d)$ at $v_{rad}(d)$ using the ICM density at $d$. We compare $P_{ram}(d)$ with the anchoring force profile of the galaxy, and calculate $f_{RPS}(d)$.  

 We plot the curve of $f_{RPS}(d)$ as a function of $d$ for both resolved RPS and non-RPS candidates in Fig.~\ref{fig:exp1}. Each curve describes the effective ram pressure strength as a function of $d$, measured by the anchoring force of the current disk. We define $d_{x}$ to be the projected cluster centric distance where $f_{RPS}(d)=x$ in the curve of a galaxy. So that $d_{0.1}$ indicates the maximum $d$ where the present $\hi$ disk can start to be stripped. From Fig.~\ref{fig:exp1}-b, there is quite a wide distribution of $d_{0.1}$ for the currently non-RPS candidates, ranging from 0.5 to 1.7 $r_{200}$, with a median value of 1.1$r_{200}$. It is consistent with the wide $d_{proj}$ range found for the r1-RPS candidates. 
 
  In the following, we will discuss questions related to the beginning and process of ram pressure stripping based on these curves. 

 %%%%%
    
 \subsection{ Time needed to strip the currently strippable HI in the RPS candidates}
For each of the resolved RPS candidates, at a projected distance of $d_{0.1}$, which by definition is higher than $d_{now}$, the ram pressure is already high enough to strip some of the currently strippable $\hi$, but the strippable $\hi$ at $d_{0.1}$ can still be partly observed at the present because the strippable $\hi$ clouds do not immediately disappear into the ICM. The parameter $d_{0.1}-d_{now}$ is an indicator ($\sim$ upper limit) of the timescale ($\tau_{RPS}$) needed for ram pressrue to remove the strippable $\hi$. 

\begin{figure*} 
\centering
\fig{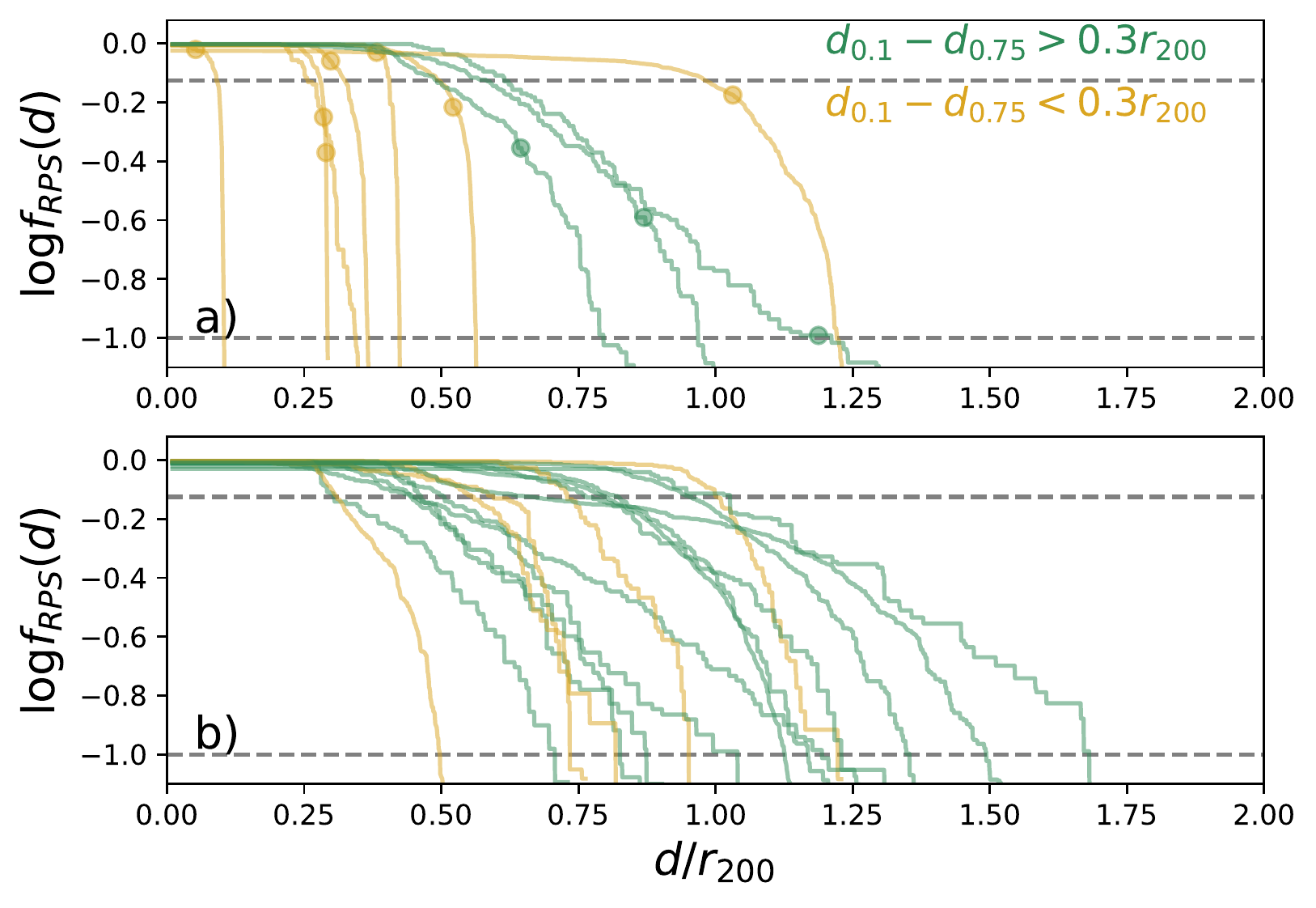}{14cm}{}
\caption{ The relation between expected fraction of strippable $\hi$ ($f_{RPS}(d)$) at different projected cluster centric distances ($d/r_{200}$) for the resolved galaxies. The steep (shallow) curves, which take less than (at least) 0.2$r_{200}$ for $f_{RPS}(d)$ to increase from 0.1 to 0.75, are in orange (green). The horizontal dashed lines mark the positions for $f_{RPS}(d)$ equaling 0.1 and 0.75. {\bf Panel a:} for the RPS candidates; their observed $d_{proj}$ and $f_{RPS}$ are marked as dots. {\bf Panel b:} for the non-RPS candidates; $d_{proj}$ are larger than the displayed maximum $d$ of each curve. }
\label{fig:exp1}
\end{figure*}

 From Fig.~\ref{fig:exp1}-a, $d_{0.1}-d_{now}<0.2r_{200}$ for all the resolved RPS candidates, and also $<0.1r_{200}$ except for two galaxies (239 and 244 in Fig.~\ref{fig:atlas}, with WALLABY ID J104059-270456 and J104142-284653 respectively). A distance of $0.1r_{200}$ takes $\sim$200 Myr for a galaxy with velocity $\sigma_C$ to travel through, suggesting that ram pressure stripping is a relatively fast mechanism (with a timescale of $\sim$ hundreds Myr) when removing the strippable gas. 

\begin{figure} 
\centering
\fig{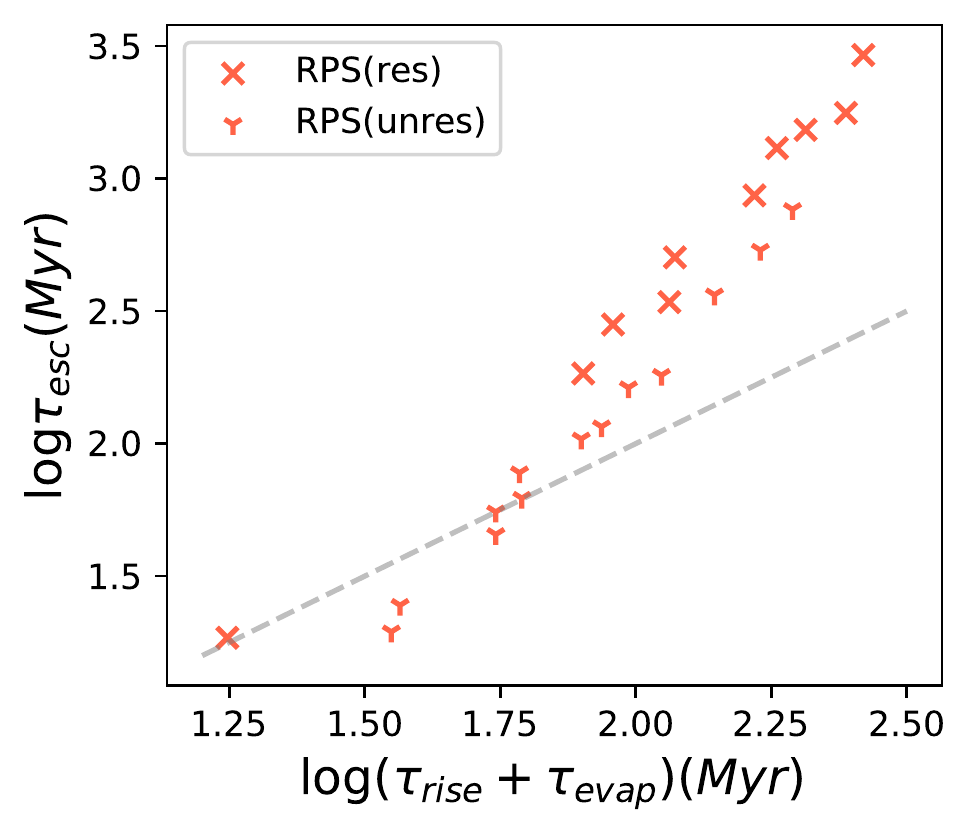}{8cm}{}
\caption{ The estimated timescales for ram pressure stripped clouds to escape ($\tau_{esc}$) and to be evaporated ($\tau_{rise}+\tau_{evap}$). The resolved (unresolved) RPS candidates are plotted in red crosses (down-tri shapes).  The dashed line marks the position of $\tau_{esc}=\tau{rise}+\tau{evap}$. }
\label{fig:timescales}
\end{figure}

We roughly estimate $\tau_{RPS}$ for the resolved RPS candidates by considering two scenarios. In the first scenario, the RPS is strong and the stripped  $\hi$ cloud quickly reaches the escape velocity. The escape timescale is estimated as 
$\tau_{esc}= \sqrt{2} v_{rot} \Sigma_{ {\rm HI},p50} /P_{ram}$ \citep{Vollmer01},
 where $\Sigma_{ {\rm HI}, p50}$ is the median $\Sigma_{{\rm HI}}$ of the currently strippable $\hi$. 
 In the second scenario, the ram pressure stripping is weak so that the stripped cloud is evaporated in the ICM after leaving the $\hi$ disk before being accelerated enough to escape. The timescale is the sum of the time needed for the cloud to rise a distance $h$ above the disk, and the time needed for the cloud to be evaporated. The former timescale is estimated as $\tau_{rise}= (2 h \Sigma_{ {\rm HI}, p50}/P_{ram})^{0.5}$ \citep{Vollmer01}, where we assume $h=5~kpc$. 
According to \citet{Cowie77}, the evaporating timescale is estimated as 
 \begin{equation} \label{eq3}
\begin{split}
\tau_{evap} & =\frac{9.07\times10^7  (N_{ {\rm HI}, p50}/10^{20}~cm^{-2} ) }{ 2.73 ( n_{{\rm ICM}} / cm^{-3})  (T/ K)^{0.5} \sigma_0^{3/8}    }  yr \\
\sigma_0 & = \frac{(T/1.54\times10^7~K)^2 }{ ( n_{{\rm ICM}} / cm^{-3}) (R_c/pc )  } 
\end{split}
\end{equation}
where $n_{{\rm ICM}}$ is the ICM density, $N_{ {\rm HI}, p50}$ is the column density corresponding to $\Sigma_{ {\rm HI}, p50}$, the Hydra ICM temperature $T= 3.76\times10^7~K$ \citep{Eckert11}, and as in \citet{Vollmer01} we assume a cloud radius $R_c=10~pc$.
Then $\tau_{RPS}$ is estimated as the lesser of $\tau_{rise}+\tau_{evap}$ and $\tau_{esc}$.
In presence of magnetic fields, the real $\tau_{evap}$ can be a few times longer, but ignoring this effect does not significantly affect our results because $\tau_{rise} \gg \tau_{evap}$ in all cases. We also ignored the gravitational force from the galaxy when estimating $\tau_{esc}$ and $\tau_{rise}$, but this uncertainty is mitigated by the fact that most of the ram pressure stripped pixels in the $\hi$ image have $P_{ram}>2F_{anchor}$ (Fig\ref{fig:atlas}).
 Nevertheless, these timescales should only be viewed as an order-of-magnitude estimate, due to uncertainties from the disk inclination and other factors. From Fig.~\ref{fig:timescales}, most of the ram pressure stripped gas is lost through evaporation in the RPS candidates, and the values of $\tau_{RPS}$ are consistent with our interpretation of the $f_{RPS}(d)$ curves of the resolved RPS candidates. The values are also consistent with theoretical predications in the literature \citep[e.g.,][]{Abadi99,Roediger06,Tonnesen19}. They support the view that ram pressure stripping is a relatively quick process when removing the strippable gas, even for cases where the ram pressure is low and thus $f_{RPS}$ is low. 

\subsection{Time needed to strip the existing $\hi$ reservoir  }
\label{sec:exp2}
Another noticeable feature from Fig~\ref{fig:exp1} is that the curves of $f_{RPS,exp}$ have distinct slopes. 
We use $d_{0.1}-d_{0.75}<(>)0.3r_{200}$ as the criteria of selecting the steep (shallow) $d_{proj,ex}$ curves. The parameter $d_{0.1}-d_{0.75}$ can be viewed as an indicator of the length of time needed to remove the majority of the existing $\hi$ reservoir in a galaxy with ram pressure. Although the existing $\hi$ mass is different from the initial $\hi$ mass upon infall, investigating the time needed to significantly deplete it is still meaningful, as most of the galaxies in our sample are still relatively $\hi$-rich. Removing $75\%$ of the $\hi$ is equivalent to decreasing $M_{\rm HI}$ by 0.6 dex. A distance of $0.3r_{200}$ takes $\sim$600 Myr for a galaxy with velocity $\sigma_C$ to travel through. If we change the indicator $d_{0.1}-d_{0.75}$ to $d_{0.1}-d_{0.9}$, the related distance (traveling time) will increase, and more of the resolved galaxies will be classified with shallow curves, but the conclusions are similar.

Galaxies with shallow $f_{RPS}(d)$ curves tend to be stripped for a relatively long period of time ($\ge600$ Myr, if we assume galaxies radially travel with an average speed of $\sigma_C$). Fig~\ref{fig:exp1} suggests that, 
the resolved galaxies on shallow curves tend to have larger $d$ than those on steep curves, with a K-S test probability of 0.01 for the distributions of $d_{proj}$ to be similar among the two sub-samples. Due to the relatively small sample size, we do not find statistically significant differences in other properties between the galaxies with shallow and steep $f_{RPS}(d)$ curves. Fig~\ref{fig:exp1}-b suggests that, most ($\sim$70\%) of the currently non-RPS candidates will start the ram pressure stripping process with shallow $f_{RPS}(d)$ curves.

\begin{figure*} 
\centering
\fig{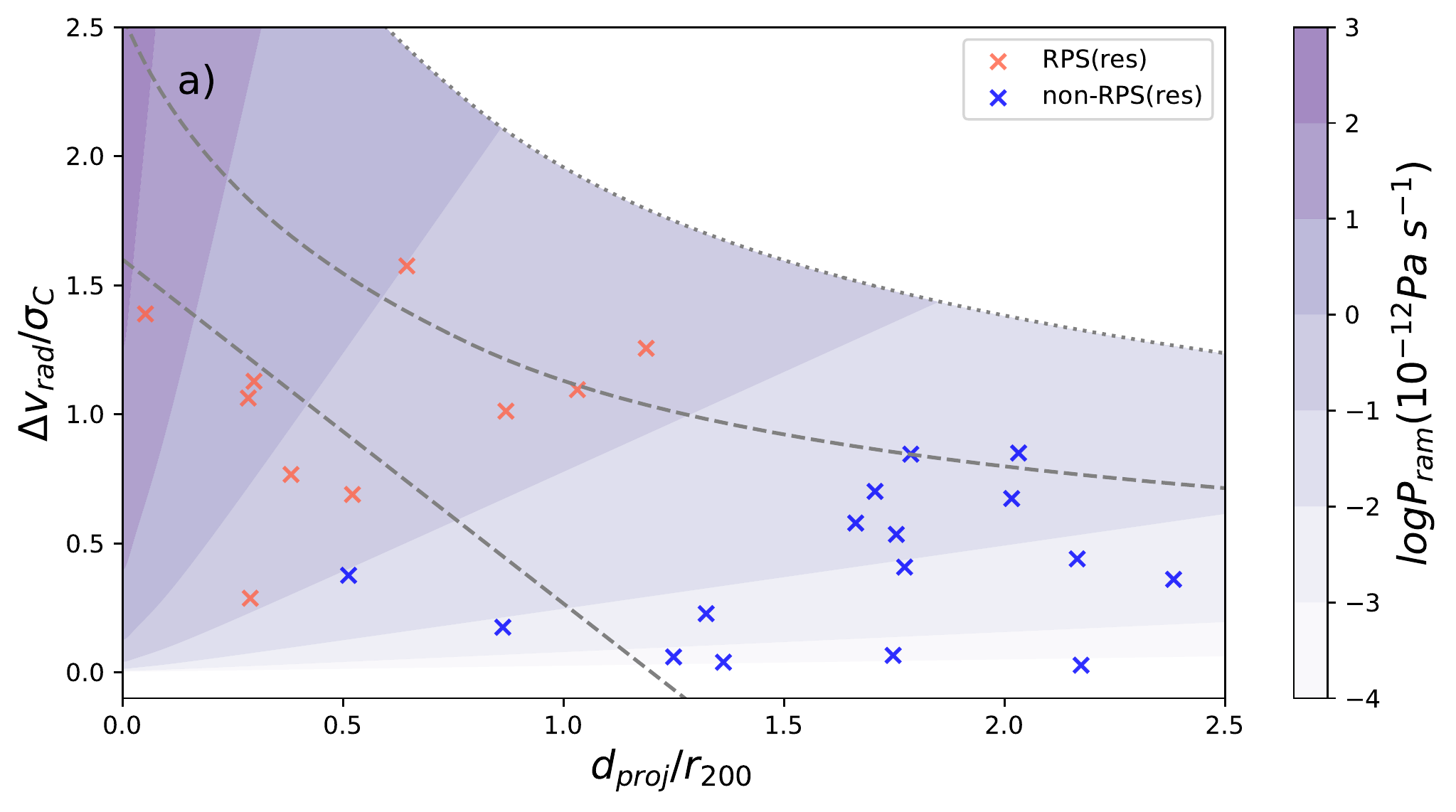}{14cm}{}

\fig{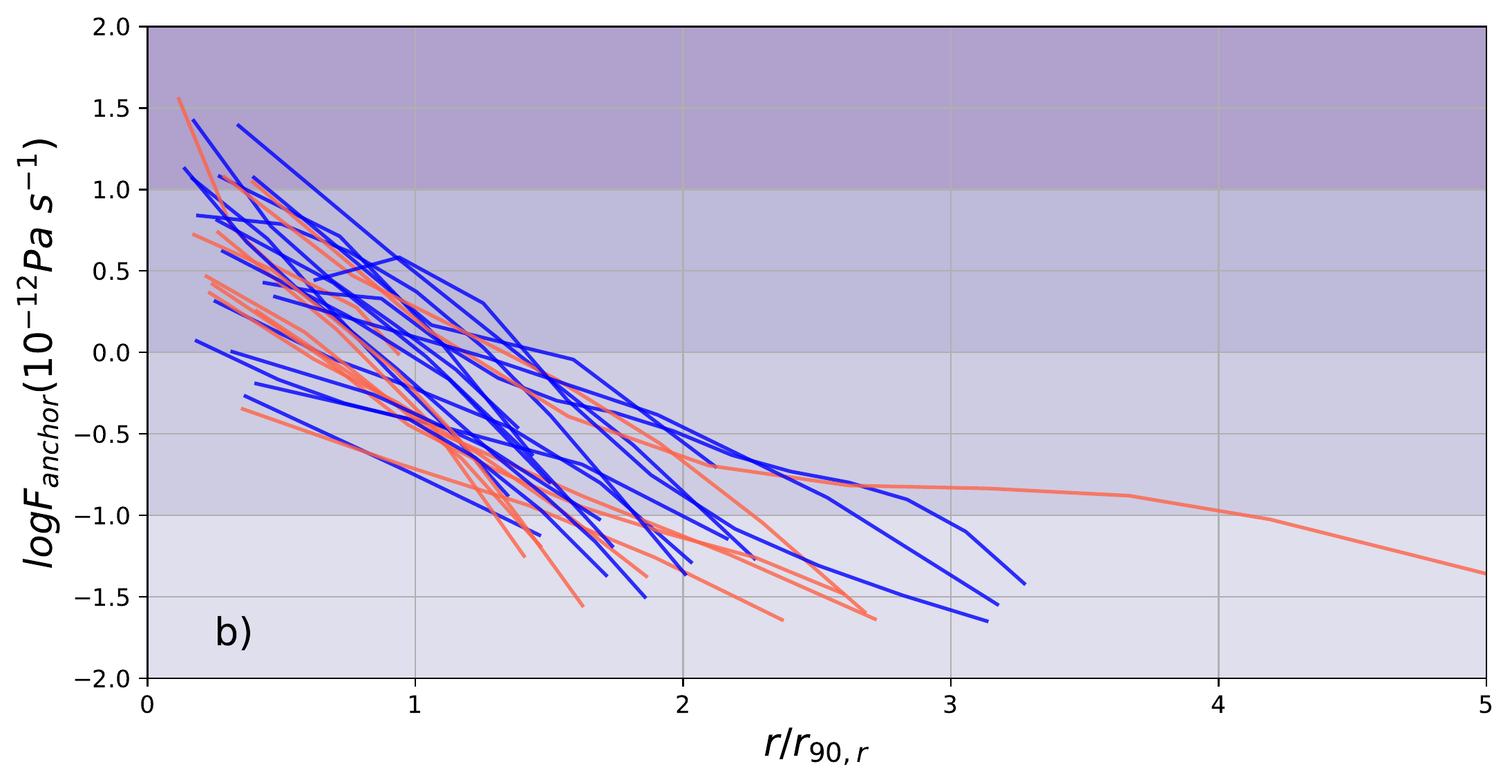}{14cm}{}
\caption{ The ram pressure level of the Hydra cluster and the anchoring force profiles of all the resolved galaxies.  
{\bf Panel a:} 
The distribution of ram pressure ($P_{ram}$) in the projected phase-space diagram. The projected phase-space diagram is the same as in Fig.~\ref{fig:phase-space diagram}. The resolved RPS and non-RPS candidates are in red and blue crosses respectively.
{\bf Panel b:}
The azimuthally averaged anchoring force ($F_{anchor}$) radial profile of galaxies. The resolved RPS and non-RPS candidates are in red and blue respectively. The radii of each profile are normalized by the 90\%-light radius of the galaxy in the $r$-band ($r_{90,r}$).
The largest radius for each galaxy is determined by the size of the $\hi$ disk. The space of $F_{anchor}$ is color coded with the same scales as $P_{ram}$ in panel a.
}
\label{fig:RPS_anchorforce}
\end{figure*}

The shallow $f_{RPS}(d)$ profiles, particularly for the non-RPS candidates in the future, are likely due to the shallow ICM density profile beyond the core region of the cluster, and the extended nature of $\hi$ disks.
Fig.~\ref{fig:RPS_anchorforce} demonstrates these two effects. In Fig.~\ref{fig:RPS_anchorforce}-a, $\log P_{ram}$ rises by three orders of magnitude when $d_{proj}<0.5r_{200}$, and only rises by two orders of magnitude when $2.5r_{200}>d_{proj}>0.5r_{200}$. In Fig.~\ref{fig:RPS_anchorforce}-b, the radial range of $F_{anchor}$ profiles ($\hi$ disks) extends far beyond the optical disk ($r_{90,r}$) in many galaxies, and the profiles rise relatively smoothly (roughly exponentially) toward the galaxy centers. In these resolved galaxies, $F_{anchor}>10^{-13}~Pa~s^{-1}$ throughout most of the radial ranges.
A ram pressure of $10^{-13}~Pa~s^{-1}$ already starts at $d_{proj}\sim1.25 r_{200}$ in the infall region (far beyond the triangle of virialized region) of the projected phase-space diagram (Fig.~\ref{fig:RPS_anchorforce}-a), consistent with where RPS and r1-RPS candidates start to be detected. 
 On the other hand, $P_{ram}$ needs to rise above $10^{-12}~Pa~s^{-1}$ (the typical $F_{anchor}$ at $r_{90,r}$) in order to effectively strip the $\hi$ within the optical disks. This requires the infalling galaxies to reach $d_{proj}<0.5r_{200}$. It needs to further rise above $10^{-11}~Pa~s^{-1}$ in order to completely strip the $\hi$ from the majority of those resolved galaxies, which requires the infalling galaxies to reach $d_{proj}<0.25r_{200}$. Indeed very few galaxies (in total 4) are detected by WALLABY in the triangular region (the ``stripping zone'' defined in \citealt{Jaffe15}) of the projected phase-space diagram where $P_{ram}>10^{-11}~Pa~s^{-1}$. The key point to make here is that, galaxies travel for a large $d_{proj}$ (hence time) between the onset and end of ram pressure stripping, emphasising the importance of the effect of weak but cumulative ram pressure stripping forces.

Thus, we have observed diversities in ram pressure stripping not only for the instantaneous strength ($f_{RPS}$), but also for the averaged speed ($d_{proj,ex,0.1}-d_{proj,ex,0.75}$). For the non-RPS candidates, there is further a diversity in $d_{proj}$ for ram pressure stripping to onset. Modern hydrodynamic simulations under a cosmological context predicted that most galaxies lose the majority of $\hi$ through ram pressure stripping in one passage of infall into massive clusters, but there is a large scatter in the speed of $\hi$ depletion \citep{Jung18, Oman21, Lotz19}, not only because of scatter in the infall orbits, but also because of scatter in the initial conditions. Our results are qualitatively consistent with the latter. The early, strong and rapid ram pressure stripping associated with large $d_{0.1}$, high $f_{RPS}$, and low $d_{0.1}-d_{0.75}$ is more likely to link with a fast depletion, while late, weak and slow ram pressure stripping associated with small $d_{proj,ex,0.1}, $low $f_{RPS}$ and high $d_{proj,ex,0.1}-d_{proj,ex,0.75}$ is more likely to link with a slow depletion. A closer comparison in distributions of $f_{RPS}$ and $d_{proj,ex,0.1}-d_{proj,ex,0.75}$ between the observed and simulated datasets may help advance our understanding of the role of ram pressure stripping in galaxy evolution, particularly when more clusters are observed in the future.

%\citep{Rasmussen12, Haines13, Paccagnella16, Boselli16, Muzzin14, Wetzel13, Haines15, Maier19}. 
The diversity in ram pressure stripping strength$\slash$speed may further link to the diversity in star forming histories of cluster galaxies. There has been a debate in the observational literature regarding the time needed to quench the SFR of galaxies infalling into clusters. Analysis of star forming histories has suggested a slow mode (2-4 Gyr, \citealt{Wetzel13,Paccagnella16}, a fast mode ($<1$ Gyr, \citealt{Muzzin14, Boselli16}), and a combination of the two modes \citep{Haines15, Maier19}. In those studies, slow quenching was typically attributed to the effect of starvation where galaxies lose their hot gas halos due to ram pressure stripping, stop gas accretion, and gradually quench when the remaining cold gas is consumed \citep{Larson80}. Fast quenching was typically linked to strong ram pressure stripping that effectively removes $\hi$ throughout the disks \citep{Jaffe15}. Our results suggest that ram pressure starts to remove $\hi$ long before it affects the whole $\hi$ disk. Although there can be considerable delay between the significant removal of $\hi$ and the final quenching of star formation \citep{Cortese09, Boselli14, Oman21}, the scatter in the timescale for ram pressure stripping to deplete most of the $\hi$ in a disk may contribute to the scatter in the quenching timescale.
The effect of weak ram pressure stripping of $\hi$ on star formation is close to the traditionally defined starvation caused by the removal of the hot gas halo \citep{Larson80}, because it gradually shrinks the extended $\hi$ disks, which serve as the reservoir instead of the direct material for star formation. The extended $\hi$ gradually flows radially inward \citep{Schmidt16} to fuel the inner disks, and the inner $\hi$ is converted to the molecular gas and then stars \citep{Bigiel08, Wang20a}. But stripping of the $\hi$ should be more efficient than starvation in quenching, as $\hi$ is a closer step to star formation than the hot gas halo. The weak ram pressure stripping of $\hi$ thus adds an intermediate step between the gentle starvation that starts at a large cluster-centric radii ($\sim3r_{200}$, \citealt{Bahe15}), and the violent strong ram pressure stripping prevalent near the cores of clusters ($<0.3r_{200}$, \citealt{Jaffe15}), possibly leading to a stronger environmental effect and less room for galactic internal mechanisms (like stellar feedbacks) in interpreting the distribution of $M_{\rm HI}$ and SFR in cluster galaxies.

\subsection{Caveats}
Because the estimates of ram pressure stripping strength suffer from uncertainties of projection, orbits and other effects, 
the analysis above should be interpreted in a statistical sense, instead of a case-by-case manner.
When interpreting our results, we ignore many physical processes including the cooling of ionized gas, the falling back of the stripped $\hi$,  the depletion due to star formation and feedback, the redistribution of $\hi$ within the disk, the change in orbital directions, external gravitational effects. Also, the analysis is not about removing all the $\hi$ from the initial conditions upon infall, but about removing what's left at the current epoch.
 Thus the analysis should be viewed as qualitative, with order-of-magnitude estimates. 
 
Quantifying the star forming history (as in \citealt{Boselli16}), using dynamic simulations combined with $\hi$-richness and star forming history to constrain the infall orbits (\citealt{Vollmer01}), and$\slash$or directly comparing the snapshots of ram pressure stripping presented in this paper with cosmological hydrodynamic simulations in the future may assist in gaining further insight into the gas depletion and star formation quenching processes in clusters.

\section{Summary}
With the WALLABY observations, we for the first time map the $\hi$ in the Hydra cluster out to 2.5$r_{200}$ (and full coverage out to 2$r_{200}$), detecting 105 members or infalling galaxies and resolving 27 of them. We quantify the extent of $\hi$ ram pressure stripping based on the $\hi$ images for the resolved galaxies, and also based on predicted $\hi$ radial distributions for all the galaxies.
We emphasize that the analysis is limited to the $\hi$ rich galaxies which are at a relatively early stage of being processed by the cluster environment. Our main results are summarized below. 

\begin{enumerate}
\item A large fraction of $\hi$-rich galaxies in and near the Hydra cluster are affected by ram pressure stripping. Within a projected distance of $1.25r_{200}$, over two thirds of $\hi$-detected galaxies are likely affected by ram pressure stripping (the r1-RPS population). 

\item A large fraction of the ram pressure affected galaxies are likely being weakly stripped at the current time. With the present level of ram pressure, $M_{{\rm HI}}$ in around one third of the r1-RPS candidates may significantly drop (by $>1$ dex), but for the remaining two thirds, $M_{{\rm HI}}$ only slightly changes (by a few 0.1 dex). The cumulative effects of weak ram pressure stripping processes need to be further investigated.

\item At least for the resolved galaxies ($M_{\rm HI}>10^{9.1}~M_{\odot}$ and $M_*>10^{8.4}~M_{\odot}$) once onset, ram pressure stripping is likely a slow process ($\gtrsim$ 600 Myr) for depleting the existing reservoir of $\hi$ in the galaxies. This is because for the $\hi$-rich galaxies with extended $\hi$ disks, ram pressure stripping can start at a relatively large cluster-centric distance ($d_{proj}\sim1.25 r_{200}$, see Reynolds et al. 2021 for a detailed analysis for one such galaxy), where the ram pressure changes slowly with cluster-centric distance. The different $d_{proj}$ for ram pressure stripping to onset, and the different speeds at which $f_{RPS}$ rises as a function of $d_{proj}$ may partly explain the scatter of quenching speed found for the SFR of cluster galaxies in the literature \citep{Wetzel13,Haines15,Boselli16}. 

\item For most of the resolved RPS candidates, stripping the strippable $\hi$ is likely a relatively quick process, on time scales of $\lesssim$200 Myr, consistent with theoretical predictions. This short timescale is supported by the short distance between the current $d_{proj}$ and the predicted larger $d_{proj}$ where part (10\%) of the current $\hi$ disk could already be stripped. It is also supported by the rough estimates of $\tau_{RPS}$ based on the cluster and galaxy properties, and simple kinematical and thermal models.
\end{enumerate}

We look forward to expanding our analysis with larger samples from future WALLABY observations, supported by hydrodynamic and semi-analytical simulations.

\acknowledgments
We thank R. White, J. Fu,  J.-Y. Tang, Z.-L. Zhang, L. Cortese, K. Oman, A. Stevens, A. Bosma for useful discussions. JW acknowledges support from the National Science Foundation of China (12073002, 11721303). Parts of this research were supported by High-performance Computing Platform of Peking University. Parts of this research were conducted by the Australian Research Council Centre of Excellence for All Sky Astrophysics in 3 Dimensions (ASTRO 3D), through project number CE170100013. This project has received funding from the European Research Council (ERC) under the European Union's Horizon 2020 research and innovation programme (grant agreement no. 679627; project name FORNAX). FB acknowledges funding from the European Research Council (ERC) under the European Union’s Horizon 2020 research and innovation programme (grant agreement No.726384/Empire). LVM acknowledges financial support from the grants AYA2015-65973-C3-1-R and RTI2018-096228- B-C31 (MINECO/FEDER, UE),  as well as from the State Agency for Research of the Spanish MCIU through the "Center of Excellence Severo Ochoa" award to the Instituto de Astrofísica de Andalucía (SEV-2017-0709). SHO acknowledges support from the National Research Foundation of Korea (NRF) grant funded by the Korea government (Ministry of Science and ICT: MSIT) (No. NRF-2020R1A2C1008706). 

The Australian SKA Pathfinder is part of the Australia Telescope National Facility which is
managed by CSIRO. Operation of ASKAP is funded by the Australian Government with support
from the National Collaborative Research Infrastructure Strategy. ASKAP uses the resources of
the Pawsey Supercomputing Centre. Establishment of ASKAP, the Murchison Radio-astronomy
Observatory and the Pawsey Supercomputing Centre are initiatives of the Australian
Government, with support from the Government of Western Australia and the Science and
Industry Endowment Fund. We acknowledge the Wajarri Yamatji people as the traditional
owners of the Observatory sites.

The Pan-STARRS1 Surveys (PS1) and the PS1 public science archive have been made possible through contributions by the Institute for Astronomy, the University of Hawaii, the Pan-STARRS Project Office, the Max-Planck Society and its participating institutes, the Max Planck Institute for Astronomy, Heidelberg and the Max Planck Institute for Extraterrestrial Physics, Garching, The Johns Hopkins University, Durham University, the University of Edinburgh, the Queen's University Belfast, the Harvard-Smithsonian Center for Astrophysics, the Las Cumbres Observatory Global Telescope Network Incorporated, the National Central University of Taiwan, the Space Telescope Science Institute, the National Aeronautics and Space Administration under Grant No. NNX08AR22G issued through the Planetary Science Division of the NASA Science Mission Directorate, the National Science Foundation Grant No. AST-1238877, the University of Maryland, E{\"o}tv{\"o}s Lor{\'a}nd University (ELTE), the Los Alamos National Laboratory, and the Gordon and Betty Moore Foundation.

\bibliographystyle{apj} % 200
{}
%\bibliography{walla_hydra}

\appendix
\section{Atlas and table, and additional information on individual resolved RPS candidates}

\begin{figure*} 
\centering
\includegraphics[width=4cm]{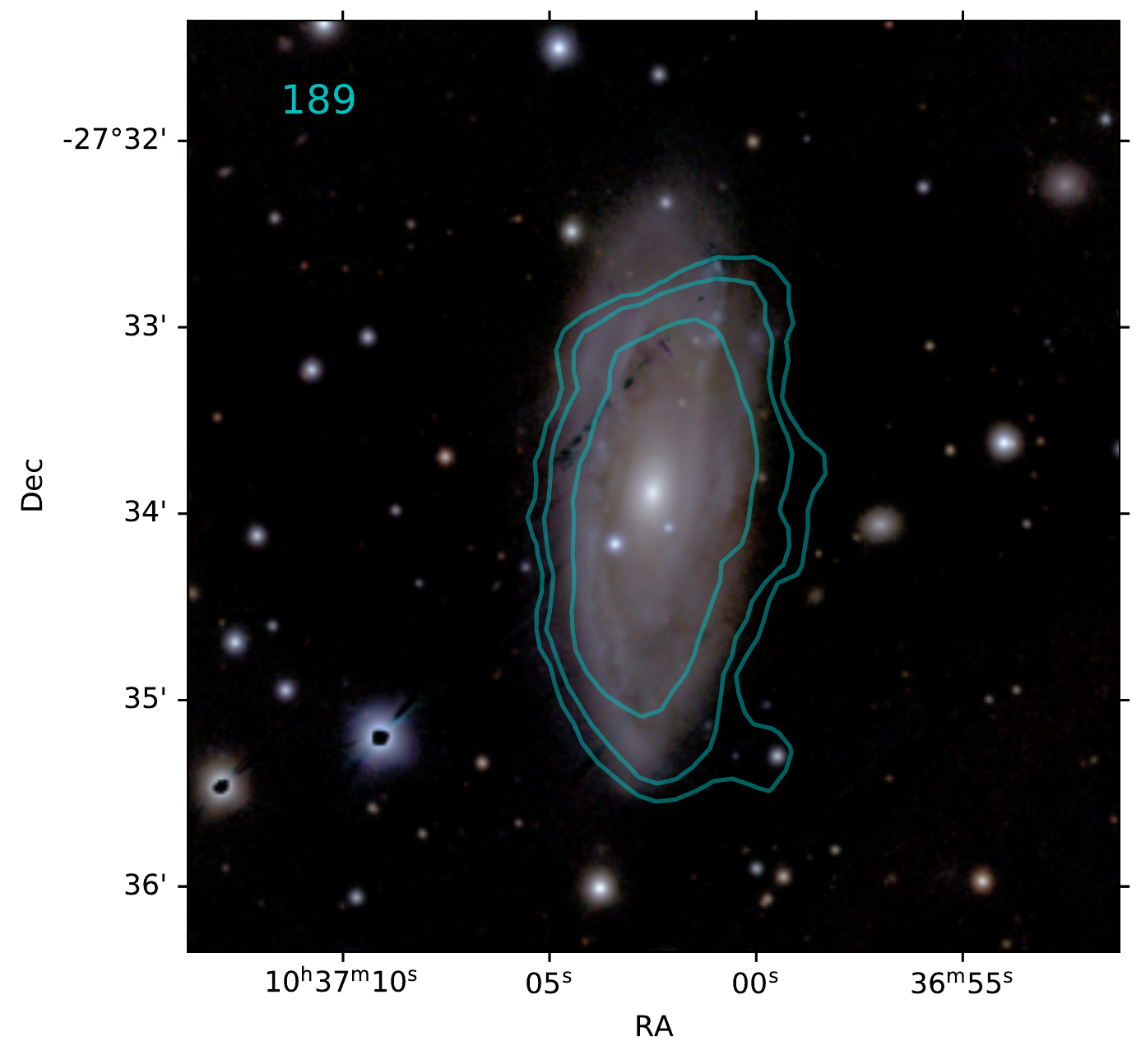}
\includegraphics[width=4cm]{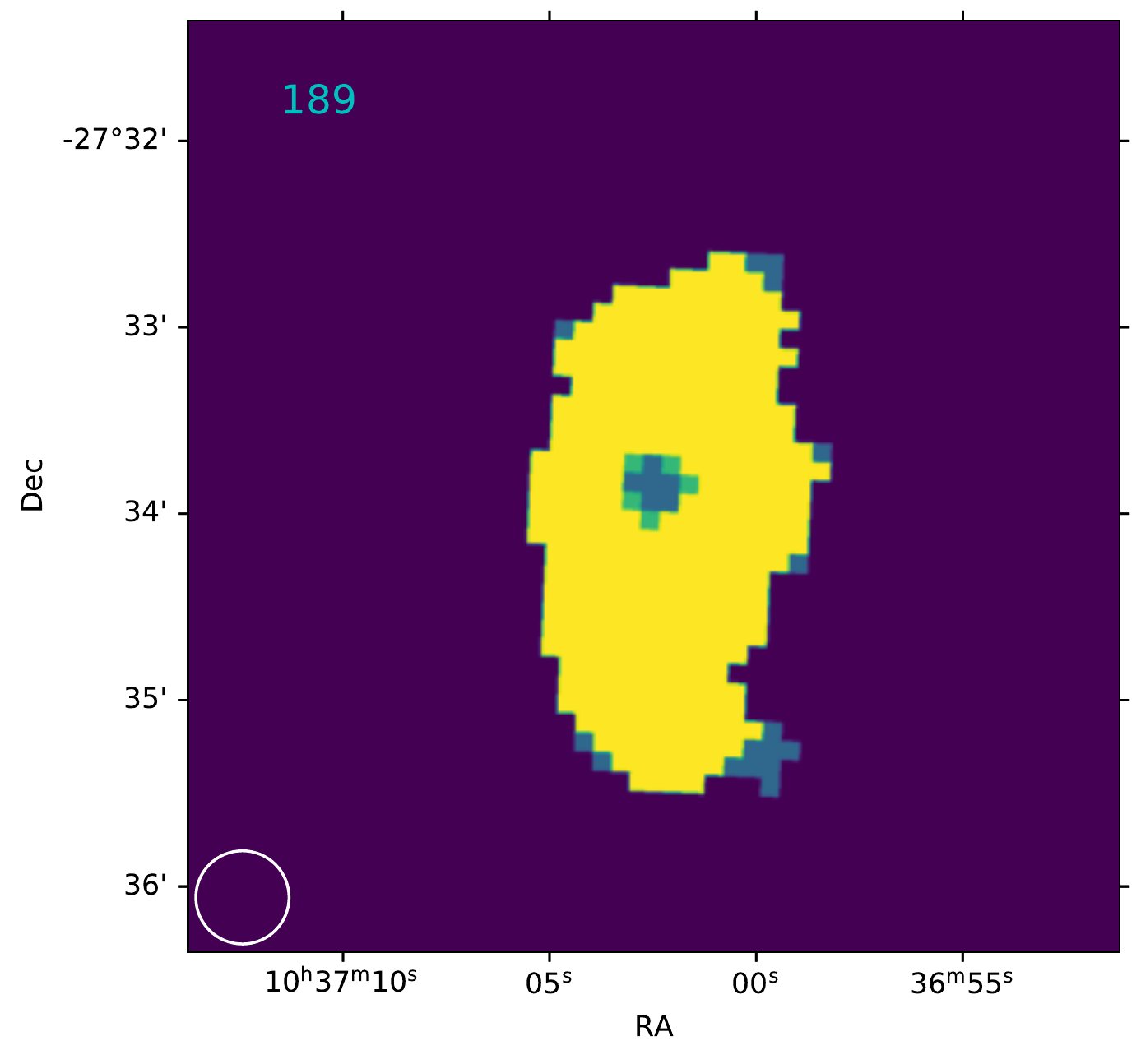}
\includegraphics[width=4cm]{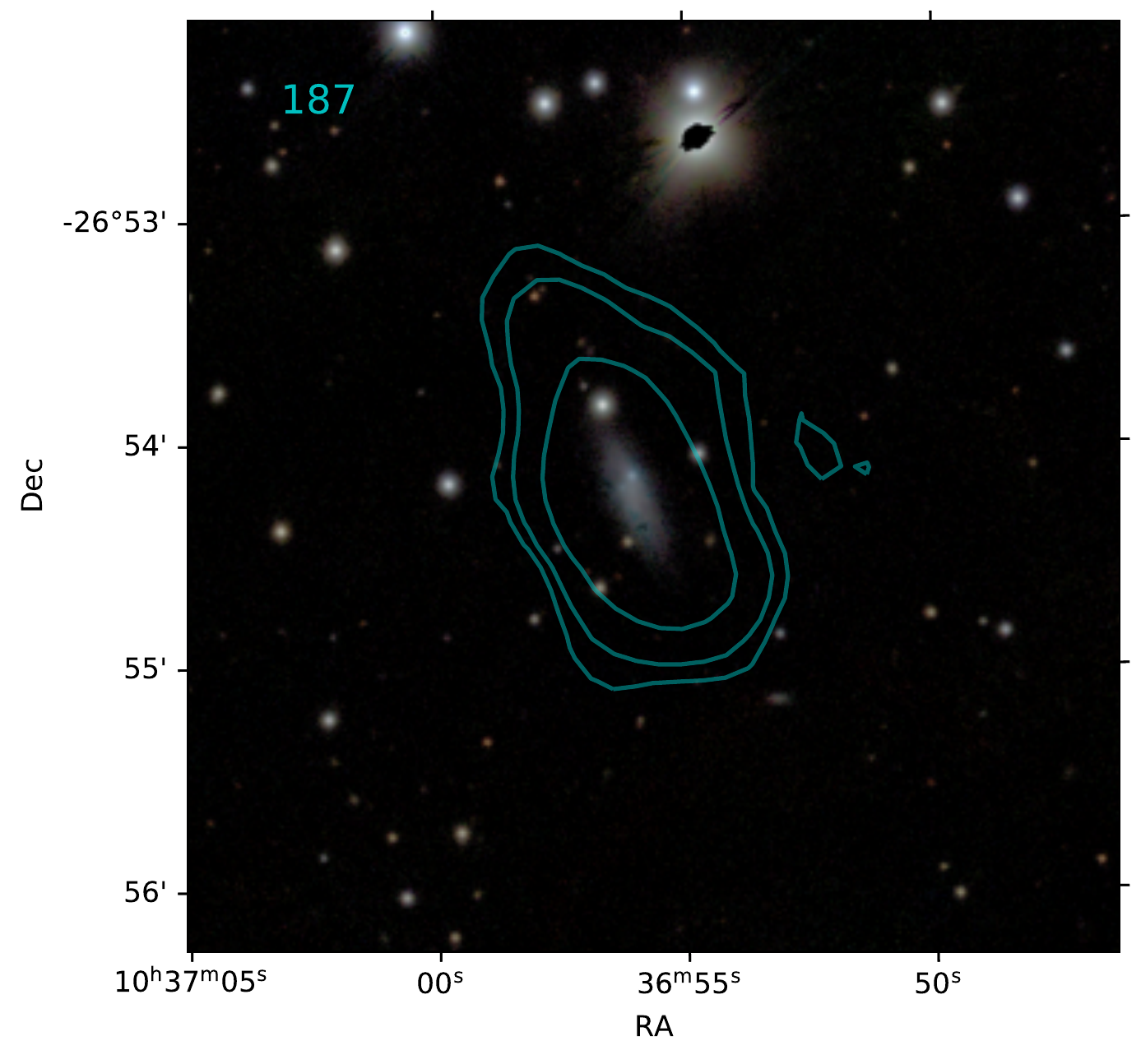}
\includegraphics[width=4cm]{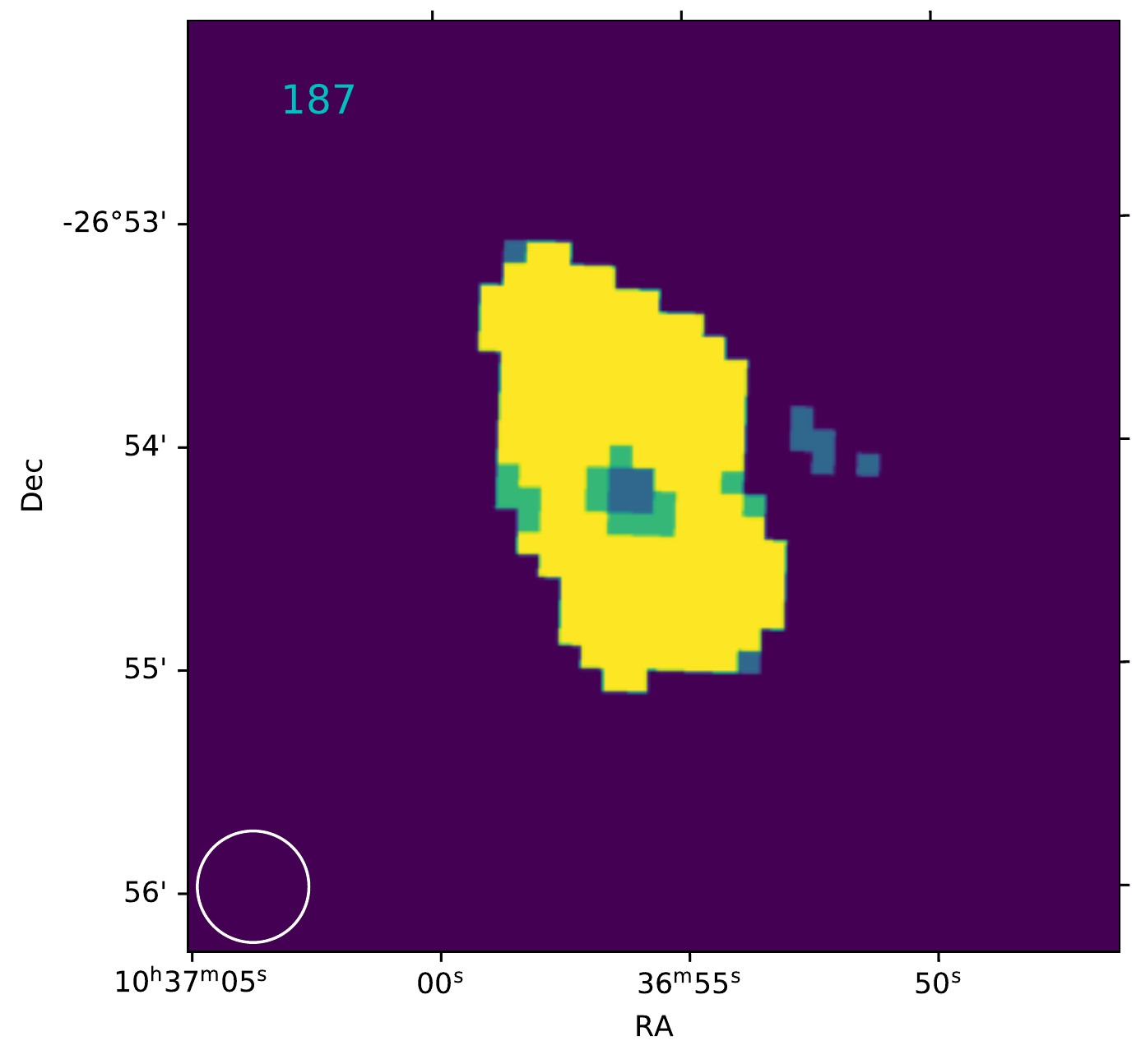}
\includegraphics[width=4cm]{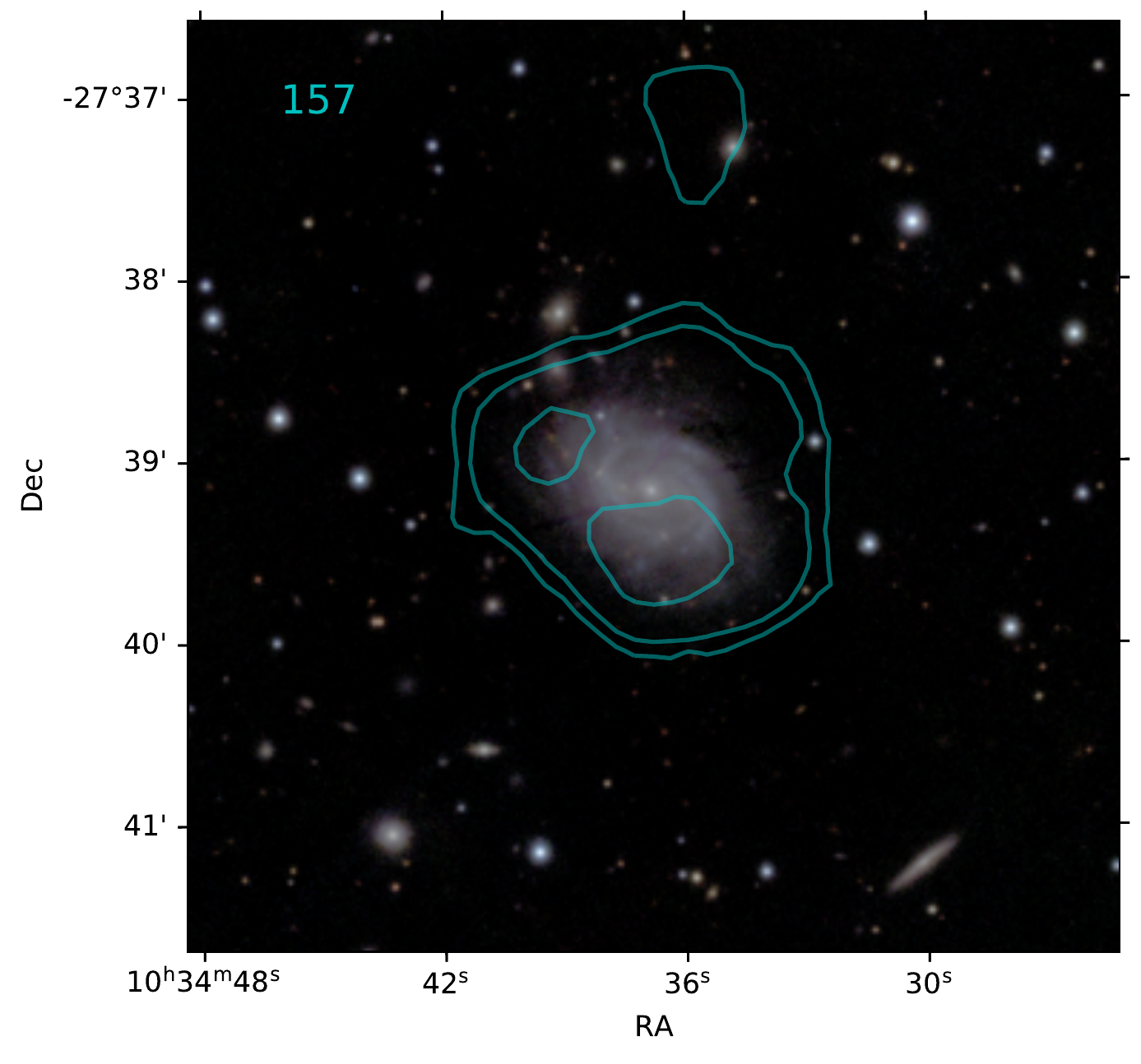}
\includegraphics[width=4cm]{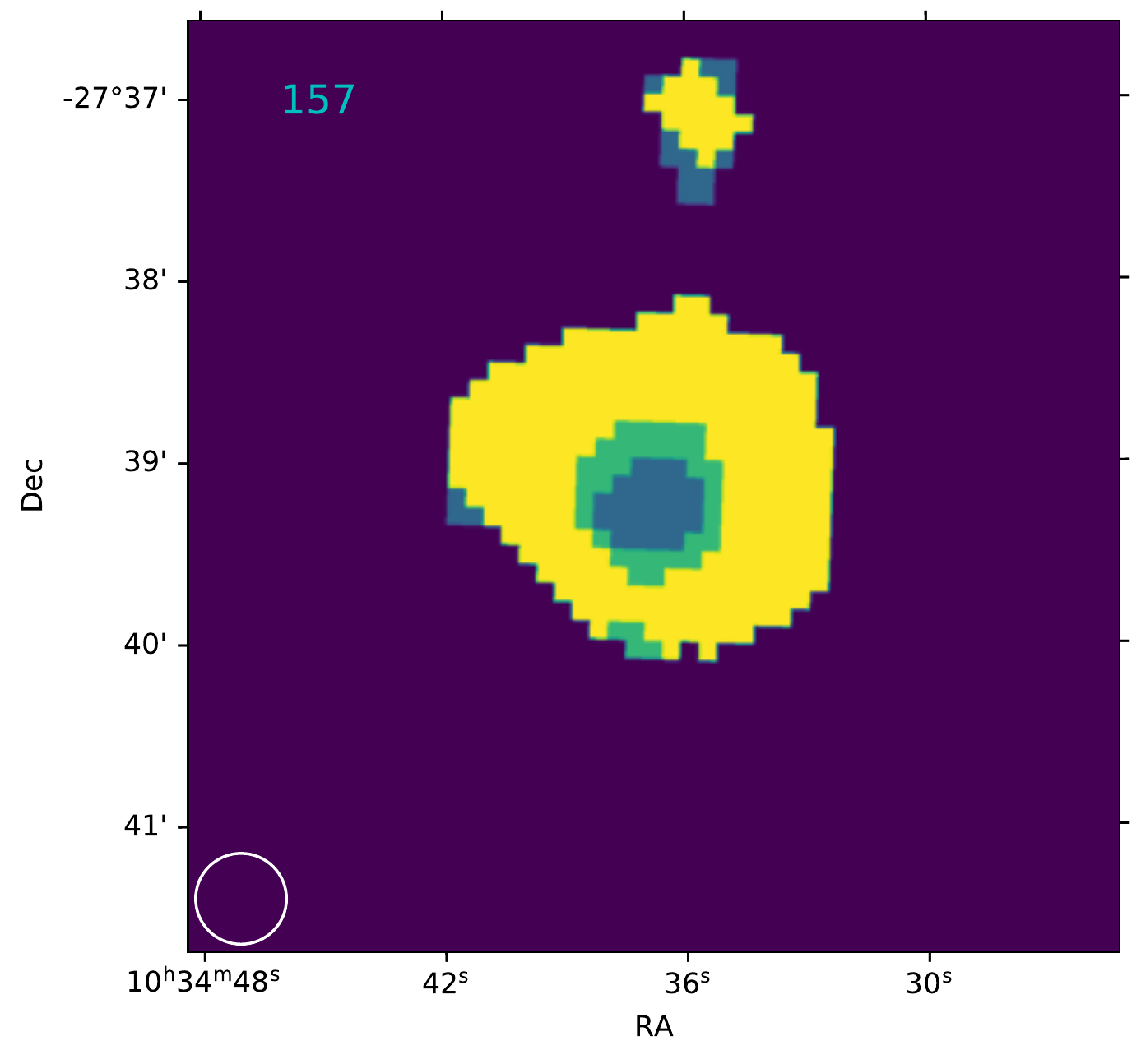}
\includegraphics[width=4cm]{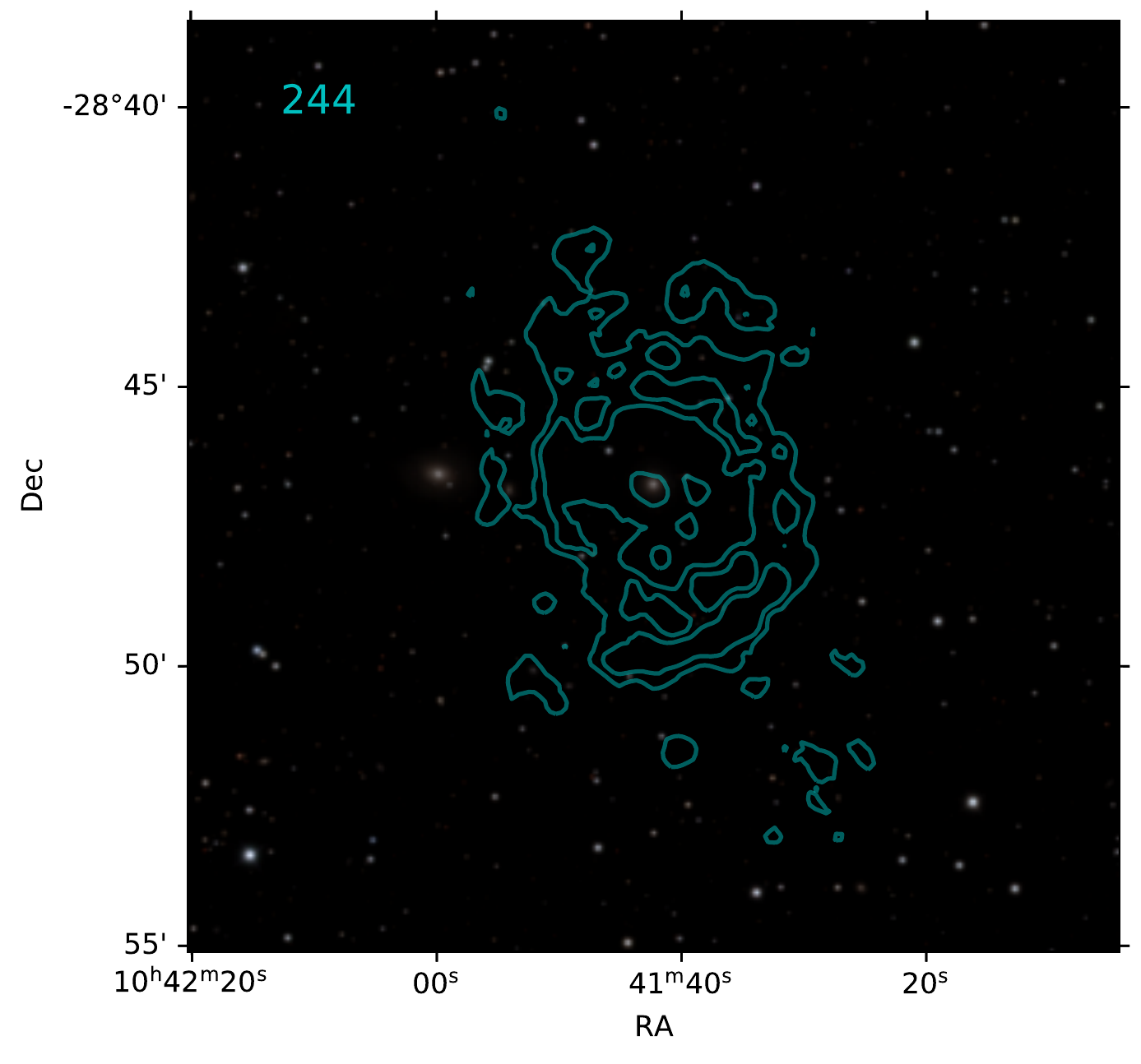}
\includegraphics[width=4cm]{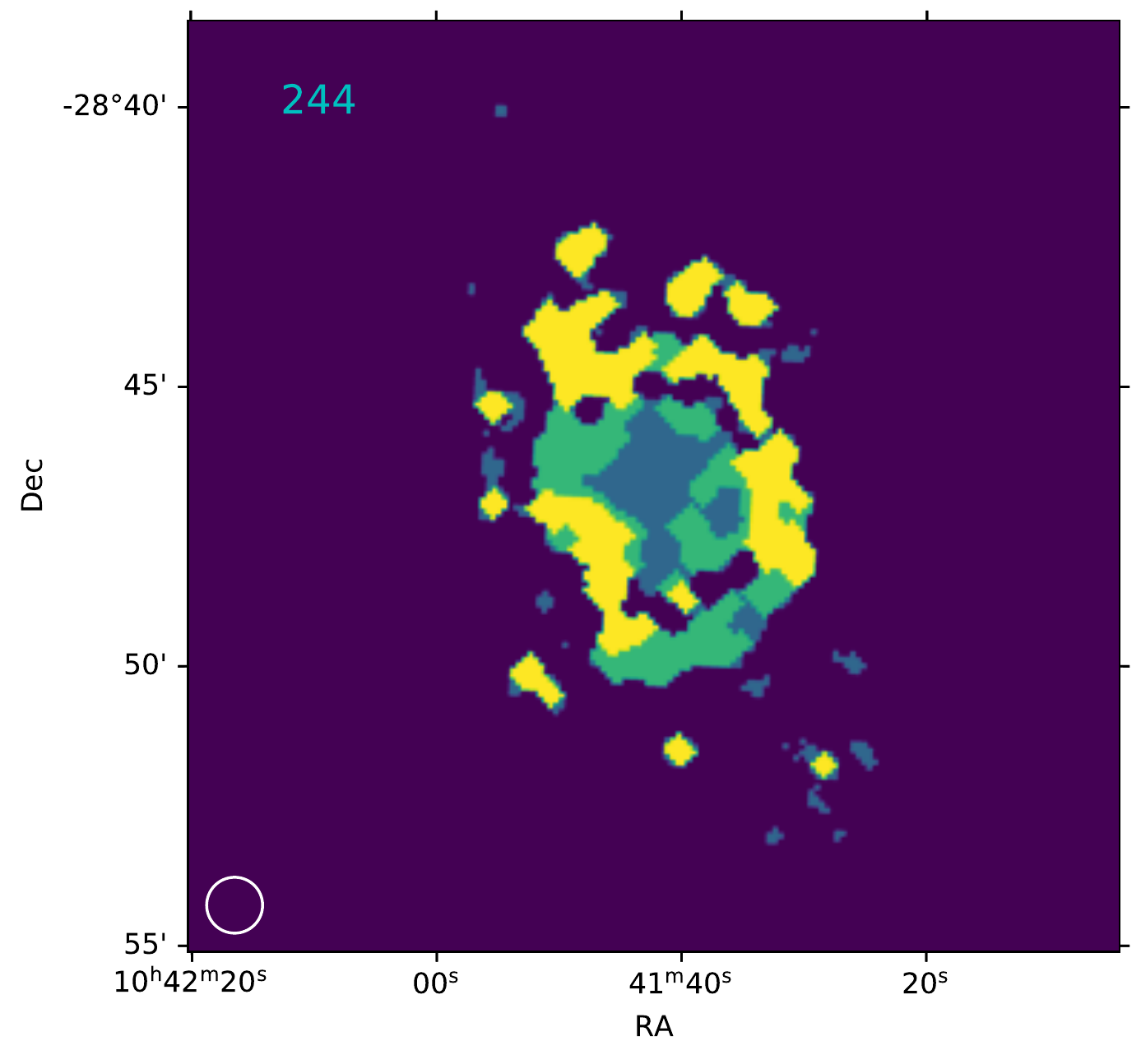}
\includegraphics[width=4cm]{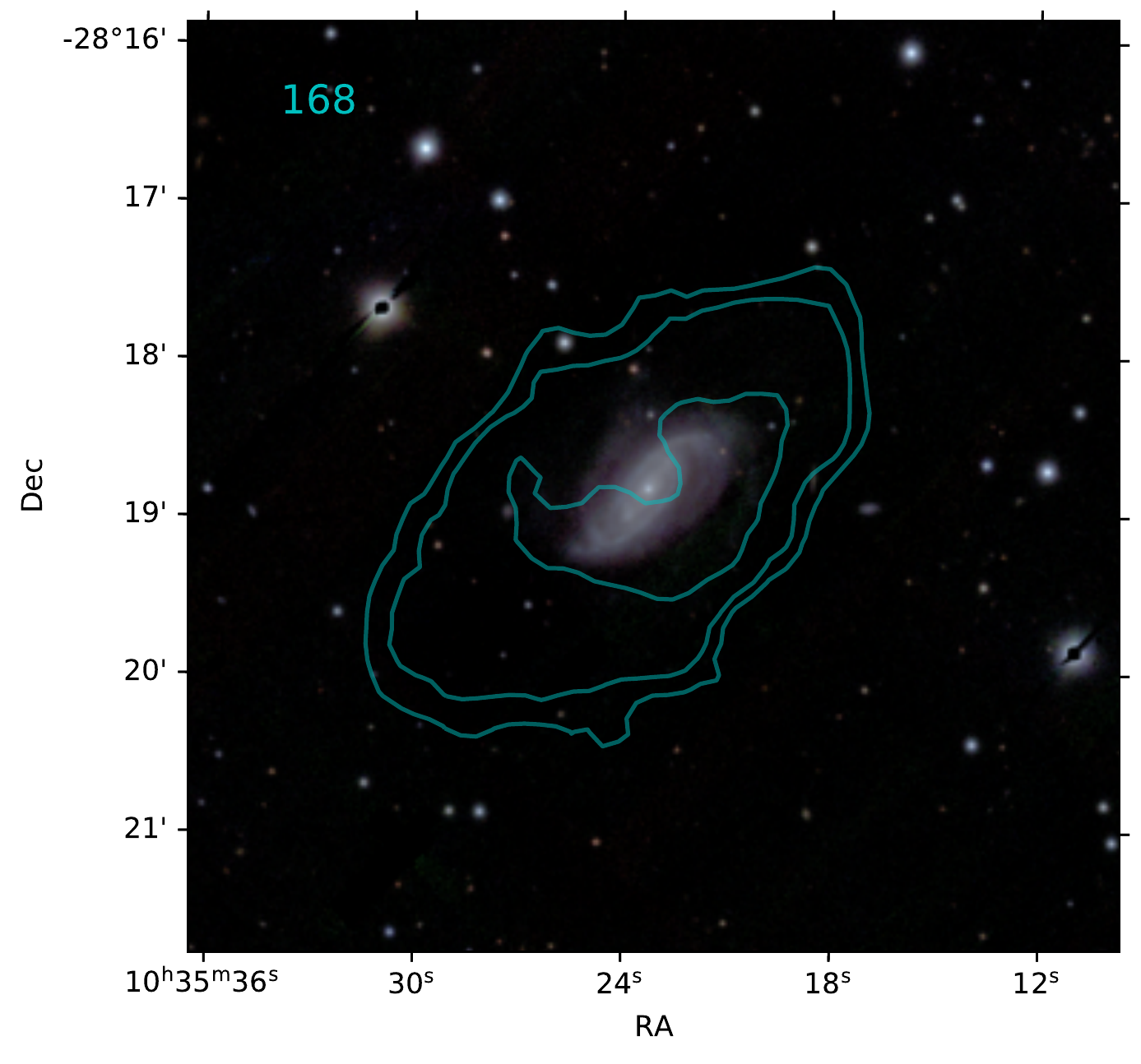}
\includegraphics[width=4cm]{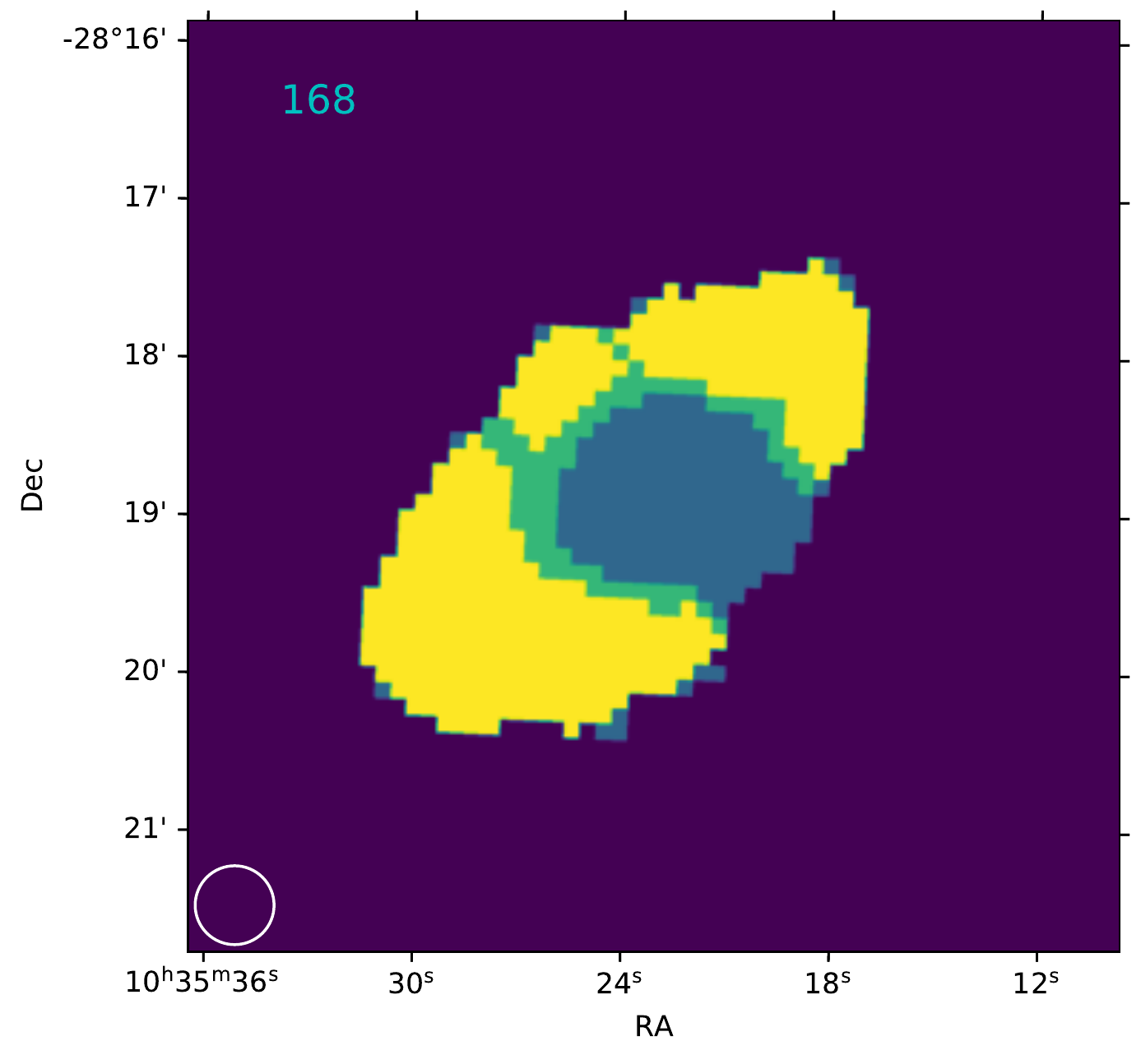}
\includegraphics[width=4cm]{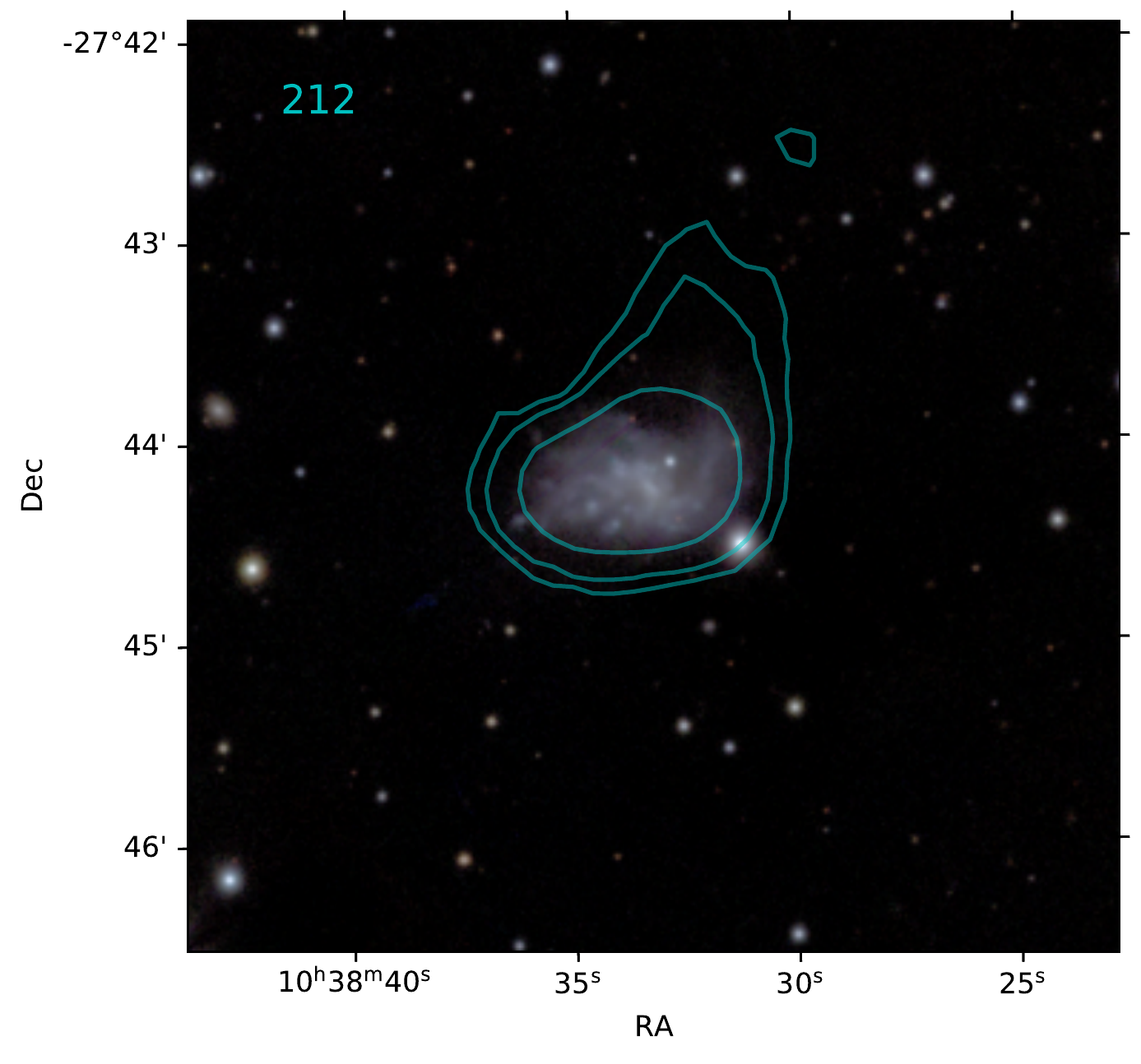}
\includegraphics[width=4cm]{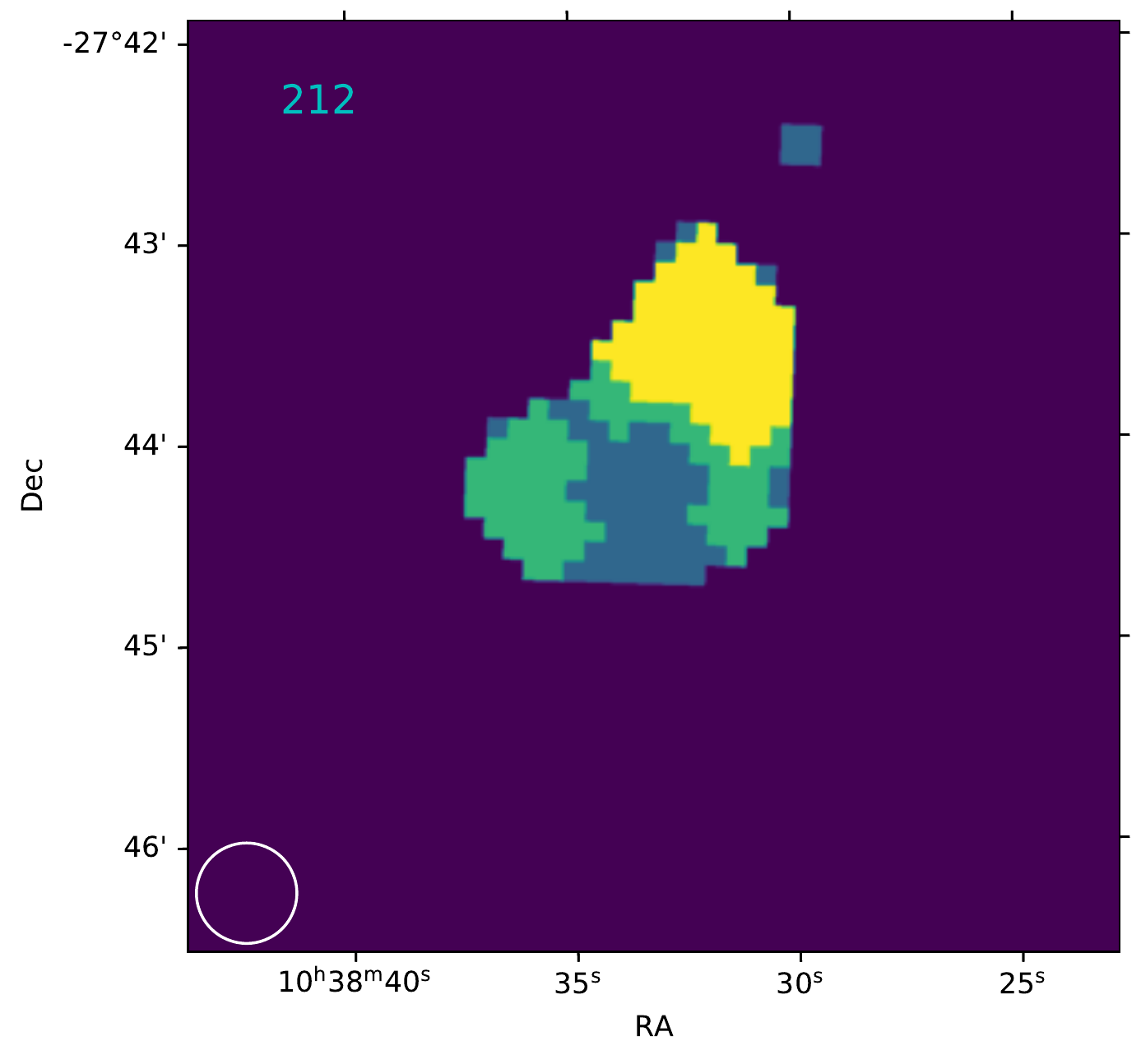}
\includegraphics[width=4cm]{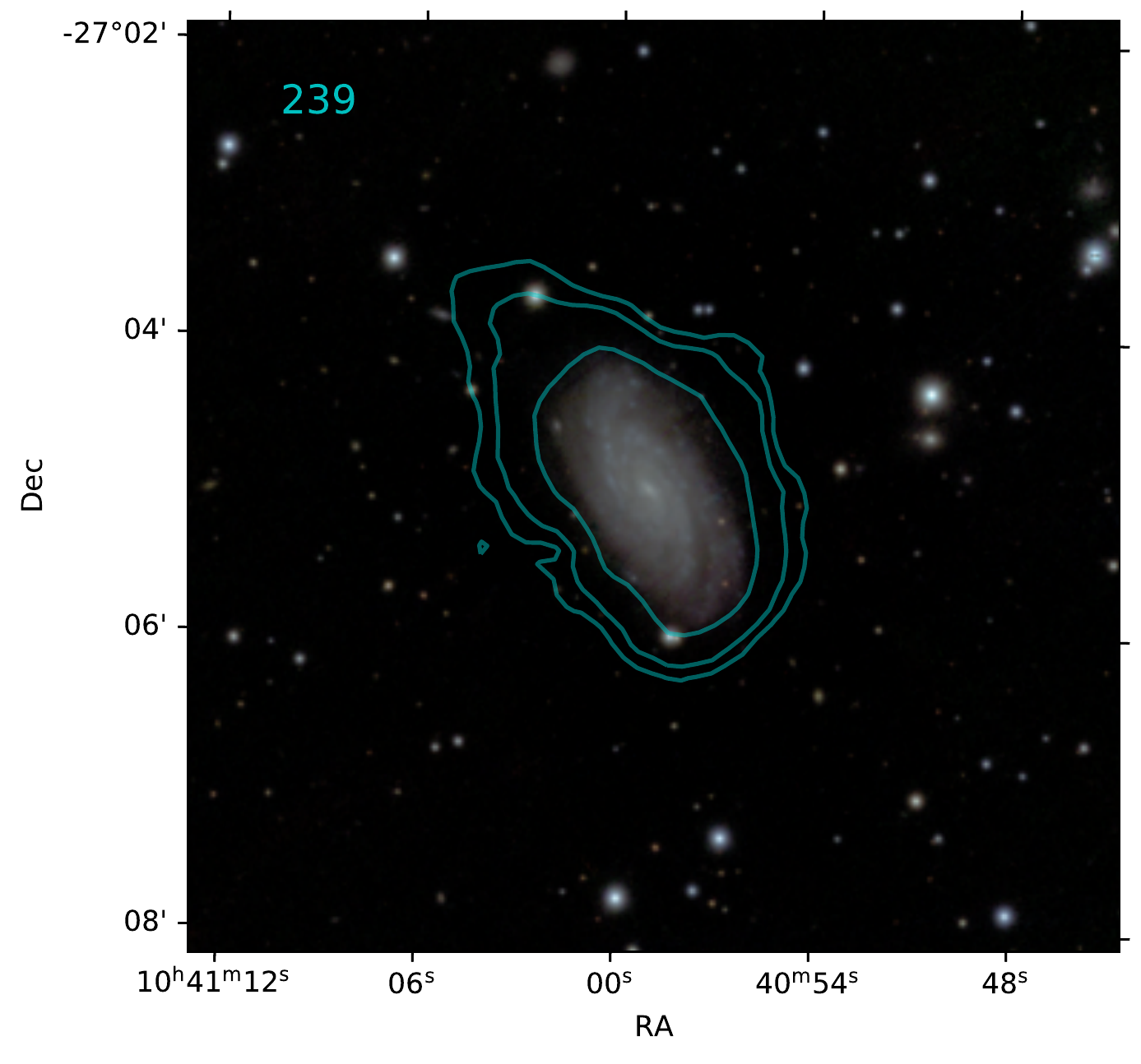}
\includegraphics[width=4cm]{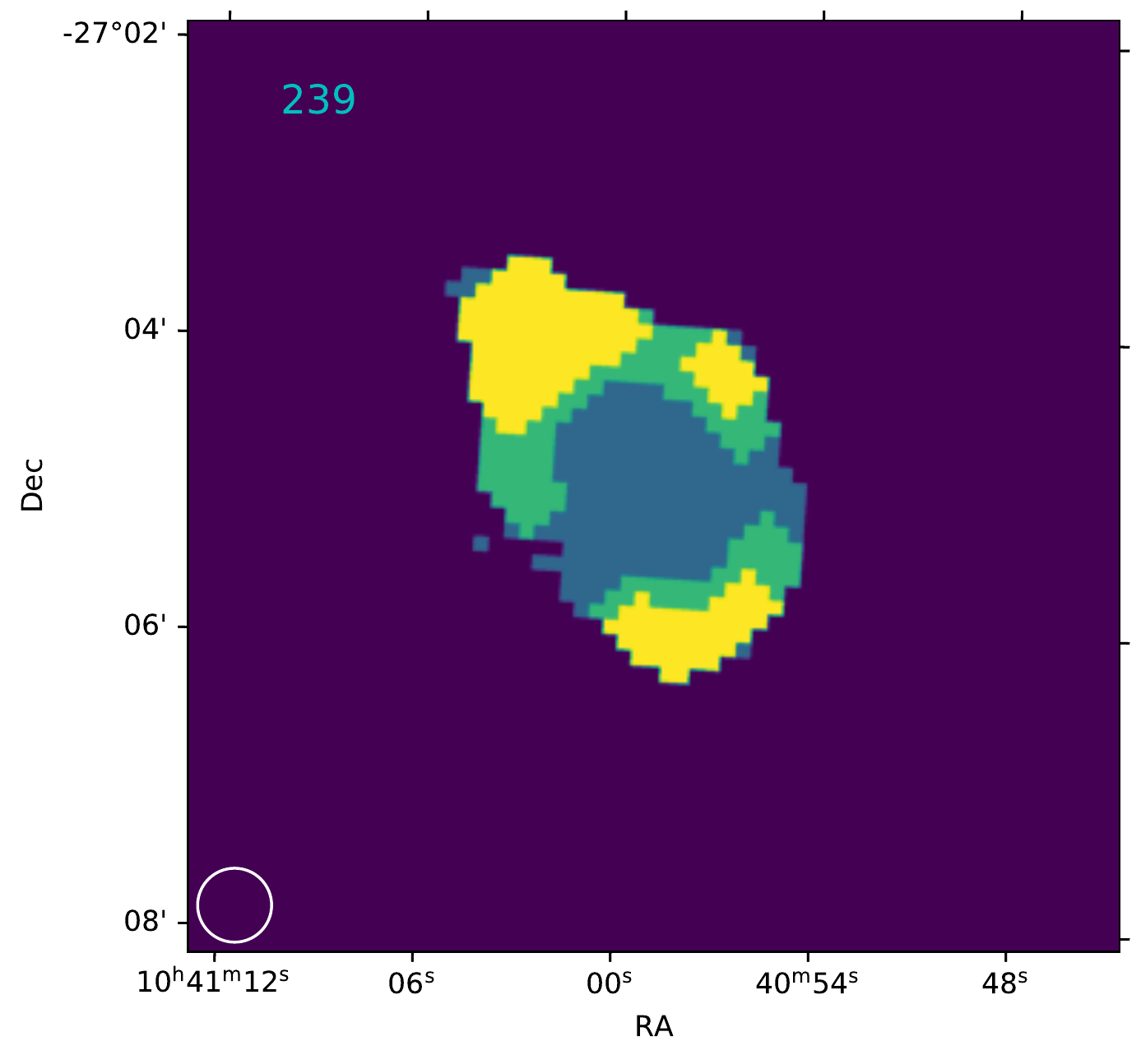}
\includegraphics[width=4cm]{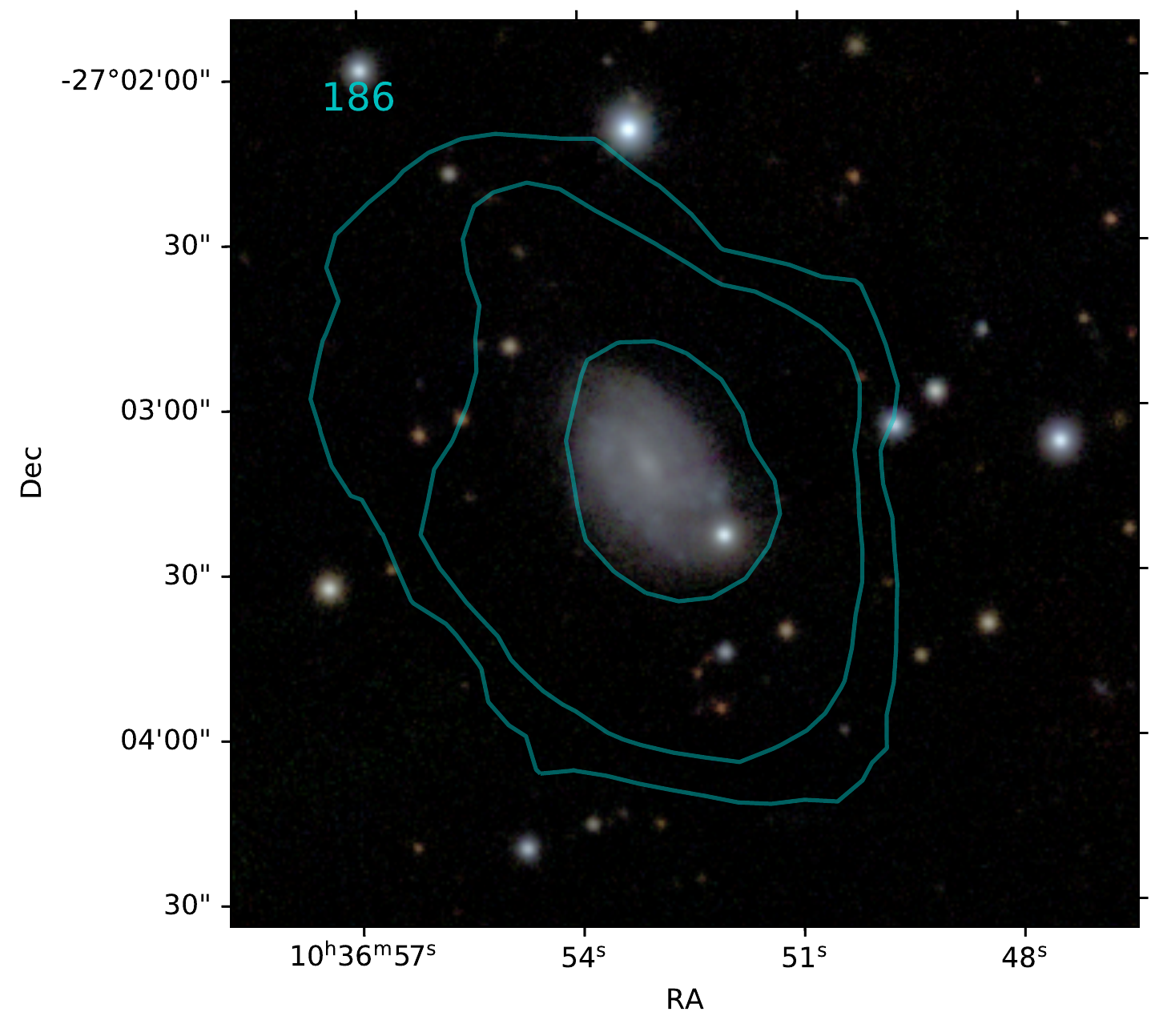}
\includegraphics[width=4cm]{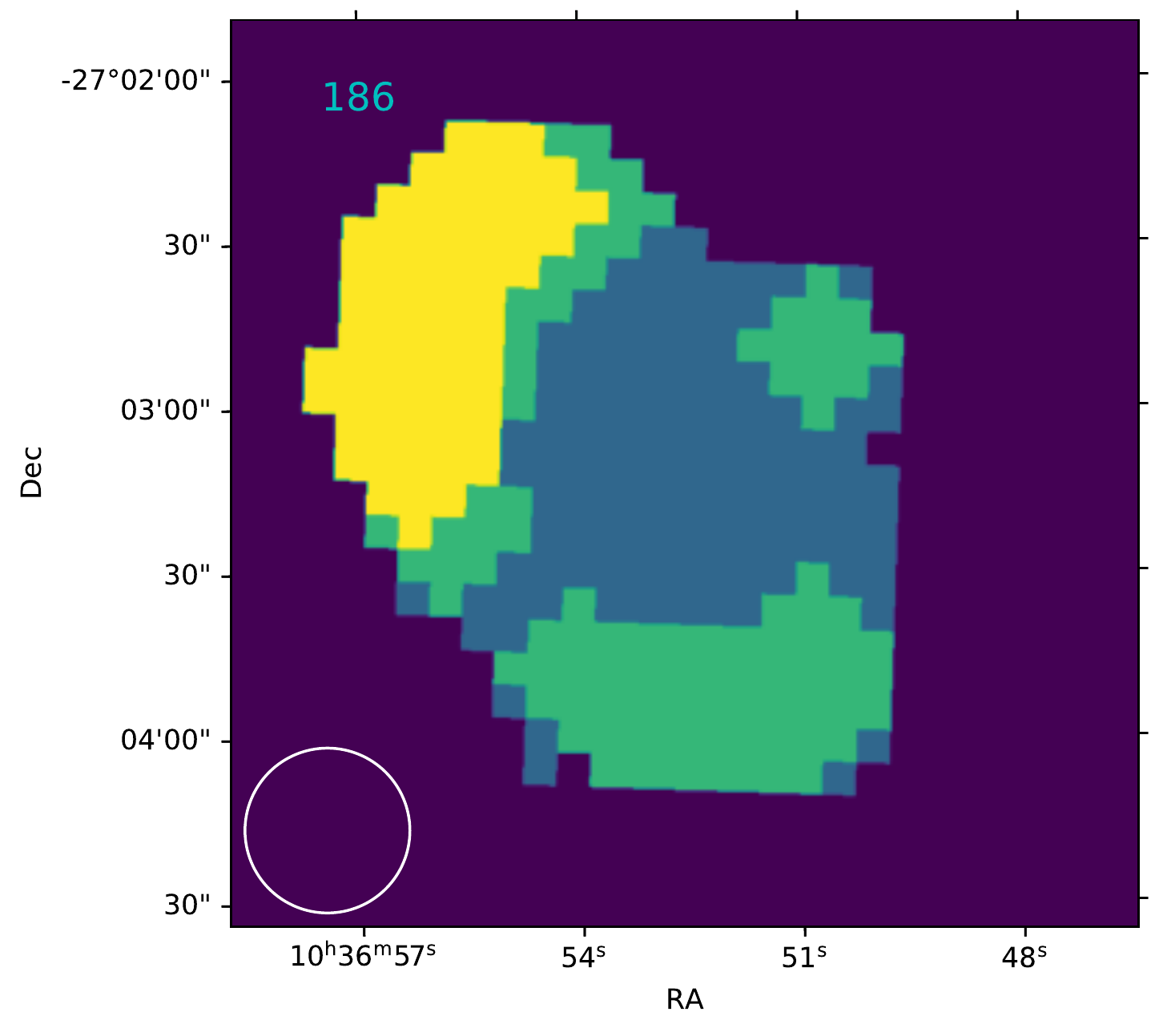}
\includegraphics[width=4cm]{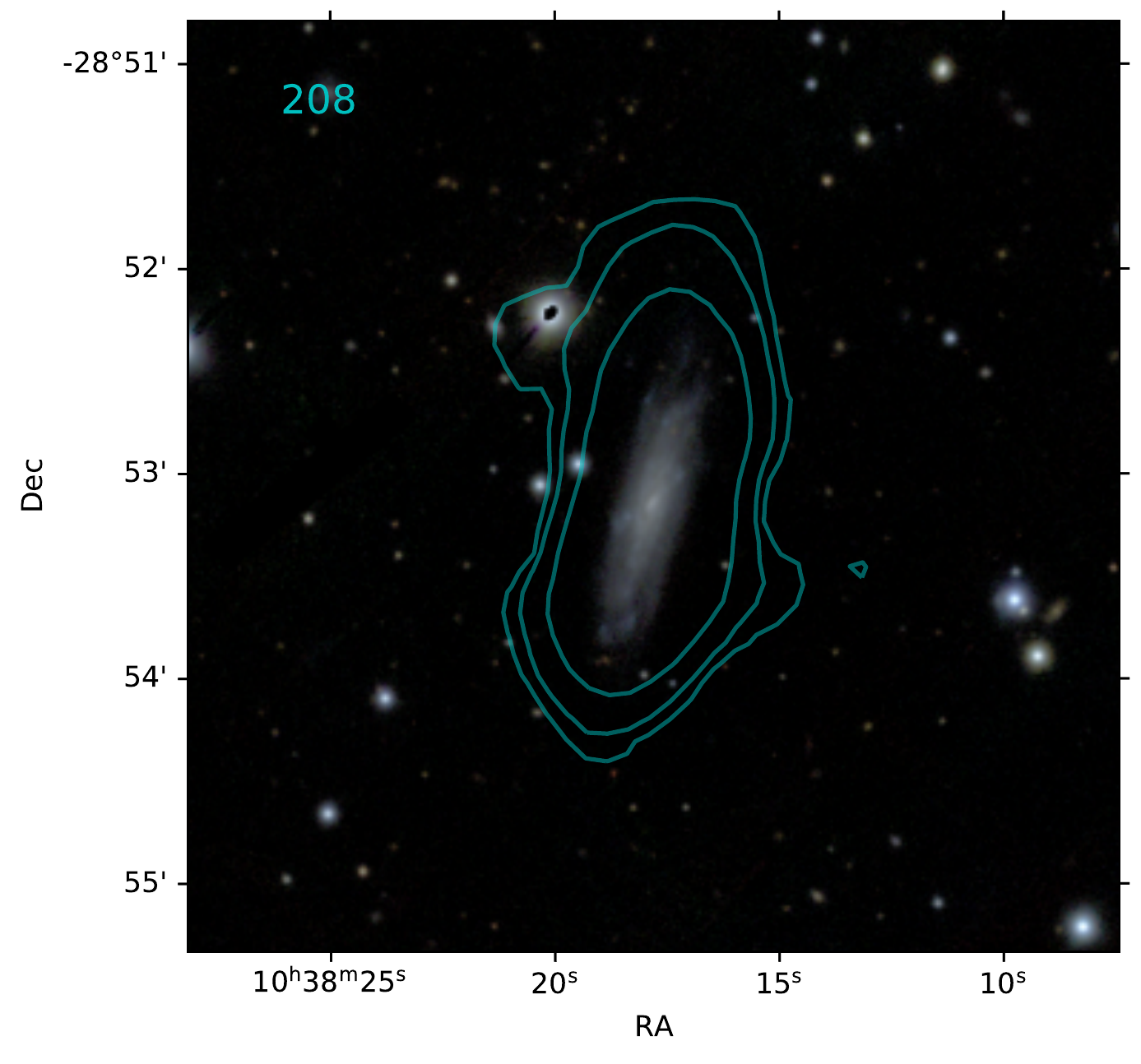}
\includegraphics[width=4cm]{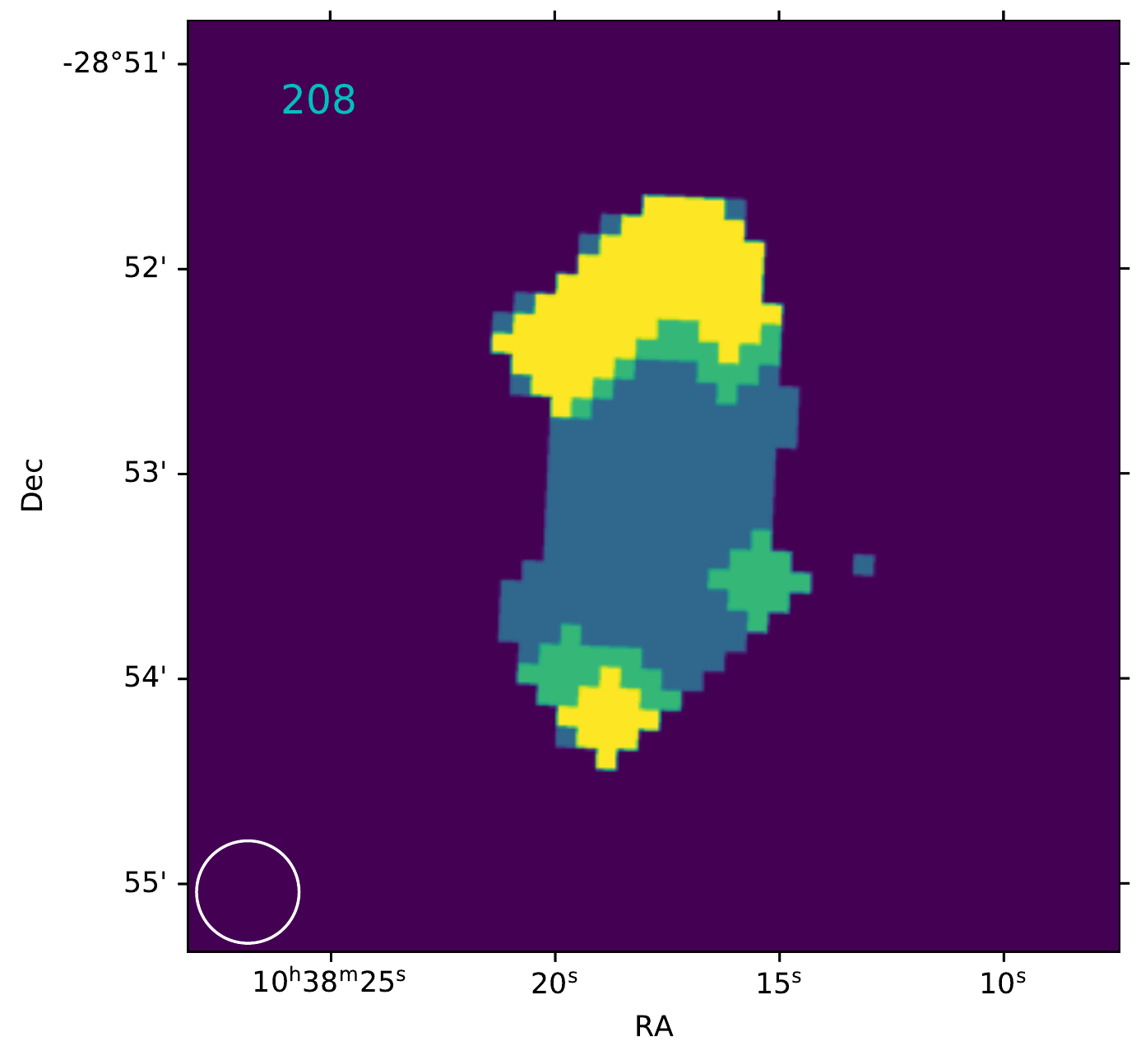}
\includegraphics[width=4cm]{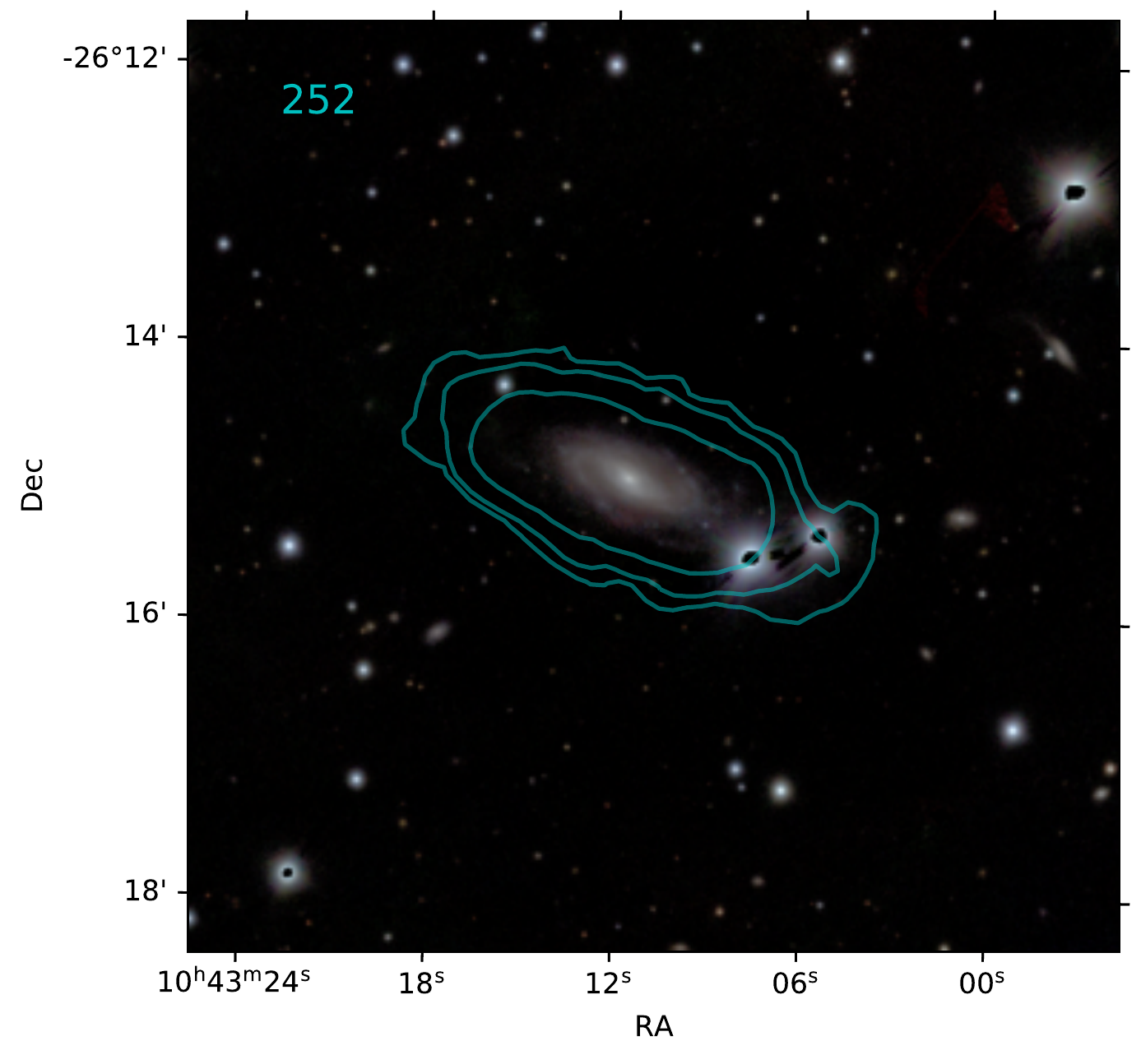}
\includegraphics[width=4cm]{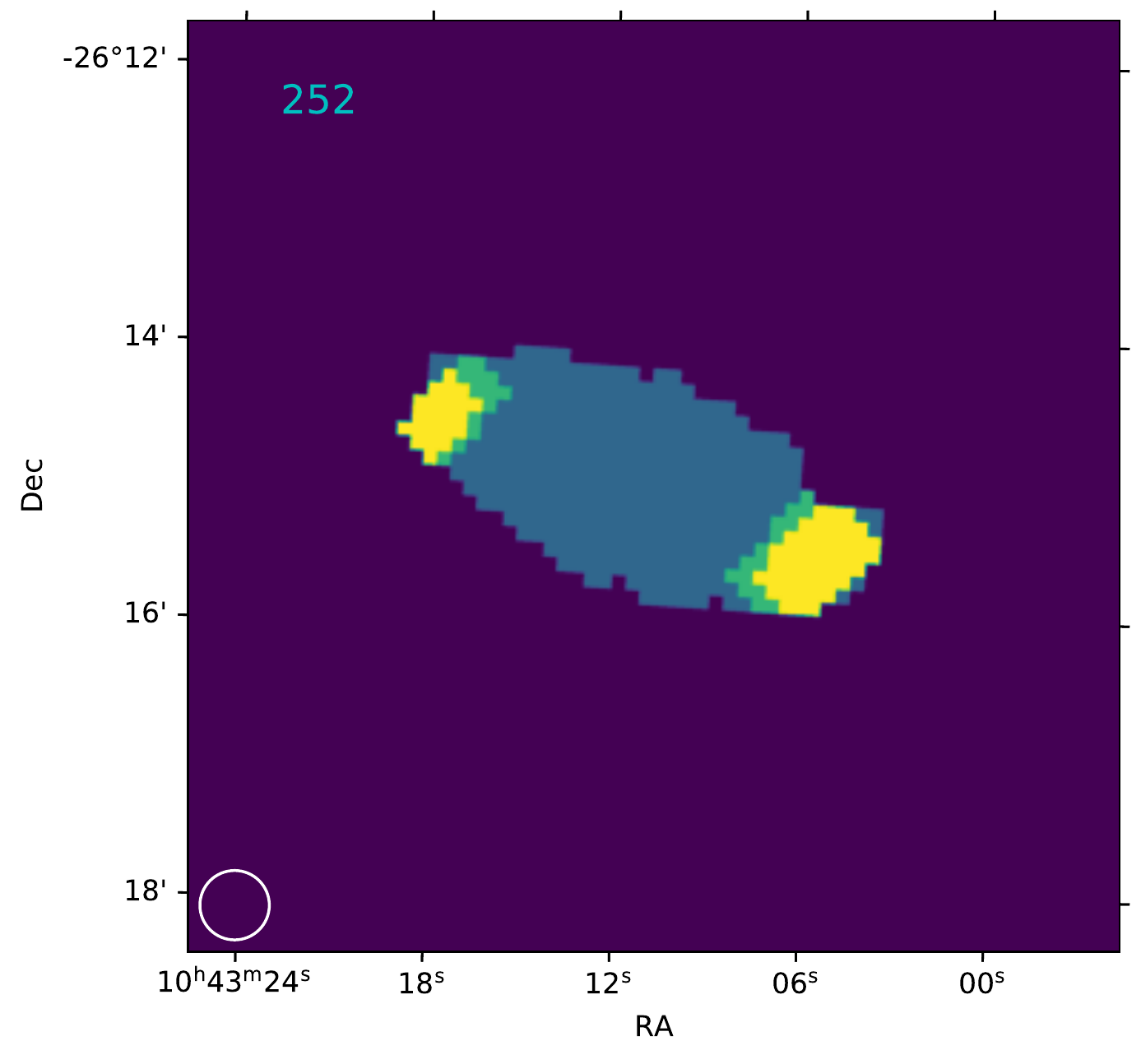}
\caption{  Atlas of resolved RPS candidates. The galaxies are ordered by decreasing fraction of strippable $\hi$, $f_{RPS}$. Each galaxy has a pair of panels: the $\hi$ contours overlaid on the PanSTARRS optical image, and the RPS map. 
In the contour map, the cyan contours mark $\hi$ column density levels of 1, 2 and 5 times 10$^{20}~cm^{-2}$. In the RPS map, the yellow and green colors mark the RPS pixels which have $P_{ram}>F_{anchor}$, and the yellow pixels are those with $P_{ram}>2F_{anchor}$. The blue color marks the ram pressure unaffected pixels. The white circle in the bottom-left corner of each RPS map shows the FWHM of the synthesis beam. }
\label{fig:atlas}
\end{figure*}

We show an atlas of the $\hi$ contours on the resolved RPS candidates (Fig.~\ref{fig:atlas}). We also list the galactic properties used in this study in Tab.~\ref{tab:cluster}.

The strongest, resolved RPS candidate, NGC 3312, (189 in Fig.~\ref{fig:atlas}, WALLABY ID: J103702-273359) was previously identified by its peculiar $\hi$ and blue light distributions (\citealt{McMahon90, Gallagher78}, but see \citealt{McMahon92}). Its neighbor, NGC 3314A also shows $\hi$ tails \citep{McMahon90}, but overlaps with another bright galaxy NGC 3314B, so has been excluded from our analysis. A detailed analysis is under way for this system (Hess et al. in prep.). 
 
One RPS candidate, ESO 501-G065 (212 in Fig.~\ref{fig:atlas}, WALLABY ID: J103833-274357) shows a hint of a tidal tail in the optical with a corresponding $\hi$ tail. It may be a case similar to the Magellanic system where gravitational effects (tidal interactions or harassment) firstly perturb the galaxy and then ram pressure is able to overide the anchoring force in the low density gas produced by the gravitational perturbation. We do not exclude this galaxy as it does not appear to experience a major merger. We leave a statistical study of the combined effects of ram pressure and tidal interactions in clusters to the future.
 
 Detailed analysis of ram pressure stripping in ESO 501-G075 (239 in Fig.~\ref{fig:atlas}, WALLABY ID: J103655-265412) can be found in Reynolds et al. (2021).

\begin{table*}
%\tiny
\centering
{
\begin{tabular}{cccccccccccccc }
ID  & WALLABY ID &  $d_{proj}/r_{200}$ &  $\Delta v_{rad}/\sigma_C$  & $\log M_*/M_{\odot}$  &  $\log M_{\rm HI}/M_{\odot}$ &  flag$_{r1-RPS}$  & flag$_{RPS}$  & $f_{RPS,pred}$ & $f_{RPS}$  \\
(1) & (2) & (3) & (4) & (5) & (6) & (7) & (8) & (9) & (10)  \\

\hline

83 &J102107-281054 &    2.16 &    0.44 &     8.5 &     9.1 &0 &0 &- &-\\
97 &J102411-285533 &    1.91 &    0.25 &     8.9 &     8.9 &0 &0 &- &-\\
102 &J102430-290904 &    1.94 &    0.03 &     8.4 &     8.6 &0 &0 &- &-\\
103 &J102439-244547 &    2.38 &   -0.36 &     8.3 &     9.0 &0 &0 &- &-\\
104 &J102439-274841 &    1.66 &   -0.21 &     6.7 &     8.8 &0 &0 &- &-\\
116 &J102621-291150 &    1.74 &   -0.05 &     8.2 &     8.9 &0 &0 &- &-\\
117 &J102629-285851 &    1.66 &   -0.03 &     7.8 &     8.6 &0 &0 &- &-\\
118 &J102636-245116 &    2.16 &    0.39 &     8.6 &     8.7 &0 &0 &- &-\\
125 &J102818-255446 &    1.52 &    0.12 &     6.8 &     8.3 &0 &0 &- &-\\
127 &J102911-302031 &    2.02 &    0.67 &    10.1 &     9.3 &0 &0 &- &-\\
129 &J102934-261937 &    1.23 &    0.32 &     8.7 &     8.7 &0 &0 &- &-\\
130 &J103002-284116 &    1.16 &    0.17 &     8.6 &     8.6 &0 &0 &- &-\\
131 &J103004-253630 &    1.49 &   -0.53 &     7.4 &     8.5 &0 &0 &- &-\\
134 &J103114-295837 &    1.69 &    0.72 &     8.5 &     8.6 &0 &0 &- &-\\
135 &J103124-295706 &    1.66 &    0.58 &     9.1 &     9.5 &0 &0 &- &-\\
137 &J103139-273049 &    0.69 &   -0.25 &     7.9 &     8.7 &0 &0 &- &-\\
141 &J103241-273137 &    0.55 &    0.14 &     9.2 &     8.5 &0 &0 &- &-\\
142 &J103244-283639 &    0.86 &   -0.17 &    10.5 &     9.3 &0 &0 &- &-\\
143 &J103248-273119 &    0.53 &   -0.19 &     9.1 &     9.0 &0 &0 &- &-\\
144 &J103250-301601 &    1.77 &   -0.41 &    10.2 &     9.3 &0 &0 &- &-\\
146 &J103258-274013 &    0.52 &   -0.94 &     8.6 &     9.0 &1 &0 & 0.59 &-\\
147 &J103259-273237 &    0.51 &    1.63 &     8.7 &     8.5 &1 &1 & 0.79 &-\\
149 &J103335-272717 &    0.43 &   -0.55 &    10.8 &     9.3 &0 &0 &- &-\\
151 &J103353-274945 &    0.43 &   -1.54 &     8.8 &     9.0 &1 &1 & 1.00 &-\\
155 &J103420-264728 &    0.56 &    0.93 &     8.4 &     8.5 &1 &0 & 0.54 &-\\
156 &J103420-265408 &    0.50 &   -0.00 &     7.9 &     8.8 &0 &0 &- &-\\
157 &J103436-273900 &    0.30 &   -1.13 &     9.9 &     9.3 &1 &1 & 0.77 & 0.87\\
160 &J103455-273816 &    0.25 &   -1.12 &     7.5 &     8.5 &1 &1 & 1.00 &-\\
161 &J103459-280440 &    0.42 &   -2.13 &     9.7 &     8.5 &1 &0 & 0.67 &-\\
162 &J103502-293019 &    1.25 &    0.02 &     9.1 &     8.8 &0 &0 &- &-\\
163 &J103507-275923 &    0.36 &   -2.05 &     8.7 &     8.8 &1 &1 & 1.00 &-\\
165 &J103521-272324 &    0.20 &   -0.94 &     7.9 &     8.6 &1 &1 & 1.00 &-\\
166 &J103521-274137 &    0.21 &   -1.33 &     8.3 &     9.0 &1 &1 & 1.00 &-\\
168 &J103523-281855 &    0.52 &   -0.69 &     9.9 &     9.5 &1 &1 & 0.44 & 0.61\\
172 &J103546-273840 &    0.15 &    1.60 &     8.1 &     8.8 &1 &1 & 1.00 &-\\
174 &J103602-261141 &    0.82 &   -0.72 &     8.1 &     8.5 &1 &0 & 0.27 &-\\
176 &J103603-245430 &    1.62 &    0.33 &     7.1 &     8.9 &0 &0 &- &-\\
178 &J103621-252235 &    1.32 &    0.44 &     9.3 &     8.9 &0 &0 &- &-\\
179 &J103627-255957 &    0.94 &   -0.78 &     7.1 &     8.2 &1 &0 & 0.30 &-\\
180 &J103644-251543 &    1.39 &   -0.11 &     8.8 &     8.8 &0 &0 &- &-\\
181 &J103645-281010 &    0.40 &   -0.38 &     8.4 &     9.0 &1 &0 & 0.34 &-\\
182 &J103646-293253 &    1.25 &   -0.26 &     8.7 &     9.0 &0 &0 &- &-\\
183 &J103650-260923 &    0.84 &   -0.12 &     8.9 &     9.1 &0 &0 &- &-\\
186 &J103653-270311 &    0.29 &   -0.29 &     9.0 &     9.1 &1 &1 & 0.34 & 0.43\\
187 &J103655-265412 &    0.38 &   -0.77 &     8.0 &     9.2 &1 &1 & 0.81 & 0.93\\
189 &J103702-273359 &    0.05 &   -1.39 &    10.9 &     9.4 &1 &1 & 1.00 & 0.96\\
190 &J103704-252038 &    1.34 &   -0.03 &     9.1 &     8.9 &0 &0 &- &-\\

\hline
\end{tabular}
}
\caption{ {\bf Galaxy properties.} To be continued.  }
\label{tab:cluster}
\end{table*}

\addtocounter{table}{-1}

\begin{table*}
%\tiny
\centering
{
\begin{tabular}{cccccccccccccc }
ID  & WALLABY ID &  $d_{proj}/r_{200}$ &  $\Delta v_{rad}/\sigma_C$  & $\log M_*/M_{\odot}$  &  $\log M_{\rm HI}/M_{\odot}$ &  flag$_{r1-RPS}$  & flag$_{RPS}$  & $f_{RPS,pred}$ & $f_{RPS}$  \\
(1) & (2) & (3) & (4) & (5) & (6) & (7) & (8) & (9) & (10)  \\
\hline

192 &J103719-281408 &    0.45 &   -0.41 &     8.3 &     8.5 &1 &0 & 0.30 &-\\
194 &J103722-273235 &    0.09 &   -1.54 &     8.0 &     8.6 &1 &1 & 1.00 &-\\
195 &J103725-251916 &    1.36 &   -0.04 &    10.5 &    10.1 &0 &0 &- &-\\
200 &J103738-281216 &    0.44 &    1.54 &     8.1 &     8.5 &1 &1 & 1.00 &-\\
203 &J103804-284333 &    0.77 &    1.13 &     8.4 &     8.8 &1 &0 & 0.55 &-\\
204 &J103805-250537 &    1.51 &    0.35 &     9.0 &     8.8 &0 &0 &- &-\\
205 &J103809-260453 &    0.91 &   -0.21 &     8.7 &     8.4 &0 &0 &- &-\\
207 &J103812-275607 &    0.33 &   -1.38 &     7.6 &     8.3 &1 &1 & 1.00 &-\\
208 &J103818-285307 &    0.87 &    1.01 &     9.2 &     9.5 &1 &1 & 0.41 & 0.26\\
210 &J103821-254126 &    1.15 &   -0.47 &     8.0 &     8.7 &0 &0 &- &-\\
211 &J103828-283056 &    0.66 &    0.98 &     8.1 &     8.8 &1 &0 & 0.55 &-\\
212 &J103833-274357 &    0.29 &    1.06 &     9.3 &     9.2 &1 &1 & 0.93 & 0.56\\
213 &J103840-283405 &    0.70 &   -0.42 &     9.8 &     8.9 &0 &0 &- &-\\
214 &J103841-253530 &    1.22 &    0.18 &     7.6 &     8.9 &0 &0 &- &-\\
215 &J103842-281535 &    0.53 &   -0.51 &     8.1 &     8.6 &1 &1 & 0.30 &-\\
220 &J103902-291255 &    1.09 &   -0.84 &     8.3 &     8.9 &1 &0 & 0.24 &-\\
222 &J103914-271511 &    0.38 &    1.26 &     8.0 &     8.4 &1 &1 & 1.00 &-\\
223 &J103915-301757 &    1.75 &    0.07 &    10.4 &     9.8 &0 &0 &- &-\\
225 &J103922-293505 &    1.32 &    0.23 &     9.3 &     9.3 &0 &0 &- &-\\
226 &J103924-275442 &    0.44 &   -0.80 &     9.0 &     8.9 &1 &0 & 0.51 &-\\
228 &J103927-271653 &    0.40 &   -0.58 &     8.3 &     8.6 &1 &0 & 0.51 &-\\
229 &J103939-280552 &    0.54 &   -0.53 &     8.1 &     8.8 &1 &0 & 0.37 &-\\
231 &J103958-301130 &    1.71 &   -0.70 &    10.1 &     9.1 &0 &0 &- &-\\
232 &J104000-292445 &    1.25 &    0.06 &     8.3 &     9.2 &0 &0 &- &-\\
233 &J104004-301606 &    1.75 &   -0.53 &     9.5 &     9.4 &0 &0 &- &-\\
234 &J104016-274630 &    0.51 &    0.38 &    10.1 &     9.8 &1 &0 & 0.25 &-\\
235 &J104026-274853 &    0.54 &    0.96 &     7.7 &     8.7 &1 &1 & 0.71 &-\\
236 &J104048-244003 &    1.85 &   -0.11 &     8.3 &     9.0 &0 &0 &- &-\\
238 &J104058-274546 &    0.60 &    0.40 &     8.1 &     8.8 &1 &0 & 0.22 &-\\
239 &J104059-270456 &    0.64 &    1.57 &    10.1 &     9.5 &1 &1 & 0.56 & 0.44\\
240 &J104100-284430 &    0.95 &    0.03 &     9.1 &     8.9 &0 &0 &- &-\\
242 &J104139-254049 &    1.32 &    0.19 &     7.8 &     8.7 &0 &0 &- &-\\
243 &J104139-274639 &    0.69 &    0.98 &     8.0 &     8.7 &1 &0 & 0.55 &-\\
244 &J104142-284653 &    1.03 &    1.10 &    10.0 &    10.0 &1 &1 & 0.41 & 0.67\\
246 &J104221-291748 &    1.34 &    1.09 &     8.1 &     9.0 &1 &0 & 0.27 &-\\
251 &J104309-300301 &    1.79 &   -0.85 &     9.1 &     9.6 &0 &0 &- &-\\
252 &J104311-261500 &    1.19 &    1.26 &     9.9 &     9.5 &1 &1 & 0.28 & 0.10\\
253 &J104326-251857 &    1.65 &    0.05 &     8.8 &     8.7 &0 &0 &- &-\\
256 &J104359-293304 &    1.59 &   -0.61 &     8.5 &     8.4 &0 &0 &- &-\\
266 &J104629-253308 &    1.82 &   -0.21 &     7.9 &     8.8 &0 &0 &- &-\\
269 &J104824-250944 &    2.17 &    0.03 &    10.7 &     9.7 &0 &0 &- &-\\
270 &J104905-292232 &    2.03 &    0.85 &     8.5 &     9.3 &0 &0 &- &-\\
272 &J105227-291155 &    2.37 &   -0.46 &     7.6 &     8.7 &0 &0 &- &-\\

\hline
\end{tabular}
}
\caption{ {\bf Galaxy properties.} Column (1): id. Column (2): WALLABY identifier. Column (3): the cluster centric projected distance. Column (4): the velocity distance from cluster center. Column (5): stellar mass. Column (6): $\hi$ mass. Column (7): flag for identification of r1-RPS candidates (the major sample studied in Sec. 3 and 4), 1 for True and 0 for False. Column (8): flag for identification of RPS candidates (both resolved and un-resolved), 1 for True and 0 for False. The identification of unresolved RPS candidates is incomplete. Column (9): the fraction of $\hi$ mass under ram pressure stripping derived from predicted $\hi$ radial profiles. Column (10): the fraction $\hi$ mass under ram pressure stripping derived from $\hi$ moment-0 maps for the resolved RPS candidates.  }
\label{tab:cluster}
\end{table*}

\end{CJK*}
\end{document}